\documentclass[dvipdfm]{USC-Thesis}
\usepackage{amsmath}
\usepackage[dvips]{graphicx}            
\usepackage{pslatex}
\usepackage{bm}
\usepackage{amssymb}
\usepackage{empheq}
\usepackage{natbib}		
\usepackage[titles]{tocloft}	

\renewcommand{\v}[1]{\ensuremath{\mathbf{#1}}} 
\newcommand{\gv}[1]{\ensuremath{\mbox{\boldmath$ #1 $}}} 
\renewcommand{\d}[2]{\frac{d #1}{d #2}} 
\newcommand{\pd}[2]{\frac{\partial #1}{\partial #2}} 
\newcommand{\pdd}[2]{\frac{\partial^2 #1}{\partial #2^2}} 
\newcommand{\pdpd}[3]{\frac{\partial^2 #1}{\partial #2 \partial #3}} 
\newcommand{\grad}[1]{\gv{\nabla} #1} 
\newcommand{\ldq}{\textquotedblleft}        	
\newcommand{\rdq}{\textquotedblright }	    	
\newcommand{\expv}[1]{\left\langle #1 \right\rangle}	

\newcommand{\m}{\item}        			

\newcommand{\addcite}[1]{[$\clubsuit$ \textbf{Add citation: #1 }]}	
\newcommand{\sidenote}[1]{} 

\newtheorem{theorem}{Theorem}[section]
\begin{document}

\title{Chaos and thermalization in the one-dimensional Bose-Hubbard model in the classical-field approximation}

\author{Amy C. Cassidy}
\major{Physics}
\month{May}
\year{2009}
\maketitle

\topmatter{Acknowlegements}

There are a many people without whose support over the last six years I would not have completed this project. First of all, I would like to thank my advisor, Maxim for all that he has taught me, for his commitment to my development and for space to grow. Others whom I would like to acknowledge are: Vanja, for asking questions; Tameem, Richard and Yi-Chun, my classmates who got me through the first year; Ilya, for guidance at a critical time; Noah and Marcos, for physics conversations and tennis matches; Katie, for countless hours working together to encourage more women to pursue physics; Sam, for perspective; Courtney, for a lifelong friendship; Karla, for reminding me to stay balanced; Stephan, for getting me started and continued support; Werner and the department at USC, for flexibility; Bala and the department at UMass Boston, for welcoming me these last two years.  Finally I would like to thank my family: my father who encouraged my intellectual curiosity; my mother, who taught me to be organized and persevere; Sara and Andrew for many words of wisdom; and Linda for encouragement and support.

\vspace*{8em}
\paragraph{\textbf{Note:}} This document contains minor corrections and revisions from the original.

\tableofcontents

\listoftables

\listoffigures

\topmatter{Abstract}  
%
%

One of the fundamental assertions of statistical mechanics is that the time average of a physical observable is equivalent to the average over phase space, with microcanonical measure. A system for which this is true is said to be ergodic and dynamical properties can be calculated from static phase-space averages. Dynamics of a system which is fully integrable, that is has as many conserved quantities as degrees of freedom, is constrained to a reduced phase space and thus not ergodic, although it may relax to a modified equilibrium.  

In this thesis, we present a comprehensive study of chaos and thermalization of the one-dimensional Bose-Hubbard Model (BHM) within the classical field approximation.  This model describes the dynamics of quantum degenerate gases in a lattice for sufficient occupation of every momentum mode and weak two-body scattering, and is of interest because of experimental advances of cooling and trapping alkali atoms in the quantum degenerate regime.

We study chaos and its relation to thermalization. Two quantitative measures are compared:  the ensemble-averaged Finite-time Maximal Lyapunov exponent, a measures of chaos and the normalized spectral entropy, a measure of the distance between the numerical time-averaged momentum distribution and the one predicted by thermodynamics.  A threshold for chaos is found, which depends on two parameters, the nonlinearity and the total energy-per-particle.  Below the threshold, the dynamics are regular, while far above the threshold, complete thermalization is observed, as measured by the normalized spectral entropy. 

We study individual resonances in the Bose-Hubbard model to determine the criterion for chaos. The criterion  based on Chirikov's method of overlapping resonances diverges in the thermodynamic limit, in contrast to the criterion parameters inferred from numerical calculations, signifying the failure of the standard Chirikov's approach.

The Ablowitz-Ladik lattice is one of several integrable models that are close to the BHM.  We outline the method of Inverse Scattering Transform and generate the integrals of motion of the Ablowitz-Ladik lattice.  Furthermore, we discuss the possible role of these quantities in the relaxation dynamics of the BHM. 


\mainmatter         
\chapter{Introduction}\label{sec:main_introduction}


\section{Introduction to Dynamical Systems}\label{sec:dynamical_systems}


The study of thermalization in nonlinear systems dates back to the early numerical studies of coupled anharmonic oscillators by Fermi, Pasta, and Ulam (FPU).  At the time it was expected that any amount of nonlinearity, no matter how small, in a system with many degrees of freedom would lead to thermal behavior.  Quite unexpectedly, thermalization was not observed.  Later work found a threshold in the nonlinear coupling strength, with energy equipartition occurring above the threshold. The absence of thermalization for small nonlinearities has been explained in terms of the presence of a threshold for the onset of widespread chaos determined by Chirikov's criterion of overlapping resonances and also in terms of the closeness to a fully integrable model, the Korteweg-de-Vries equation. Additional studies on thermalization and approach to equilibrium have been carried out in many classical field theories.


Fermi, Pasta, Ulam (FPU) and Tsingou performed one of the first numerical experiments in the 1950's \citep{fermi_studies_1974}.  Their discovery led to significant work and important discoveries in the field of nonlinear dynamics. The motivation for the experiment was to study the approach to equilibrium of a nonlinear system.  The contemporary expectation was that an system with a large number of degrees of freedom would exhibit thermal behavior in the presence of any non-linearity, no matter how small.  In particular they were looking to observe the Fourier heat law for conduction. They numerical integrated a system of one-dimensional anharmonically coupled oscillators with quadratic and cubic forces. 

The results found by FPU were completely suprising. In order to understand the prevalent view at the time and the significance of their results, let us take some time to review important concepts in Hamiltonian systems and statistical mechanics.

\section{\label{sec:dynamical_sys}Dynamical Systems}
Given a system whose state is completely known at some time $t_0$, what can be said about the state at some future time $t>t_0$? What about its state at some time in the past, $t<t_0$?  Classical mechanics addresses these questions. We begin with a short review of Hamiltonian dynamics, referring to \citet{tabor_chaos_1989}. 

\sidenote{Start with single degree of freedom and move to more degrees of freedom? or start with many degrees of freedom?}

\paragraph{\label{sec:hamiltonian_sys}Hamiltonian Systems}
Consider a generic Hamiltonian of an N-particle system, $H(\{q_i,p_i\})$ where $\{q_i\}$ are the coordinates of the N-particles and $\{ p_i\}$ are the corresponding momentum.  The equations of motion are given by
\begin{equation}
\dot{q}_i= - \pd{H}{p_i}, \qquad \dot{p}_i = \pd{H}{q_i}.
\end{equation}
These equations, called the Hamilton's equations of motion, uniquely determine the time evolution of the state, from the initial state, which is specified by a complete set of values $ \{q_i,p_i\}$. Note that canonical coordinates and momentum satisfy Liouville's theorem, so that a volume element in phase space moves like an incompressible fluid under the Hamiltonian flow. 
\paragraph{\label{sec:poisson_brackets}Poisson Brackets}
For any dynamical quantities $f,g$ in a system with canonical coordinates $\{q_i,p_i\}$, the Poisson brackets are defined as 
\[ \{f,g\} \equiv \sum_i \pd{f}{p_i}\pd{g}{q_i} - \pd{f}{q_i}\pd{g}{p_i}. \]

From this one can write the time dependence of a function $f(p,q,t)$ as 
\begin{align*}
 \d{f}{t} &=\sum_i \left(\pd{f}{q_i}\pd{q_i}{t} + \pd{f}{p_i}\pd{p_i}{t}  \right) + \pd{f}{t}\\
	  &=\sum_i \left(\pd{f}{q_i}\pd{H}{p_i} - \pd{f}{p_i}\pd{H}{q_i}  \right) + \pd{f}{t}\\
	  &=\{H,f\}+ \pd{f}{t}.
\end{align*}
Thus any quantity that does not explicity depend on time is a constant of motion if the Poisson bracket of that quantity vanishes.  The Poisson brackets are antisymmetric and satisfy the identity
\[ \{f,gh\} = g\{f,h\} + \{f,g\}h. \]

\paragraph{\label{sec:canonical_trans}Canonical Transformations}
Oftentimes it is more convenient to work in one coordinate system than another.
\sidenote{For example, consider a particle moving in a spherically symmetric potential.  It is more convenient to describe the system in the spherical coordinates $(r,\phi,\theta)$ rather than the Cartesian coordinates $(x,y,z)$. Maybe this is a bad example, because it is not canonical.  Two harmonic oscillators?  Or two body problem - center of mass motion vs. relative motion}
A canonical transformation is one in which the canonical form of Hamilton's equations is preserved. \sidenote{Symplectic structure} The phase volume is preserved under a canonical transformation.  Thus the Jacobian, must be unity.  Consider a transformation from one set of phase-space variables $(p_i,q_i)$ to a new set of variables $(P_i,Q_i)$ the preservation of phase volume is expressed as
\[\int dp_1 dq_1\int dp_2 dq_2 ...\int dp_N dq_N = \int dP_1 dQ_1\int dP_2 dQ_2 ...\int dP_N dQ_N  \]
and the Jacobian must satisfy the condition
\[ \pd{(p_1,...,p_N,q_1,...,q_N)}{(P_1,...,P_N,Q_1,...,Q_N)} = \pd{(P_1,...,P_N,Q_1,...,Q_N)}{(p_1,...,p_N,q_1,...,q_N)}= 1\]
\citep{tabor_chaos_1989}.

One of the representations of the canonical transformation, is through the type 2 generating function $\Phi(\theta, \tilde{I})$, which transforms $I,\theta \rightarrow \tilde{I},\tilde{\theta} $.  The Hamiltonian is transformed according 
to  
\begin{equation}\begin{split}
  \widetilde{H}(\tilde{I},\tilde{\theta}) =& H(I,\theta) + \frac{\partial \Phi}{\partial t}\\
  I=&\frac{\partial \Phi}{\partial\theta} \\
  \tilde{\theta}=&\frac{\partial \Phi}{\partial \tilde{I}}.
\end{split}\end{equation}

\sidenote{\paragraph{Action angle variables and Hamilton-Jacobi equation}}

\subsection{\label{sec:integrable}Integrable Systems}
An integrable system has as many independent constants of the motion as degrees of freedom. Stated another way, a systems with $n$ degrees of freedom is completely integrable is there exist $n$ integrals of motion which are in involution.
\begin{align*} 
\{F_i,H\}   &=0, \qquad i=1\cdots n, \phantom{j=1\cdots n} 	\qquad \{ F_i \} \text{ are constants of the motion}\\
\{F_i,F_j\} &=0, \qquad i=1\cdots n, j=1\cdots n  		\qquad \{ F_i \} \text{ are in involution}
\end{align*}

Ford makes the distinction that these must be in fact well-behaved integrals of motion \citep{ford_fermi-pasta-ulam_1992}.  All one dimensional systems are integrable \citep{tabor_chaos_1989}. Examples of integrable systems with many degrees of freedom include the Toda lattice (infinite-dimensional countable), the Korteweg-de-Vries equation, sine-Gordon model and continuous nonlinear Schr\"odinger equation (infinite-dimensional continuous). 

\subsection{\label{sec:chaos}Chaos}

The term chaos is commonly used to describe systems where the motion appears random, erratic or unpredictable.  It is sometimes called \ldq deterministic chaos'' to emphasize that it refers to dynamical systems in which the motion is governed by deterministic equations of motion. A key feature of chaotic systems is extreme sensitivity to initial conditions.  Imagine two identical systems that are governed by the same equations of motion and with initial conditions that are only slightly different.  In a chaotic system, the initial different will grow rapidly and after some time the states of the two systems will be entirely different from one another.  In contrast, in a regular system, the difference will grow slowly, so that two remain highly correlated. The difference between chaotic and regular motion is not only quantitative, but qualitative. The very functional form of the divergence is different: chaotic trajectories diverge exponentially while regular trajectories diverge linearly.

\paragraph{Lyapunov Exponents}
A chaotic system is characterized by local instability, specifically by exponential divergence of trajectories neighboring in phase space.  This rate of divergence of is measured by the Lyapunov exponent. The maximal Lyapunov Exponent (MLE) of a system is defined as 
\begin{equation}
 \lambda(\mathbf{x}_{0})= \lim_{t\rightarrow \infty}
\lim_{\widetilde{\mathbf{x}}_{0}\to\mathbf{x}_{0}}\;
\frac{1}{t}\ln \frac{\|\widetilde{\mathbf{x}}(t)-\mathbf{x}(t)\|}
{\|\widetilde{\mathbf{x}}_{0}-\mathbf{x}_{0}\|} \nonumber
\end{equation}
where $\mathbf{x}(t)$ and $\widetilde{\mathbf{x}}(t)$ are two phase space trajectories \citep{tabor_chaos_1989}.
Chaotic motion is characterized by positive MLE, $\lambda > 0$, in which case trajectories that are initially close in phase space will diverge exponentially.  On the other hand, a zero MLE, $\lambda = 0$, indicates regular motion and linear divergence.
The Lyapunov exponent is a function of an initial state and is thus a measure of local chaos.  A system may have a mixed phase space, consisting of both chaotic and regular regions \citep{zaslavsky_chaotic_1999}.  Thus initial states in the regular regions will have zero Lyapunov exponent while initial states in chaotic regions have a positive Lyapunov exponent.  As a system transitions from regular to globally chaotic behavior, the volume of phase space corresponding to chaotic regions grows while the regular regions shrink.  When the chaotic regions dominate the phase space, there can still exist regular regions, called islands of stability, which are surrounded by the chaotic sea.

\section{\label{sec:stat_mech}Statistical Mechanics}

Thermodynamics is a phenomenological theory that has very successfully described the equilibrium properties of isolated systems with many degrees of freedom. In this section we give an overview of basic concepts of statistical mechanics.
  The field arose in the study of macroscopic systems of atoms with $N\sim 10^{23}$ particles and correspondingly large volumes. One of the great success of thermodynamics is the Maxwell-Boltzmann distribution for ideal gases.  It can be derived in two ways: (1) through the Boltzmann transport equation and (2) through the most probable distribution.  Systems are described not by position and momentum of ever particle in the system, but by probability distribution functions from which macroscopic properties such as pressure and temperature and volume can be determined.  The aim of statistical mechanics is to derive the laws of thermodynamics from molecular dynamics. 

For a system with $N$ particles with canonical coordinates $q_1, q_2, ..., q_{3N}$ and conjugate momenta $q_1,q_2,...,q_{3N}$, the $6N$ dimensional space spanned by these coordinates is the \textit{phase space} or $\Gamma$ space.  A \textit{representative point} is a point in this $\Gamma$ space that specifies the position and momentum of each of the $N$ particles.  The representative point will also be called a \textit{microstate}.  For each microstate, there is a corresponding macrostate that specifies properties of  the system such as density, temperature, and pressure. Many microstates and in fact, an infinite number, correspond to the same macrostate.

The dynamics of the individual $N$ particles is governed by the Hamiltonian, $H(p,q)$ where $(p,q) = (p_1,p_2,...,p_{3N},q_1,q_2,...,q_{3N})$.  The equations of motion are given by the Hamilton's equations of motion,
\begin{align*}
 \dot{q}_i &= \pd{H}{p_i}\\
 \dot{p}_i &= - \pd{H}{q_i}.
\end{align*}
These equations become quickly intractable for systems with large numbers of degrees of freedom.  Instead systems will be studies by various ensembles which give the distribution of representative points in phase space.

\subsection{Statistical Ensembles}
A statistical ensemble is a collection of microstates which correspond to the same macrostate. Geometrically the ensemble can be described by the distribution of representative points in the $\Gamma$ space with the density distribution function $\rho(p,q,t)$ which is defined so that
\[ \rho(p,q,t)d^{3N}\!p \;d^{3N}\!q\]
is the number of representative points in phase space volume $d^{3N}\!p \;d^{3N}\!q$ at time t. 

\paragraph{Liouville Theorem}
\begin{equation*}
  \d{\rho}{t} + \sum_{i=1}^{3N} \left(\d{\rho}{p_i}\dot{p}_i+\d{\rho}{q_i}\dot{q}_i \right) = 0
\end{equation*}
The interpretation of the Liouville theorem is that the distribution of representative points in phase space move like an incompressible fluid.

\paragraph{Ensemble averages}
The ensemble average of an observable $O$ is give by
\[ \expv{O} = \frac{\int d^{3N}\!p d^{3N}\!q O(p,q) \rho(p,q,t)}{\int d^{3N}\!p d^{3N}\!q \rho(p,q,t)}\]
The time dependence of $O$ comes from the time-dependence of $\rho$. 

\paragraph{Postulate of Equal a Priori probability} For a system in thermodynamic equilibrium, it is equally likely to be in any microstate (of the same phase volume) that satisfies the macroscopic conditions of the system.

For an isolated system, the distribution is described by the \textit{microcanonical ensemble}, which corresponds to all microstates with the same energy, volume and number of particles.  The density of representative points in phase space is given by
\[ 
\rho(p,q) = \left\lbrace \begin{array}{l}
		\text{const.}  \qquad E< H(p,q) < E + \Delta\\
                0 \qquad \qquad  \text{otherwise}
             \end{array} \right.
\]

The question arises, what ensemble describes a system that is not in isolation, but instead is in equilibrium with another, larger system?  A system in contact with a heat reservoir such that the temperature, volume and number of particles are constant is described by the \textit{canonical ensemble}. Is is also possible to have a system where particles can be exchanged with a larger system.  Such a system, that is in contact with a heat reservoir and a particle reservoir is described by the \textit{grand canonical ensemble}, in which temperature, volume and chemical potential are kept constant.

The most probable value of an observable is given by the value of $O$ that the most members of the ensemble have. The most probable value and ensemble average are close if the mean square fluctuation
\[ \frac{\expv{O^2} - \expv{O}^2}{\expv{O}^2} \ll 1\]
is small.

\paragraph{Ergodic theorem}
Under certain conditions, a representative point in phase space will pass arbitrarily close to any other point in the accessible phase space, if one waits a sufficiently long time.  Following from the postulate of equal a priori probability, the time average of some physical observable is equal to the ensemble average over phase-space, that is
\[ \lim_{T\rightarrow \infty} \frac{1}{T} \int_0^T O(x,t) dt = \expv{O(x,t)} \]

Statistical mechanics does not specify whether a system is ergodic or not and there is no generic test for ergodicity.  For an ergodic system, a single trajectory will uniformly cover the phase space. An integrable system will not be ergodic in the full phase space due to the additional conserved quantities that act as constraints.  Ergodicity says nothing about the time-scale involved in covering the entire phase space.  For typical thermodynamic systems, the size of the phase space is immense.  A stronger condition than ergodicity is \textit{mixing} \citep{tabor_chaos_1989}.  In the long time limit, the values of a macroscopic observable in a system that is mixing will equal the ensemble average, without a need to time average over the trajectory as in an ergodic system.

\section{FPU Model and Results}

Let us now return to the numerical experiments of FPU.  Recall that the expectation was that a nonlinear system with many degrees of freedom would exhibit thermal behavior for any non-linearity, no matter how small. The stystem studies was a chain of one-dimensional anharmonically coupled oscillators with quadratic and cubic forces. The Hamiltonians of these two models can be written as a sum of 
\begin{equation}
 H=H_0+H_1,
\end{equation}
where the integrable Hamiltonian
\begin{equation}
H_0 = \sum_{n=1}^{N-1} p_n^2 + \sum_{n=1}^{N-1}\left(x_{n+1} - x_{n}\right)^2 ,
\end{equation}
is weakly perturbed by the non-integrable Hamiltonian $H_1$, which for the $\alpha-$model is
\begin{equation}
H_1 = \frac{\alpha}{3}\sum_{n=1}^{N-1}\left(x_{n+1} - x_{n}\right)^3
\end{equation}
and for the $\beta-$model is
\begin{equation}
H_1 = \frac{\beta}{4}\sum_{n=1}^{N-1}\left(x_{n+1} - x_{n}\right)^4.
\end{equation}
The displacement of particle $n$ from equilibrium is $x_n$ and $\alpha$ and $\beta$ are nonlinear interaction parameters. The resulting equations of motions for the $\alpha$-model are
\begin{equation}
\ddot{x}_n = (x_{n+1}-2x_n + x_{n-1}) + \alpha[(x_{n+1} - x_n)^2 - (x_n - x_{n-1}
)^2]
\end{equation}
and for the $\beta$-model
\begin{equation}
\ddot{x}_n = (x_{n+1}-2x_n + x_{n-1}) + \beta[(x_{n+1} - x_n)^3 - (x_n - x_{n-1})^3].
\end{equation}
The problem can be analyzed in terms of normal modes, $A_k(t)$, which are the Fourier modes of the displacement, $x_n(t)$,
\begin{equation}
 A_k(t) = \sqrt{\frac{2}{N}}\sum_{n=1}^N x_n(t)\sin\left(\frac{\pi kn}{N}\right)
\end{equation}
In terms of the normal modes, $H_0$ is a sum of harmonic oscillators and $H_1$ is a perturbing term that couples the oscillators.
\begin{align*}
 H_0 &= \frac{1}{2}\sum_k \left(\dot{A}_k^2 + \omega_k^2 A_k^2\right)\\
 \omega_k&= 2 \left| \sin\left(\frac{\pi k}{N}\right)\right|
\end{align*}
\sidenote{Approximations made in normal modes - dropping higher order terms.}
\sidenote{\citep{Add normal mode reference.}} For the linear system, that is when $\alpha=\beta=0$, there is no interaction between normal modes, so that modes that are initially populated remain populated and modes that are initially unpopulated will remain unpopulated.  
What happens when nonlinearity is added?  Would the coupling between the modes cause the energy to spread from a single mode to all of the other modes in the system? For the systems studied, thermalization would be marked by equipartition of energy among all of the modes, $A_k E_k=A_k^\prime E_k^\prime$ for all $k,k^\prime$. The expectation was that thermal behavior would be observed.
\begin{figure}
\begin{center}
\includegraphics[scale=0.5]{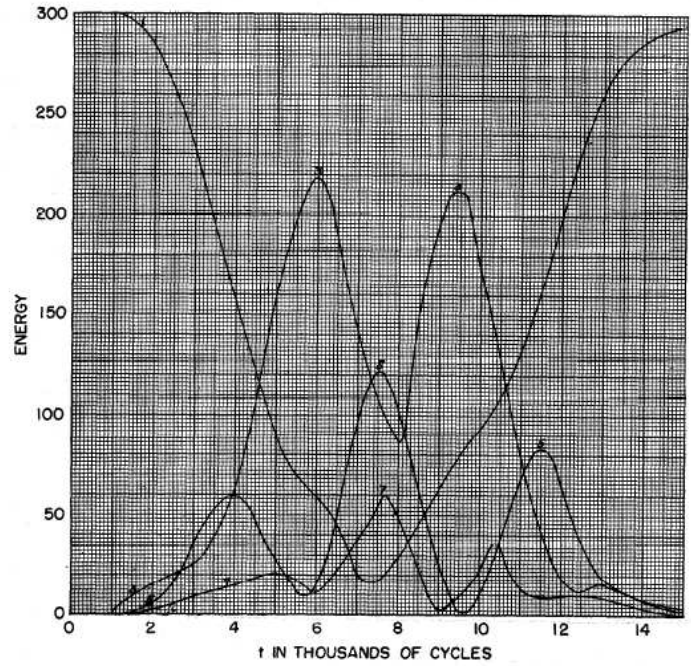}
\end{center}
\caption{\label{fig:fpu_beta} Time dependence of normal modes for cubic forces with $\beta=8$. \citep{fermi_studies_1974}}
\end{figure}
In Fig.~\ref{fig:fpu_beta} the time dependence of the normal modes is plotted for $N=32$ oscillators with cubic forces and $\beta=8$.  The initial condition is given by a sine wave and the velocity is zero. The ends are fixed.  The model preserves symmetry so that the effective number of particles is 16 and even modes have zero energy.  As can be seen from the plot only a few modes are active in the dynamics and there are recurrences. Additionally the period of recurrence was found to decrease with increasing nonlinearity.  Later work found a super period, where almost all of energy ($99\%$ ) returns to the initial modes after about 80 000 $T_1$. \citep{tuck_superperiod_1972}.

The failure of FPU to thermalize was very puzzling when first discovered and it was some time before explanations came out to explain the observed phenomena. The first explanation has to do with closeness to integrable systems and the second has to do with a stochasticity threshold with respect to the linear system.

\subsection{\label{sec:KAM_theorem}KAM Theorem}
A theorem outlined by Komologorov and subsequently proved by Arnold and Moser provided one resolution to the apparent paradox of FPU \citep{kolmogorov_preservation_1954, moser_invariant_1962, arnold_small_1963}. \sidenote{mention small divisors problem of classical perturbation theory?} Consider an integrable Hamiltonian that is weakly perturbed: $H=H_0(\v{I}) + \epsilon H_1(\v{I},\v{\theta})$ where $H_0$ is integrable with constants of motion $\v{I}$ and frequencies $\omega_i = \pd{H_0}{I}$and $H_1$ is periodic in the original angle variables, $H_1(\v{I},\v{\theta}+2\pi)=H_1(\v{I},\v{\theta})$.  The motion of the unperturbed Hamiltonian corresponds to motion on an $n-$dimensional torus.  It is assumed that the Hamiltonian is analytic on the complex domain and that the unperturbed Hamiltonian is non-degenerate,
\[ \det \left| \pdpd{H_0}{I_i}{I_j}\right| \neq 0. \]

Consider a frequency vector of the unperturbed Hamiltonian, $\v{\omega^\ast}$ that is incommensurate ($ \v{\omega}^\ast \cdot \v{k} \neq 0$ for all integer $k_i$). The corresponding motion of the unperturbed system is on the torus $T_0(\v{\omega}^\ast)$.

One statement of the KAM theorem is 
\begin{theorem}{KAM Theorem} If $H_1$ is small enough, then for almost all $\omega^\ast$, there exists an invariant torus $T(\v{\omega}^\ast)$ of the perturbed system such that $T(\v{\omega}^\ast)$ is close to $T_0(\v{\omega}^\ast)$. \citep{arnold_ergodic_1968}.
\end{theorem}

Stated informally, KAM showed that if the perturbation is sufficiently small, then almost all of the tori of the unperturbed motion are preserved and the resulting motion is quasi-periodic.

\subsection{\label{sec:solitons}Solitons and the Korteweg-de-Vries Equation}

Soon after the proof of the KAM theorem, Zabusky and Kruskal discovered solitary wave solutions to the Korteweg-de-Vries (KdV) equation, which they termed \ldq solitons''\citep{zabusky_interaction_1965}. Solitons are solitary wave that preserve their shape both under free propagation and after collisions. Zabusky and Kruskal also show that that the continuum limit of the $\beta-$ FPU model is close to the KdV equation. In KdV the speed of the solitons depends only on their amplitude. The KdV equation is given by
\begin{equation}
 u_t + u u_x + \delta^2 u_{xxx} = 0
\end{equation}
and was first found as a description of the motion of shallow water waves. 
\begin{figure}[ht]
\begin{center}
\includegraphics[scale=0.15]{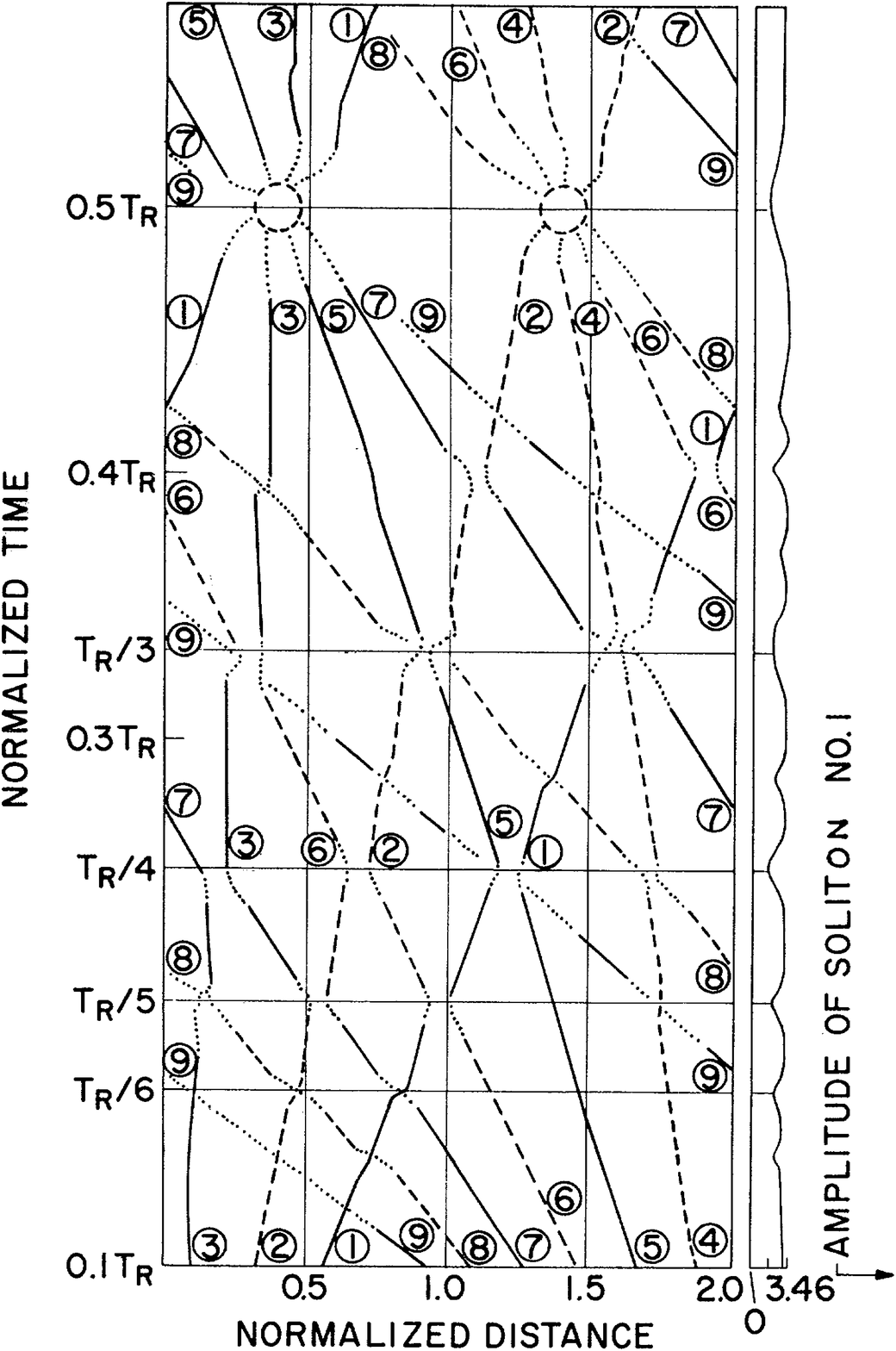}
\end{center}
\caption{\label{fig:solitons} Space-time diagram of soliton trajectories. $u_0 = \cos(\pi x)$. $\delta = 0.022$. $T_R$ is the recurrence time \citep{zabusky_interaction_1965}.}
\end{figure}
Starting with a cosine pulse, the negative slope regions of $u$ steepen, and then oscillations develop on the steep front and grow in amplitude, and finally each solitary wave or \ldq soliton'' moves at a constant speed, which is proportional to the amplitude. The trajectories of interacting solitons are shown in Fig.~\ref{fig:solitons}.  The solid (dashed) lines represent the  odd-(even-)numbered solitons. From the trajectories, it is clear that when solitons interact, they emerge with the same speed and thus shape. The dotted lines represent interactions, during which the joint amplitude is less than the sum of the individual amplitudes due to the non-linear interaction. At the recurrence time $T_R$, all of the solitons arrive at one point in space and almost reconstruct the initial state.

\sidenote{Properties of solitons:  velocity depends only on amplitude.  When solitons interact, they emerge unchanged in shape and speed, except for a phase shift. How does this relate to integrability? Solitons are conserved (amplitude, shape,...).  Put it in better words.}

KdV was later shown to be fully integrable by the method of Inverse Scattering Transform \citep{gardner_method_1967}.  The discovery of the solitons of KdV provides one route for explaining the absence of thermal behavior in FPU. The KAM theorem provides another explanation. The nonlinearities of the FPU studies were insufficient to break the KAM tori and thus the motion remained quasi-periodic.  Indeed later work showed that for larger nonlinearities, FPU did exhibit thermal behavior \sidenote{\addcite{}.}

\subsection{\label{sec:chirikov_criterion}Chirikov Criterion}

A method for predicting the criterion for the onset of chaos in the FPU experiments was the theory of overlapping resonances developed by Chirikov in the context of plasma dynamics, and later applied to FPU \citep{izrailev_statistical_1966}.

Consider a one-dimensional non-linear oscillator perturbed by an external periodic force. Write the unperturbed system in action-angle variables $(I,\theta)$ and the external field as a Fourier series.
\[ H=H_0(I) + \epsilon\sum_{m,n} V_{mn}(I)e^{i(m\theta+n\phi)}\]
where $H_0$ is the unperturbed Hamiltonian with frequencies $\omega(I)=\pd{H_0}{I}$, $\phi=\Omega t + \phi_0$ is the phase of the external force and $\theta=\omega t + \phi^\prime $. Resonances occur for the set of $I_r$ such that $\omega(I_r) l = \Omega k$.  The main ingredients are then to:
\begin{enumerate}
 \item Assume that when the system is near a resonance, the resonance dominates the motion.  Thus resonance are studied in isolation.
 \item Make a canonical transformation from $(I,\theta) \rightarrow (J,\psi)$ into the rotating reference frame of the resonance.  ($\psi$ measures deviations from resonance).  The old and new coordinates are related by
\[ J =\frac{1}{l} (I-I_r), \qquad \psi = l\theta - k\phi. \]
 \item Integrate out the fast phase $\phi$ motion.
 \item Expand the Hamiltonian about $I=I_r$ and keep terms up to second order in $H_0(I)$ and zeroth order in $V_{nm}(I)$.
\end{enumerate}

The resulting Hamiltonian of this process is,
\[ H_r = \frac{J^2}{2M} + \epsilon V_{l,-k}\cos(\psi) \]
which is the Hamiltonian of a simple pendulum, with \textquotedblleft mass'' $M \equiv l^2\left(\pdd{H_0}{I}\right)_{I=I_r}$.  
\begin{figure}[ht]
\begin{center}
\includegraphics[scale=0.7]{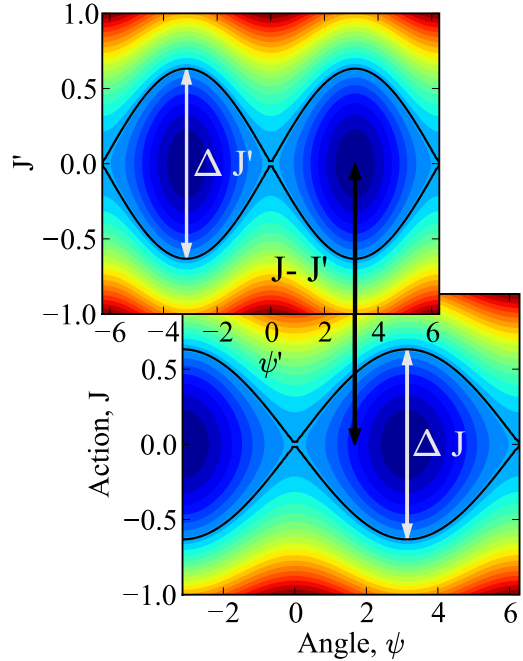}
\end{center}
\caption{\label{fig:chirikov_pendulum} Schematic of Chirikov's criterion of the onset of chaos for the kicked rotor. $J,\psi$ and $J^\prime,\psi^\prime$ are coordinates generated by two different canonical transformations.  $\Delta J$ and $\Delta J^\prime$ are width of the separatricies. $J-J^\prime$ represents the distance between the resonances}
\end{figure}
In Fig.~\ref{fig:chirikov_pendulum} a schematic is plotted for $J,\psi$ and $J^\prime,\psi^\prime$ coordinates generated by two different canonical transformations corresponding to two different resonances. The separatricies, which separate bounded and unbounded motion, are given by the black contour lines. An initial state inside the separatrix is captured by the resonance and both the action variable and the angle are bounded in phase-space.  Outside of the separatrix, the angle is unbounded.  When the resonances are well-separated in phase-space, that is $J^\prime-J \gg \Delta J$, the motion near an individual resonance is dominated by that resonance. As the strength of the external drive increases, the width of the separatrix, $\Delta J$, grows. Eventually the width of the two separatricies becomes comparable to the distance between the resonances. 

For two neighboring resonances, with resonant values $I_A$ and $I_B$, the \textquotedblleft distance''  between the resonances is given by $I_B - I_A$.  For separatrix width $\Delta I$, the ratio of these two quantities,
\begin{equation*}
 \mathcal{K} \equiv \frac{\text{separatrix width}}{\text{distance between resonances}} = \frac{\Delta I}{I_{B}- I_{A}},
\end{equation*}
gives the condition for onset of chaos in the system
\begin{center}
$\mathcal{K} \gtrsim 1 $.
\end{center}
Crossing the threshold corresponds to the overlap of the separatrices of the two resonances in phase space,
 so that the action variable is free to travel between resonances and thus explore all of phase space.
Complete chaos emerges when all of the neighboring resonances become coupled.  

Chirikov applied this criterion to the FPU system.  For a chain of N oscillators, and an excitation of momentum mode $k$, the conditions for chaos are given by:
\paragraph{Low modes}
\begin{equation}
 k \ll N:\qquad 3 \beta\left( \frac{\partial x}{\partial z}\right)^2_m  \sim \frac{3}{k}
\end{equation}
\paragraph{High modes ($k \approx N$)}
\begin{equation}
 N-k \ll N:\qquad 3 \beta\left( \frac{\partial x}{\partial z}\right)^2_m  \sim \frac{3 \pi^2}{N^2}\left(\frac{k}{N}\right) ^2
\end{equation}
This criterion predicts that for initial excitation of low modes, there is only have stochasticity for large perturbations, while for the high mode have stochasticity even for small non-linearities when $N$ is large. Data from FPU experiments show a threshold for thermalization that is close to the prediction of Chirikov. 

\section{Relationship between Chaos and Thermalization} 

Statistical mechanics dictates the of equilibrium state of a system, but it does not tell us whether a system will thermalize or not. 
How is it that statistical behavior can arise from dynamics? 

Consider a system in equilibrium, with some constraint.  Lift the constraint and let it evolve.  What will the equilibrium state be, if in fact it reaches equilibrium?  According to the second law of thermodynamics, when a constraint is lifted, the system moves to a state of greater entropy. To explain this further, consider a simple example a box of volume $V$ with $N$ particles, initially confined to $1/2$ of the box. When Boltzmann first derived statistical mechanics for an ideal gas through kinetic equations, there were several objections that were raised \citep{zaslavsky_chaotic_1999}:

\paragraph{Objection 1: Zermelo's Paradox (Recurrence)} Poincar\'e's recurrence theorem states that after sufficiently long times, any trajectory will pass arbitrarily close to any phase space point, including the initial state.  This would contradict the expected increase in entropy. \sidenote{ [except measure zero, for bounded system ...]}

\paragraph{Objection 2: Loschmidt's Paradox (Reversibility) } The equations of motion are invariant under time-reserval. If one simply reverses the velocities of particles, one goes back to the original state, a process that decreases entropy. 

\paragraph{Microscopic Origins of Macroscopic Irreversibility }
The second objection to the can be restated, how is it that the microscopic equations of motion are reversible, but the macroscopic behavior is irreversible? 
Lebowitz argues that this question was satisfactorily settled years ago by Thomson, Maxwell and Boltzmann \citep{lebowitz_statistical_1999}.  The essential ingredients of understanding this, according to Lebowitz are 
\begin{enumerate}
 \item the vast difference in scales between microstates and macrostates
 \item initial conditions are special
 \item significance of probabilities
\end{enumerate}

Consider an isolated classical system of $N$ particles.  Let $X$ be the microstate that completely specifies the system, $X=(\v{r}_1,\v{p}_1,\v{r}_2,\v{p}_2,\ldots,\v{r}_N,\v{p}_N)$ in phase space $\Gamma$. Let $M$ represent the macrostate of a system and let $\Gamma_M$ be the the region of the phase space $\Gamma$ that correspond to macrostate $M$. Note that there are many microstates $X$ that correspond to the same macrostate $M$. The number of microstates that correspond to is so large as to make certain macrostates extremely unlikely. Consider the example of the gas of particles initially confined to half of a box. The initial state is indeed a special state.  Once the constraint is lifted, the volume in phase space corresponding to all of the particles in one-half of the box is vastly smaller than the volume of phase space that corresponds to a roughly equal distribution of the particles between the two halves of the box. Thus, the probability that it will return to the initial state is effectively zero. Boltzmann made a rough estimate of the Poincar\'e recurrence times and found them to be much larger than the lifetime of the universe and thus irrelevant. This interpretation of the second law of thermodynamics explains why FPU expected to see thermal behavior in their system. However, it fails to account for the thermalization threshold found in FPU. Additionally, thermal behavior has been observed in systems with at few degrees of freedom, where this reasoning does not apply.

Zavslasky argues that chaotic dynamics introduce mixing properties in a system \citep{zaslavsky_chaotic_1999} that well resolves these paradoxes.  Furthermore, he calculates the distribution of Poincar\'e recurrence times and concludes that they are irrelevant.  While there is not a current consensus on the origins of statistical laws, this discussion highlights some relevant questions, namely,
\begin{itemize}
 \item Is chaos necessary for thermalization?
 \item What is the role of chaos in thermalization in systems with many degrees of freedom?
\end{itemize}

\subsection{Thermalization in Classical Field Models}
Since  the FPU studies on anharmonic oscillators, further studies on thermalization and approach to equilibrium have been carried out in several classical field theories, including recent studies on the classical $\phi^4$ model
\citep{boyanovsky_approach_2004}, nonlinear Klein-Gordon equation (NLKG) \citep{gerhardt_sudden_2002},
nonlinear Schr\"{o}dinger equation (NLSE) \citep{villain_fermi-pasta-ulam_2000,herbst_numerically_1989}, discrete nonlinear Schr\"{o}dinger equation (DNLS) \citep{herbst_numerically_1989,ablowitz_numerical_1993}
equivalent to the Bose-Hubbard model, and Integrable Discrete Non-Linear Schr\"{o}dinger equation (IDNLS)\citep{herbst_numerically_1989}.

No conventional thermalization is expected in the NLSE and IDNLS, which are both integrable.
In NLKG, like in FPU, the ability of the system to reach thermal equilibrium in the course of time evolution emerges only when the degree of nonlinearity exceeds a certain critical value (see \citep{izrailev_statistical_1966,livi_equipartition_1985} for the thermalization threshold in FPU). On the contrary, the $\phi^4$ model eventually reaches equilibrium regardless of how small the nonlinearity is.

There are several other studies on thermalization and chaos in system with a large number of degrees of freedom that are highly relevant to the work presented here. \citet{livi_equipartition_1985} investigated the equipartition threshold in the FPU $\beta$ model   in the thermodynamic limit. For $N$ oscillators, $64 \le N \le 512$, the thermodynamic limit is simulated by initially exciting a block of modes, $\Delta n$, such that $\frac{\Delta n}{N}$ remains constant. The threshold for equipartition of energy is found to be independent of the number of degrees of freedom with respect to the relevant control parameter, the energy density, with a critical value of $\epsilon_c \simeq 0.35$. They also calculate the Asymptotic Reynolds number, R, given by 
\[ \expv{\frac{O(NL)}{O(L)}}_{space} = \frac{\beta}{N} \sum_{i=1}^{N-1}(\phi_{i+1} - \phi_i)^2 \rightarrow R \qquad {t\rightarrow \infty}\]
which is a measure of ratio of strength of nonlinear to linear terms. Again, there is universal behavior with $R_c \simeq 0.03$, consistent with findings for energy density. There is evidence that the threshold energy is independent of the mode excited, when a narrow range of energies is initially excited. Note that very long equipartition times $(t \rightarrow \infty)$ are not ruled out and they conclude that the results are relevant for long, but finite times. It is significant to note that the result that the threshold for FPU remains in the thermodynamics limit contradict the predictions of Chirikov's criterion of overlapping resonances.

Another study by the same group focuses on the relationship between chaotic dynamics and statistical mechanics in two nonlinear Hamiltonian systems, the FPU model of nonlinearly coupled oscillators and coupled rotators \citep{livi_chaotic_1987}. For both systems  thermodynamic quantities are computed analytically using ensemble theory and compared with dynamical results from numerical simulations. For the FPU model, there is qualitative agreement between ensemble-averages and time-averages, independent of the stochasticity.  That is the system is ergodic in both the chaotic and regular region. For the rotator model, there is good agreement between the ensemble-averages and time-averages at low temperatures but not at high temperatures where the system is strongly chaotic. This result is explained in terms of localized chaos. In conclusion it is possible that a system (a) is not chaotic, but is ergodic for some physically \textquotedblleft relevant'' quantities and also (b) is chaotic, but some observable are not ergodic. This study highlights the open questions in the relationship between stochasticity and thermalization, particularly in systems with many degrees of freedom. 


\section{Quantum Degenerate Gases - Ultracold Atoms}\label{sec:ultracold_atoms}


Advances in the cooling and trapping of alkali atoms into the quantum degenerate regime has led to an explosion of experimental and theoretical studies of ultracold atoms. In recent years Nobel prizes have been awarded for advances in laser cooling techniques \citep{phillips_laser_1982, chu_three-dimensional_1985, aspect_laser_1988} and the subsequent observation of Bose-Einstein condensation (BEC) \citep{davis_bose-einstein_1995,anderson_observation_1995}. The manipulation of atoms by electric and magentic fields offers unprecedented control over parameters in the system and the ability to address fundamental questions in physics. Numerous proposals have been put forth for quantum simulators and applications have arisen in precision measurements. Studies in ultracold atoms have led to fruitful collaborations across fields, such as condensed matter and quantum information.  One application of these advances is using matter-wave interferometers based on ultracold atomic systems for high precision sensing of accelerations and gravitational fields \citep{gustavson_rotation_2000, durfee_long-term_2006, weiss_precision_1994, wicht_preliminary_2002}.  Fundamental questions in physics related to out-of-equilibrium dynamics and thermalization in classical and quantum integrable systems have also been studied in one-dimensional ultracold atoms. Quasi-one-dimensional systems have been realized in optical lattices \citep{paredes_tonks-girardeau_2004, kinoshita_quantum_2006} and on atom chips, where BEC's have been created and manipulated \citep{esteve_observations_2006, schumm_matter-wave_2005,wang_atom_2005}. 

\subsection{\label{sec:bec}Bose-Einstein Condensation}

\begin{figure}
\includegraphics[scale=0.6]{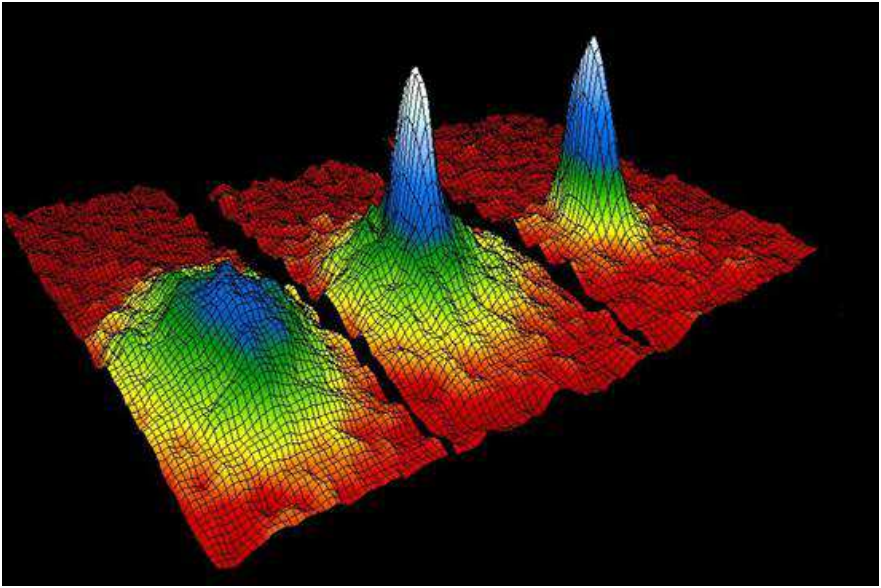}
\caption{\label{fig:bec} Velocity distribution of rubidium atoms. Left: prior to condensation. Center: just after condensation. Right: after further cooling \citep{anderson_observation_1995}}
\end{figure}

Advances in laser cooling and trapping led to the realization of Bose-Einstein condensation (BEC) in alkali atoms \citep{davis_bose-einstein_1995, anderson_observation_1995, bradley_evidence_1995}. BEC is a phase of matter, first proposed by \citet{bose_1924} and \citet{einstein_1925}, in which there is  macroscopic 
occupation of a single quantum state.  Seventy year later BEC was created in Rb${}^{87}$ gas
\citep{anderson_observation_1995}, sodium \citep{davis_bose-einstein_1995} and Li${}^7$ \citep{bradley_evidence_1995}. Since the initial experiments, BEC has been observed in twelve species of alkali atoms as well as in Bose molecules \citep{yukalov_cold_2009}. BEC was created by confining and cooling atoms to microkelvin temperatures with a magneto-optical trap (MOT), followed by evaporative cooling to nanokelvin temperatures.  In Fig.~\ref{fig:bec} the velocity distribution of rubidium atoms is show prior to and after condensation.
The velocity distribution of rubidium atoms is measured by turning off 
the confining trap, allowing the atoms to expand and performing a time-of-flight 
measurement.  The leftmost plot shows the velocity distribution just before 
condensation.  The center plot is the velocity distribution just after 
condensation, where the sharp peak in velocity distribution clearly indicates the
presence of the condensate.  In rightmost plot, the system has been cooled 
further such that most of the atoms are in the condensate. The presence of the condensate was confirmed by the anisotropic velocity distribution due to the magnetic trap in contrast with the isotropic, thermal velocity distribution. BEC's demonstrate long-range phase coherence, confirmed experimentally by the observation of interference between two independent condensates \citep{andrews_observation_1997}. 

Historically BEC was defined for a uniform, ideal gas as the macroscopic occupation of a single quantum state in the thermodynamic limit,
\[ \lim_{\substack{N \rightarrow \infty,\\ V \rightarrow \infty,\\ N/V \rightarrow \text{const}}} \frac{N_0}{V} > 0 \]
 where $N$ is the total number of particles, $V$ is the volume and $N_0$ is the occupation number of a single quantum state \citep{yukalov_cold_2009}. The question arises as to to define BEC in a non-uniform system such as in presence of trap. Penrose and Onsager propose that a condensate is present when the largest eigenvalue of the single-particle density matrix is extensive \citep{penrose_bose-einstein_1956}. The Penrose-Onsager scheme is more general than idea of off-diagonal long-range order and applicable to both uniform systems and trapped systems. 
There is no true condensate in finite or 1D systems, which will be the focus of this work, however there can be a quasi-condensate in 1D, when the coherence length is much larger than the de Broglie wavelength \citep{castin_simple_2004}. 

\subsection{\label{sec:opt_latt}Bosons in Optical Lattices}

The versatility offered by optical lattices allows one to control parameters such as the interaction strength, lattice spacing and the dimensionality of the system.  In particular, one-dimensional systems in cold atoms have been realized in optical lattices by tight confinement in two dimensions.

The dynamics of ultracold bosons in optical lattices can be described the Bose-Hubbard model (BHM). The Bose-Hubbard Hamiltonian \citep{jaksch_cold_1998} is
\[ H = -J\sum_{\expv{i,j}}\hat{b}_i^\dagger \hat{b}_j+\sum_i \epsilon_i\hat{n}_i + \frac{1}{2}U\sum_i\hat{n}_i(\hat{n}_i -1). \]
To derive this Hamiltonian, one begins with the Hamiltonian for bosonic atoms in an external potential
\begin{align*} H =& \int d^3x \psi^\dagger(\v{x})\left( -\frac{\hbar^2}{2m}\grad{}^2 + V_0(\v{x}) + V_T(\v{x}) \right)\psi(\v{x}) \\
& + \frac{1}{2} \frac{4\pi a_s \hbar^2}{m}\int d^3x\psi^\dagger(\v{x})\psi^\dagger(\v{x})\psi(\v{x})\psi(\v{x})
\end{align*}
where $\psi(\v{x})$ is a boson field operator, $V_0(\v{x})$ is the potential of the optical lattice, $V_T(\v{x})$ is the trapping potential and $a_s$ is the s-wave scattering length.  For single atoms, energy eigenfunctions are Bloch wave functions.  If the energies involved are much less than the excitation energy to the second band then a single band model is justified.  Wave functions localized at an individual lattice site, $w(\v{x})$, which are called Wannier wave function are introduced and the energy eigenfunctions are expanded in the Wannier basis
\[\psi(\v{x}) = \sum_i b_i w(\v{x}-\v{x}_i).\]
The Bose-Hubbard Hamiltonian follows from this expansion, where the hopping energy between matrix elements is given by
\[J = \int d^3x w^\ast (\v{x}-\v{x}_i)\left[ -\frac{\hbar^2}{2m}\grad{}^2 + V_0(\v{x})\right]w(\v{x}-\v{x}_j),\]
the on-site repulsion is
\[U = \frac{4\pi a_s \hbar^2}{m}\int d^3x |w(\v{x})|^4, \]
and the energy offset due to the lattice is
\[\epsilon_i = \int d^3x V_T(\v{x})|w(\v{x}-\v{x}_i)|^2 \approx V_T(\v{x_i}).\] In the last step the trapping potential is assumed to be approximately constant over the spatial variation of a single Wannier function.

One expects a zero-temperature quantum phase transition from the superfluid (SF) state to the Mott insulator (MI) state and as the depth of the lattice is increased for integer fillings. The SF state supports long-range phase coherence while in the MI state, the atoms are localized and there is no phase coherence. This transition was observed in ultracold atoms by \citet{greiner_quantum_2002}.

\subsection{\label{sec:interfere}Atom Chips}
Quasi-one-dimensional systems have also been realized on atom chips.  The miniaturization and integration of matter-wave optics has led to the development of atom chips \citep{folman_microscopic_2002,fortagh_magnetic_2007}. It is now possible to confine, manipulate and measure atoms on a single device using electric, magnetic and optical fields. Bose-Einstein condenstation has been created in magnetic microtraps \citep{ott_bose-einstein_2001, hansel_bose-einstein_2001}. The traps are highly elongated and the one-dimensional regime is realized when the transvere confining potential, $\hbar\omega_{\bot}$ is much greater than the relevant energy scales of the system, the thermal energy, $k_BT$ and chemical potential, $\mu$.  \citet{esteve_observations_2006} realized both the ideal Bose Gas as well as the quasicondenstate in a quasi-one-dimensional trap on an atom chip. Theoretical work has investigated the transition from the 1D Bose gas to the quasicondensate \citep{bouchoule_interaction-induced_2007} as well as the growth of the quasicondensate \citep{proukakis_quasicondensate_2006}.  Other experiments in one-dimensional traps include the demonstration of the first phase-preserving matter-wave beam-splitter on an atom chip  \citep{schumm_matter-wave_2005} and of an atom Michelson interferometer on an atom chip \citep{wang_atom_2005}.

\subsection{\label{sec:classical_field_approx}Classical Field Model of Bose Gas}

In this thesis we numerically study the Bose-Hubbard model, presented earlier, within the classical-field approximation. The classical field approximation is equivalent to the first-order mean-field approximation. In this section we outline the validity of the classical field approach for studying the dynamics of interacting Bose gases.

The dynamics of a BEC can be well-described by the Gross-Piteavskii equation (GPE) \citep{gross_structure_1961, pitaevskii_vortex_1961}
\[ i\hbar\pd{\Psi(\v{r},t)}{t} = \left(-\frac{\hbar^2}{2m}\grad + V(\v{r}) + g |\Psi(\v{r},t) |^2 \right)\Psi(\v{r},t),\]
where the coupling constant is given by $g = 4 \pi \hbar^2a/m$ ($a=$ scattering length, $m =$ mass). The Gross-Pitaevskii equation is a mean-field approximation and is equivalent to the continuous nonlinear Schr\"odinger equation which is integrable. The GPE has been used extensively to describe the dynamics of the condensate in three-dimensional systems.

Several studies have looked at the applicability of the mean-field or classical field description beyond the dynamics of the condensate.  Kagan and Svistunov studied the evolution of an interacting Bose gas from a strongly non-equilibrium state towards condenstation. They demonstrated that the classical-field description accurately describes a weakly interacting Bose gas \textit{in the absence of a condensate} provided that the occupation numbers of the initially occupied state are much greater than unity \citep{kagan_evolution_1997}. Given this condition, the time evolution of a state can be accurately described by the diagonal elements of the statistical matrix in the coherent state representation.

Castin studies the classical field model for one-dimensional weakly interacting Bose gases \citep{castin_simple_2004}.
The classical field model is generated by replacing the quantum mechanical operator $\hat{\psi}(z)$ with a complex field $\psi(z)$. 
For the interacting Bose gas, the state of the classical field is governed by a single parameter,
\[ \chi = \frac{\hbar^2\rho^2}{mk_B T}\frac{\rho g}{k_B T}\]
where $\rho$ is the mean density, $T$ is the temperature, $m$ is the mass, and $g$ is the interaction parameter. Castin calculates the correlation functions $g_1(z)=\expv{\hat{\psi}^\dagger(z)\hat{\psi}(0)}$ and $g_2(z)=\expv{\hat{\psi}^\dagger(z)\hat{\psi}^\dagger(0)\hat{\psi}(0)\hat{\psi}(z)} $. The contrast, $C\equiv g_2(0)/g_1^2(0)$ drops off quickly as $\chi$ increases and then slowly approaches unity for $\chi \gg 1$, indicating that density fluctuations are suppressed for large $\chi$.

The conditions for the validity of the classical field model for $\chi \gg 1$ are summarized as : 
\begin{enumerate}
 \m large occupation numbers,   $k_B T \gg \mu$
 \m gas is degenerate,  $\rho\lambda \gg 1$
 \m weakly interacting regime, $\rho\xi \gg 1 $
\end{enumerate}
where $\xi=(\hbar/m\mu)^{1/2}$ is the healing length. Note that the condition that $\chi \gg 1 $ automatically satisfies conditions (2) and (3).

In summary, the classical field model is a good approximation for weakly-interacting particles of a degenerate gas for large occupation number, in which case fluctuations are suppressed.

Mishmash and Carr study the correspondence between the mean-field and the fully quantum BHM in the dynamics of atoms in 1D optical lattices \citep{mishmash_ultracold_2008}. The mean-field BHM is equivalent to the discrete nonlinear Schr\"odinger equation (DNLS). They numerically investigate the analogs of dark soliton of DLNS in BHM and use the time-evolving block decimation algorithm (TEBD) developed by Vidal \citep{vidal_efficient_2004} to carry out the full quantum calculations.

\subsection{\label{sec:integ_quant}Chaos and Integrability in Quantum Systems}

Access to one-dimensional systems of ultracold atoms in optical lattices has led to realization of some known integrable models and observed effects of integrability in the dynamics of these models.
We focus on the effects of integrability in bosonic systems in optical lattices. 
The Lieb-Liniger model is a completely integrable quantum description of one-dimensional bosons with two-body $\delta$-interactions \citep{lieb_exact_1963,lieb_exact_1963-1}. The Hamiltonian is
\begin{equation}
 \hat{H}=-\sum_{i=1}^N \frac{\partial^2}{\partial x_i^2} + 2c \sum_{\langle i,j \rangle}
\delta(x_i-x_j),
\end{equation}
where the interaction, governed by the parameter $c$ is repulsive. The Lieb-Liniger model has been solved via Bethe Ansatz. In the limit of infinitely strong $\delta-$ repulsions, $c\to \infty$, the hard-core bosons, also known as a Tonks-Girardeau gas, map to non-interacting fermions \citep{girardeau_relationship_1960}.  While these models were proposed a half-century ago, the Tonks-Girardeau gas was only recently realized experimentally in Rb${}^{87}$ atoms that were strongly confined in two directions in an optical lattice to create one-dimensional tubes \citep{paredes_tonks-girardeau_2004}. By applying a shallow lattice in the longitudinal direction, the effective mass and thus interaction strength were increased in order to reach the Tonks-Giradeau regime.

Later experiments observed the effects of integrability on thermalization in a one-dimensional Bose gas.  \citet{kinoshita_quantum_2006}  demonstrate the first experimental evidence for the lack of thermalization in a many-body system with a large number of degrees of freedom for bosons in optical lattices. A gas of interacting bosons was prepared out-of-equilibrium by applying a laser pulse to a one-dimensional Bose-Einstein condensate in an optical lattice.  For both strongly- and weakly-interacting bosons, the expanded momentum distribution retains the initial double peak structure.
Even with the background harmonic potential, the system is integrable in the limit of infinite-strength repulsion. It was expected that a system with finite interactions, which is believed to be non-integrable in the presence of a harmonic trap, would reach thermal equilibrium.  However, the absence of thermalization occurred even for finite interactions.
\begin{figure}
\centering
\includegraphics[scale=0.25]{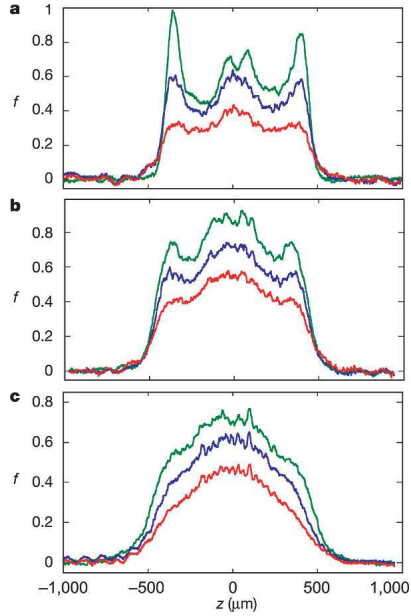}
\caption{\label{fig:weiss_mom_dist} $f(p_{\text{ex}})$, momentum distribution for 
(a) $\gamma_0$=4, $\tau$=34 ms. $t_{blue}=15\tau$, $t_{red}=30\tau$
(b) $\gamma_0$=1, $\tau$=13 ms.$t_{blue}=15\tau$, $t_{red}=40\tau$
(c) $\gamma_0$=0.62, $\tau$=13 ms. $t_{blue}=15\tau$, $t_{red}=40\tau$.  Green is the initial momentum distribution averaged over the first period. \citep{kinoshita_quantum_2006}}
\end{figure}
Figure~\ref{fig:weiss_mom_dist} shows the expanded momentum distribution for three different coupling 
strengths.  From the three peak structure in Fig.~\ref{fig:weiss_mom_dist}(a)-(b) it is clear that the gas has not thermalized for the Tonks-Girardeau limit ($\gamma_0=4$) and the intermediate regime
($\gamma_0=1)$ even after thousands of collisions have occurred between atoms.  

\subsection{Constrained equilibrium}
One question that arises from these experiments on hard-core bosons is: do integrable systems, which don't relax to the usual thermodynamic equilibrium distribution attain some other steady state? Numerical studies on one-dimensional hard-core bosons in a lattice addressed the relaxation dynamics of a fully-integrable quantum system \citep{rigol_relaxation_2007}.

The method to derive the steady-state distribution is to maximize the entropy subject to the constraints of the system, which include all of the conserved quantities \citep{jaynes_information_1957,jaynes_information_1957-1}. In this approach the many-body density matrix is given by 
\[\hat{\rho} = Z^{-1}exp\left[ - \sum_m \lambda_m \hat{I}_m \right] \]
where $\{ \hat{I}_m \} $, $\{ \lambda_m \}$ are the Lagrange multipliers which are determined by the initial conditions. This distribution is called the generalized Gibbs ensemble or fully-constrained thermodynamic ensemble.

One-dimensional hard-core bosons on a lattice can be mapped to free-fermions via a Jordan-Wigner transformation. The conserved quantities are moments of the fermionic momentum distribution.  \citeauthor{rigol_relaxation_2007} solve analytically for the density matrix with the constraints from the fermionic momentum distribution. The results of numerical simulations confirmed that when the system is prepared in the ground state of a small box and then allowed to expand in a larger box, it reaches a steady state, which is in agreement with the analytic results for the fully constrained system, rather than the grand canonical thermodynamic distribution. Additionally for an initial state with two momentum peaks, the two peaks structure remains after many oscillations, in agreement with the experiment performed by \citet{kinoshita_quantum_2006}.

\section{Outline of Thesis}
In this work we present a comprehensive study of chaos and thermalization in the 1D Bose-Hubbard model within the classical-field approximation. We study the threshold for chaos and its relation to thermalization. Two quantitative measures of thermalizability are compared:  the Finite-time Maximal Lyapunov exponents (FTMLE) and the normalized spectral entropy (NSE).  The FTMLE, averaged over phase space, converges to the maximal Lyapunov Exponent, the standard measure of chaos.  A positive MLE indicates that points that are initially close in phase-space diverge exponentially, rather than linearly. The spectral entropy measures the distance between the time-averaged momentum distribution of the numerical results and the momentum distribution predicted by thermodynamics, within the independent mode approximation. We investigate the dependence of the averaged FTMLE and normalized spectral entropy on a dimensionless nonlinearity parameter and the energy-per-particle, both of which are finite in the thermodynamic limit. The BHM is found to have a threshold for chaos which depends on the nonlinearity and the energy-per-particle. We study the size scaling of the Lyapunov exponent and normalized spectral entropy.

Furthermore we study resonances in the Bose-Hubbard model to find the Chrikov criterion for chaos. The criterion predicted by the Chrikov criterion is different from the one inferred from numerical calculations, signifying the failure of the standard Chirkov's approach.

There are at least three near-by integrable models: the Ablowitz-Ladik lattice, the continuous nonlinear Schr\"odinger equation and the noninteracting model.  We outline the method of Inverse Scattering Transform and generate all of the integrals of motion of the closely related, fully integrable model of Ablowitz-Ladik.  Furthermore, we discuss the possible role of these conserved quantities in relaxation in the BHM.  We conjecture that the presense of quasi-conserved quantities may alter the scaling of the chaos criterion.

%
%


\chapter{Thermalization and Chaos in the 1D Bose-Hubbard Model}
\label{chap:therm&chaos_1DMFBH}
%
%
%
%
%


\section{Introduction}\label{sec:intro}

One of the fundamental assertions of statistical mechanics is that the time average of a physical observable is equivalent to the average over phase-space, with microcanonical measure. A system for which this is true is said to be ergodic and one can calculate dynamical properties of the system from static phase-space averages. While this is believed to be true, because of the success of statistical mechanics in accurately predicting experimental results, many open questions remain. Is ergodicity sufficient to ensure the accuracy of statstical mechanical predictions for times that are relevant for observations? \sidenote{What if the time scale over which the system is ergodic is the age of the universe?}

\section{BHM: Hamiltonian and Equations of Motion}\label{sec:eq_motion}
We study the dynamics of an interacting one-dimensional Bose gas on a lattice (1D Bose-Hubbard model (BHM)) \citep{jaksch_cold_1998} with periodic boundary conditions in the classical field approximation. The Hamiltonian of the system of interest can be studied in many different forms through canonical transformations.  Each equivalent representation has a different Hamiltonian, canonical coordinates and equations of motion.  A well-choosen canonical transformation can pose the problem in a way which is more intuitive. Here we present three different representations: \sidenote{which will be used somewhere in this thesis}  The real-space representation, the momentum-space representation in terms of classical fields and the action-angle representation. 

\subsection{Real-Space Hamiltonian} In real space, the Hamiltonian is 
\begin{equation}
H=-J\sum_{s}\tilde{\psi}_s^\ast\left(\tilde{\psi}_{s+1} + \tilde{\psi}_{s-1}\right)
+ \frac{\mu_0N_{\mbox{\scriptsize s}}}{2} \sum_{s}|\tilde{\psi}_s|^4. 
\end{equation}
The equations of motion are given by \sidenote{I want to set $\hbar$ equal to 1.}
\begin{align}\label{eqn:bhm_eqm_real}
\pd{}{t}\tilde{\psi}_s = -\frac{i}{\hbar}\pd{H}{\tilde{\psi}_s^\ast} &= 
-i \left[ -J\left(\tilde{\psi}_{s+1} + \tilde{\psi}_{s-1}- 2\tilde{\psi}_s\right)
+ \mu_0N_{\mbox{\scriptsize s}}|\tilde{\psi}_s|^2 \tilde{\psi}_s \right]\\
\pd{}{t}\tilde{\psi^\ast}_s = \phantom{-}\frac{i}{\hbar}\pd{H}{\tilde{\psi}_s} &= \phantom{-}i\left[ -J\left(\tilde{\psi}^\ast_{s+1} + \tilde{\psi}^\ast_{s-1}- 2\tilde{\psi}^\ast_s\right)
+ \mu_0N_{\mbox{\scriptsize s}}|\tilde{\psi}_s|^2 \tilde{\psi}^\ast_s \right],
\end{align}
and the canonical pairs are ${\cal Q}_{s} = \psi_{s}^{}, \, {\cal P}_{s} = i\hbar\psi_{s}^{\ast}$.
These equations of motion are equivalent to the discrete nonlinear Schr\"odinger equation (DNLS). The time-evolution of the fields is carried out in real-space in all of the calculations.

\subsection{Momentum-Space Representation}
Another set of canonical coordinates, the momentum-space fields, $\psi_n=\psi(k_n)$, are related to the real-space field $\tilde{\psi_s}=\tilde{\psi}(x_s)$, by 
\begin{equation}\begin{split}
 \psi_n =& \frac{1}{\sqrt{N_{\mbox{\scriptsize s}}}}\sum_{s=1}^{N_{\mbox{\scriptsize s}}} \tilde{\psi}_s e^{-ik_n \cdot x_s}\\
 \tilde{\psi}_s =& \frac{1}{\sqrt{N_{\mbox{\scriptsize s}}}}\sum_{n=1}^{N_{\mbox{\scriptsize s}}} \psi_n e^{ik_n \cdot x_s}.
\end{split}\end{equation}
where $k_n=2\pi n/L$ and $x_s=sa$. 

In the momentum-space representation, the Hamiltonian is
\begin{equation}\label{eqn:hamiltonian}
H = \sum_m \left( \hbar\omega_m|\psi_m|^2 - \frac{\mu_0}{2} |\psi_m|^4 \right)
  + \frac{\mu_0}{2}\sum_{l, i, j}
    \psi_{l}^\ast \psi^\ast_{i} \psi_{j} \psi_{l+i-j}
\end{equation}
where the sum carries the restrictions: $j \neq l,i $. Several canonical transformations have been performed in order to write the Hamiltonian in this form.  \sidenote{Full details are given in Appendix.}

The canonical pairs are ${\cal Q}_{n} = \psi_{n}^{}, \, {\cal P}_{n} = i\hbar\psi_{n}^{\ast}$.
 and the equations of motion are given by
\begin{align}
\pd{}{t}\psi_n &= -\frac{i}{\hbar}\frac{\partial H}{\partial \psi_n^\ast} = -i\left(\omega_n - \frac{\mu_0}{\hbar}|\psi_n|^2 \right)\psi_n - i\frac{\mu_0} {\hbar}\sum_{i,j}\psi_i^\ast\psi_j\psi_{n+i-j}\\
\pd{}{t}\psi_n &= -\frac{i}{\hbar}\frac{\partial H}{\partial \psi_n^\ast} = -i\left(\omega_n - \frac{\mu_0}{\hbar}|\psi_n|^2 \right)\psi_n - i\frac{\mu_0} {\hbar}\sum_{i,j}\psi_i^\ast\psi_j\psi_{n+i-j} 
\end{align}
with the restrictions $j \neq n,i$ on the sums and the indices span the range $n,\,i,\,j = 0,\,\pm 1,\, \pm 2,\,\ldots,\,\pm \frac{N_{\mbox{\scriptsize s}}-1}{2}$
($N_{\mbox{\scriptsize s}}$ is supposed to be odd).

The bare frequency of each momentum mode is given by
\begin{equation}
 \hbar \omega_n=\frac{\hbar^2}{ma^2}\left(1-\cos\left(\frac{2 \pi n}{N_{\mbox{\scriptsize s}}} \right)\right) = 2J\left(1-\cos\left(\frac{2 \pi n}{N_{\mbox{\scriptsize s}}} \right)\right).
\end{equation}
The coupling constant is $\mu_0 = U N_{\mbox{\scriptsize a}}/N_{\mbox{\scriptsize s}}$.
Here $J$ and $U$ are the nearest-neighbor site-hopping and on-site repulsion constants
of the standard Bose-Hubbard model respectively, and $N_{\mbox{\scriptsize a}}$ is the number of atoms.

Time propagation is performed in real space, while the output and analysis of the numerical calculations focus on the momentum fields.

\begin{table}
\caption{\label{tab:variables}List of Parameters and Variables}
\centering
\renewcommand{\arraystretch}{1.5}
\begin{tabular}{|r @{ = } l|}
\hline $a$ & lattice spacing \\
\hline $J$ & B-H nearest-neighbor kinetic energy\\
\hline $\kappa$ & nonlinearity parameter \\
\hline $L$ & length of lattice\\
\hline $\mu_0$ & coupling constant\\
\hline $N_{\mbox{\scriptsize s}}$ & number of lattice sites = number of momentum modes \\
\hline $N_a$ & number of atoms\\
\hline $\tau_{\text{tal}}$ & Talbot time\\
\hline $U$ & B-H on-site repulsion energy\\
\hline $\hbar \tilde{\omega}_1$ & ground state energy of non-interacting model with quadratic dispersion\\
\hline $\xi$ & size-dependent nonlinearity parameter \\
\hline 
\end{tabular} 
\end{table}
\subsection{Action-Angle Representation}
Equivalently, the momentum-space Hamiltonian can be written in terms of action-angle variables, by performing serveral canonical transformation on the momentum-space representation.  It is this action-angle representation that will be the starting point of the resonant approximations and studies of individual resonances. The Hamiltonian is \sidenote{add details about canonical transformations here to get to a-a representation}
\begin{equation}
H = \sum_m \left(\omega_m I_m - \frac{\mu_0}{2\hbar^2} I_m^2 \right)
  + \frac{\mu_0}{2\hbar^2}\sum_{m,l,i,j}
    \left(I_m I_l I_i I_j \right)^{1/2}
   e^{i(\theta_m+\theta_l-\theta_i-\theta_j)}
\end{equation}
where the sum carries the restrictions: $m+l = i+j$; $m \neq i, j$;  $l \neq i, j$. The momentum wavefunction canonical variables, $\{(\psi_k, i\hbar\psi_k^\ast)\}$ are 
related to the action-angle canonical variables $\{(I_k, \theta_k)\}$ by 
$\psi_k=\sqrt{\frac{I_k}{\hbar}}e^{-i\theta_k},\;
\psi_k^\ast=\sqrt{\frac{I_k}{\hbar}}e^{i\theta_k}$.
In this form, the Hamiltonian can be seen as the sum of an integrable term and a perturbation, $H(\{I_k,\theta_k\})=H_0(\{I_k\}) + V(\{I_k,\theta_k\})$, where the integrable Hamiltonian is 
\begin{equation}\label{eqn:integrable_hamiltonian}
H_0 = \sum_m \left(\omega_m I_m - \frac{\mu_0}{2\hbar^2} I_m^2 \right) = \sum_m \left( \hbar\omega_m|\psi_m|^2 - \frac{\mu_0}{2} |\psi_m|^4 \right).
  \end{equation}

Throughout the wavefunction $\psi_n$ is normalized to unity: $\sum_{n} |\psi_n|^2 = 1$.

\paragraph{Dimensionless nonlinearity parameter} We define the dimensionless nonlinearity parameter,
\begin{equation}
\boxed{\kappa \equiv \frac{\mu_{0}}{J} \equiv \frac{U (N_{a}/N_{\mbox{\scriptsize s}})^2}{J (N_a/N_{\mbox{\scriptsize s}})}}
\label{kappa}
\end{equation}
whose physical meaning is the ratio between the typical interaction energy per site $U(N_{\mbox{\scriptsize a}}/N_{\mbox{\scriptsize s}})^2$  and the
kinetic energy per site $J N_{\mbox{\scriptsize a}}/N_{\mbox{\scriptsize s}}$. Note that this parameter governs both the strength of the nonlinearity and the strength of the perturbation from the integrable Hamiltonian (\ref{eqn:integrable_hamiltonian}).

\begin{table}
\caption{\label{tab:parameters}Relationship between Parameters}
\centering
\begin{empheq}[box=\fbox]{align*} 
\hbar\tilde{\omega}_1 &=  \frac{\hbar^2}{2m}\left(\frac{2\pi}{L}\right)^2= J\left(\frac{2\pi}{N_s}\right)^2\\
 \cline{1-2} \tau_{\text{tal}} &= \frac{2\pi}{\tilde{\omega}_1}\\
 \cline{1-2} \mu_0 &=\frac{gN_{a}}{L}\\
 \cline{1-2} J &= \frac{\hbar^2}{2ma^2}\\
 \cline{1-2} U &= \frac{g}{a} = \mu_0\frac{N_s}{N_{a}}\\
 \cline{1-2} U/J  &= \frac{2mag}{\hbar^2}=\frac{2m L^2}{\hbar^2N_sN_{a}}\mu_0\\
 \cline{1-2} \kappa &= \frac{\mu_0}{J} \equiv \frac{U (N_{a}/N_s)^2}{J (N_a/N_s)}\\
 \cline{1-2} \xi &\equiv  \frac{\mu_0}{\hbar\tilde{\omega}_1} = \kappa\left(\frac{N_s}{2\pi}\right)^2\\
 \cline{1-2} \hbar\omega_n &= \frac{\hbar^2}{ma^2}\left[1-\cos\left(\frac{2\pi n}{N_s}\right)\right]=2J\left[ 1-\cos\left(\frac{2\pi n}{N_s}\right)\right]
\end{empheq}
\end{table}
In Tables~\ref{tab:variables} and \ref{tab:parameters}, variables are listed and the relationship between relevant parameters are summarized.
\subsection{Validity of the Classical-Field Theory}
Based on the studies of the validity of the classical-field theory for Bose gases \citep{castin_simple_2004, kagan_evolution_1997} discussed earlier, the classical-field approximation will
apply for the lattice site occupations satisfying 
\[ \frac{N_{\mbox{\scriptsize a}}}{N_{\mbox{\scriptsize s}}} \gg \mbox{max}(\kappa,\, 1)\, \mbox{max}\left[\left(\frac{N_{\mbox{\scriptsize s}}}{\Delta n}\right),\, 1\right],\]
where $\Delta n$ is the typical width of the momentum distribution.
We note that the Mott regime, $ N_{\mbox{\scriptsize a}}= \mbox{integer}\times N_{\mbox{\scriptsize s}}$, $\Delta n = N_{\mbox{\scriptsize s}}$, $ U/J\geq 2.2\, N_{\mbox{\scriptsize a}}/N_{\mbox{\scriptsize s}}$ \citep{hamer_accurate_1979}, lies well outside of the above criteria.

\subsection{Nearby Integrable Models}
There are several known intergrable models that are limiting cases of the BHM. These include
\begin{enumerate}
 \item \textbf{Continuous Nonlinear Schr\"odinger Equation:}  In the contiuum limit, the 1D BHM becomes
	 \begin{equation*}
 	H=\int_0^L dz \Psi^\ast \left\lbrace-\frac{\hbar^2}{2m}\Delta\right\rbrace\Psi + \int_0^L dz \frac{1}{2}g|\Psi|^4.
 	\end{equation*}
 \item \textbf{Linear Model: } In the non-interacting limit, $\kappa \rightarrow 0$, the 1D BHM becomes a sum of harmonic oscillators
	\[ H = \sum_m \hbar\omega_m|\psi_m|^2.\]
 \item \textbf{Independent Mode: } If the interating term of vanishes, the 1D BHM becomes a sum of decoupled nonlinear oscillators,
	\[H = \sum_m \hbar\omega_m|\psi_m|^2 - \frac{\mu_0}{2} |\psi_m|^4 \]
 with nonlinear frequencies given by $\Omega_m = \hbar\omega_m - \frac{\mu_0}{2} |\psi_m|^2$.
 \item \textbf{Ablowitz-Ladik Lattice: } An alternate discretization of the NLS yields the Ablowitz-Ladik lattice, with Hamiltonian,
	\[ H = -\sum_n  \left(q_n q^\ast_{n+1} + q^\ast_n q_{n+1} \right) - \frac{4}{\sigma} \sum_n \ln\left(1-\frac{\sigma}{2}|q_n|^2\right),\]
which will be discussed further in later chapters.
\end{enumerate}

\sidenote{Add Hamiltonians for each of these?}

\section{Time Dynamics}\label{sec:time_dynamics}
First we study the time dynamics of the 1D BHM on a lattice with 
$N_{\mbox{\scriptsize s}}=21$ modes.  The system is prepared in a state that is narrowly distributed in momentum space and evolves according to the classical equations of motion. 
Initially the lowest three momentum modes are occupied, the minimum number of modes required by selection rules for non-trivial processes leading to population of initially unoccupied modes. 
\begin{figure}[ht]
\includegraphics[scale=.6]{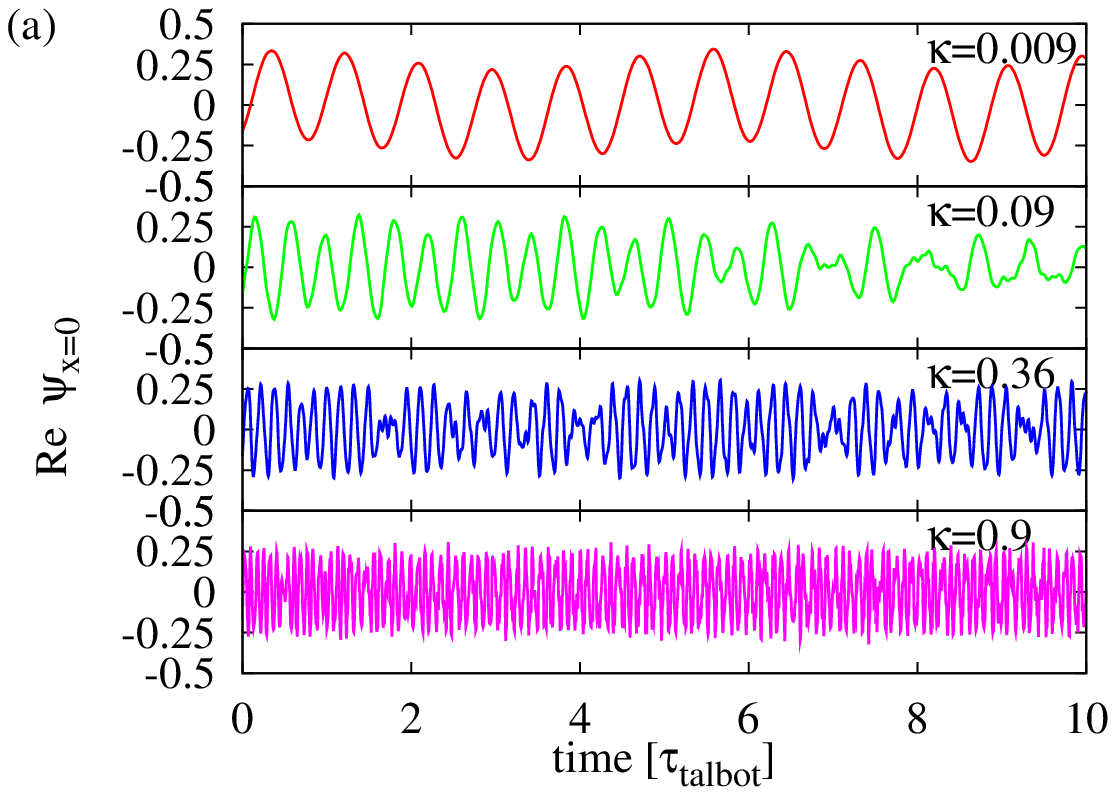}
\includegraphics[scale=.6]{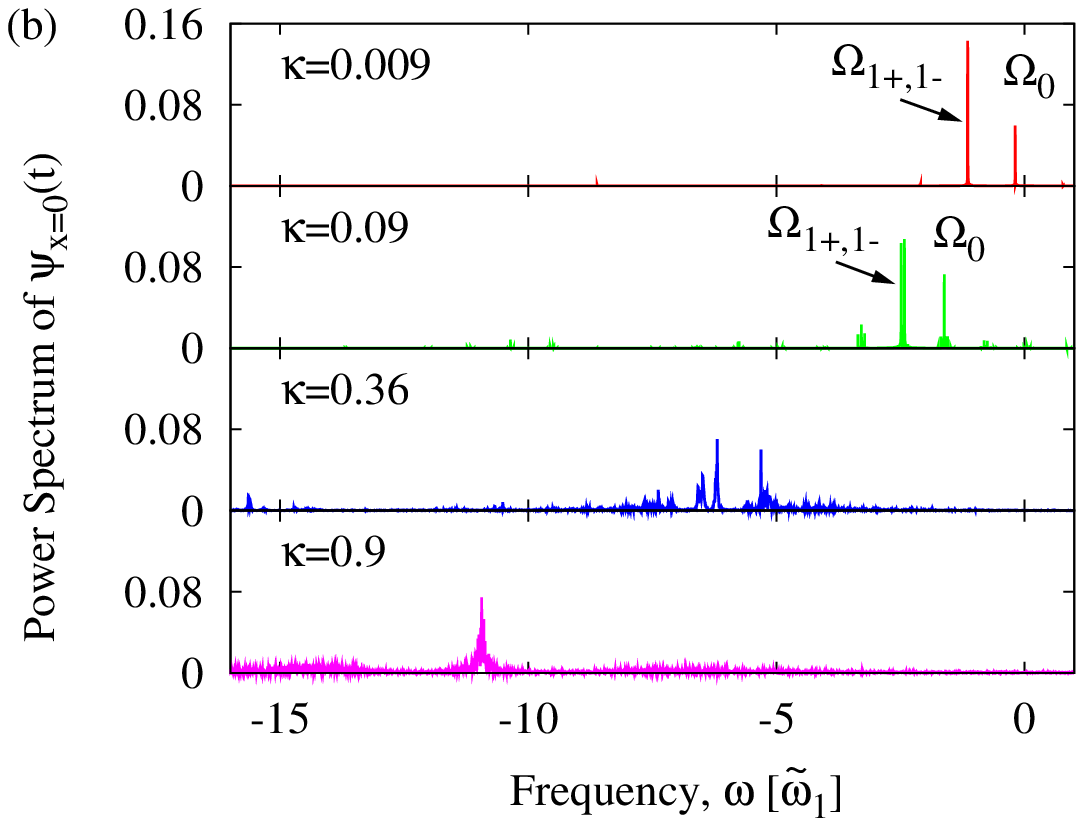}
\caption{
\label{fig:time_dyn} Time evolution of the wavefunction at the center of the box 
for a typical run for $\kappa=0.009,0.09,0.36,0.9$ with identical initial conditions.  (a) Time dynamics of the real part of the wavefunction, Re $\psi_{x=0}(t)$.
(b) Frequencies of the real-space wavefunction, $|\text{FT}[\psi_{x=0}(t)]|^2$. For $\kappa=0.009,0.09$ the descendants of the unperturbed frequencies generated 
by the first, \ldq integrable\rdq term of the Hamiltonian (\ref{eqn:hamiltonian}) are labeled.
        }
\end{figure}
In Fig.~\ref{fig:time_dyn}, the time dynamics and power spectrum of the 
wavefunction at the center of the box, $\psi_{x=0}(t)$, are plotted 
for various interaction strengths for a typical initial state.  
As seen in Fig.~\ref{fig:time_dyn}(a), the time evolutions of the zero 
momentum mode is quasi-periodic for weak interactions, with a few  
easily identifiable frequencies entering the dynamics, which is confirmed by the power spectrum in Fig.~\ref{fig:time_dyn}(b).
As the nonlinearity increases, more frequencies determine the dynamics and for sufficiently 
large nonlinearity the motion loses its quasi-periodic character and appears to be chaotic.

The clear distinction between quasi-periodic and seemingly chaotic behavior of the time 
dynamics leads to the following questions: 
\begin{enumerate}
 \item Is the motion really chaotic?
 \item If it is, where is the chaos threshold as one increases the nonlinearity $\kappa$? 
 \item When chaotic, does the system reach thermal equilibrium?
\end{enumerate}
In order to answers these questions, it is necessary to define appropriate measures of chaos and thermalization.

\sidenote{Add corresponding initial and final momentum distributions.}

\section{Chaos: Calculating Lyapunov Exponents}\label{sec:MLE}
The standard signature of the chaotic nature of a region in phase-space is that the separation between trajectories that are initially close grows exponentially with time,
for typical trajectories, as captured by a positive maximal Lyapunov exponent (MLE). In regular regions the separation 
grows linearly \citep{chirikov_universal_1979}, resulting in zero MLE. As we increase $\kappa$ in our system, we expect the phase space to change from being dominated 
by regular regions for small $\kappa$ to being dominated by chaotic regions for large $\kappa$. In the present section, we use the MLEs to quantify this 
transition to chaos, which, as we will see in the subsequent section, coincides with a relatively broad change from unthermalizability to complete thermalizability.

Consider two trajectories $\bm{x}(t)$ and $\widetilde{\bm{x}}(t)$ with initial points $\bm{x}_{0}$ and $\widetilde{\bm{x}}_{0}$, respectively. The separation 
$\delta\bm{x}(t)=\widetilde{\bm{x}}(t)-\bm{x}(t)$ initially satisfies a linear differential equation, and the duration of this linear regime grows without bound as the initial separation 
$\widetilde{\bm{x}}_{0}-\bm{x}_{0}$ goes to zero. The finite-time maximal Lyapunov exponent (FTMLE) corresponding to the phase-space point $\bm{x}_{0}$ \citep{eckhardt_local_1993,voglis_invariant_1994} is given by
\begin{eqnarray}
\lambda_{t_{\mbox{\scriptsize fin}}}(\bm{x}_{0})=
\lim_{\widetilde{\bm{x}}_{0}\to\bm{x}_{0}}\;
\frac{1}{t_{\mbox{\scriptsize fin}}}
\ln \frac{\|\widetilde{\bm{x}}(t_{\mbox{\scriptsize fin}})-\bm{x}(t_{\mbox{\scriptsize fin}})\|}{\|\widetilde{\bm{x}}_{0}-\bm{x}_{0}\|}
\,.
\label{lyapunov_exponent}
\end{eqnarray}
The limit $t_{\mbox{\scriptsize fin}}\to \infty$ gives the MLE, $\lambda_{\infty}(\bm{x}_{0})$, but the FTMLE are themselves of intrinsic interest \citep{eckhardt_local_1993,voglis_invariant_1994,contopoulos_number_1978,contopoulos_fast_1997}. 
We chose a convenient quantum 
mechanical metric,
\mbox{$\|\widetilde{\bm{x}}-\bm{x}\|^{2} =$} $\sum_{n} |\widetilde{\psi}_{n}-\psi^{}_{n}|^2$. This metric becomes Euclidian under the canonical transformation 
\begin{align*}
{\mathsf{Q}}_{n} &= (2\hbar)^{1/2} \mathsf{Re}(\psi_{n}^{})\\
{\mathsf{P}}_{n} &= (2\hbar)^{1/2} \mathsf{Im}(\psi_{n}^{})\\
\|\widetilde{\bm{x}}-\bm{x}\|^{2} &= \frac{1}{2\hbar} \sum_{n}  \left[(\mathsf{Q}'_{n}-\mathsf{Q}_{n})^2 + ({\mathsf{P}}'_{n}-{\mathsf{P}}_{n})^2 \right]
\end{align*}
where the sum runs from $n=-(N_{\mbox{\scriptsize s}}-1)/2 $ to $n=(N_{\mbox{\scriptsize s}}-1)/2$.

\section{\label{sec:thermodynamics}Thermalization: Calculating Spectral Entropy}

In order to measure thermalization, it is necessary to make thermodynamic predictions and compare the dynamics from the propagation of the equations of motion, with the thermodynamic state.  In this section a method for calculating the thermodynamic state, within the Hartree-Fock approximation, is laid out.  A full account is given in Appendix A.  Additionally the spectral entropy is defined, which is a quantitiative measure of the difference between the time-averaged dynamical state and the expected thermodynamic state.

\subsection{Conserved Quantities}\label{sec:conserved_quantities}
A treatment of the thermodynamic state must take into account all conserved quantities. The known conserved quantities of the 1D BHM are the energy and norm. Conservation of norm, $\sum_{n=1}^N |\psi_n|^2$, is associated with $U(1)$ symmetry in the real/imaginary plane represented by the tranformation $\lbrace\psi_n,\psi_n^\ast\rbrace \rightarrow \lbrace\psi_n e^{i\theta},\psi_n^\ast e^{-i\theta}\rbrace$. Unlike the continuous NLSE, the total momentum is not conserved.

\subsection{Hartree-Fock}\label{hartree_fock}

Within Hartree-Fock the form of the density distribution function is taken to be Gaussian and the thermal expectation value of the Grand Potental,
\begin{equation}
 \langle F\rangle=\langle H\rangle-T\langle S \rangle-\mu \langle N_a \rangle,
\end{equation}
is minimized, where $N_a$ is the norm. The density distribution function with two-body interactions has the form, 
\begin{equation}
 \sigma_{HF}=\frac{1}{Z}\exp\left( \sum_{n,n^\prime} - \hbar \alpha_{n,n^\prime} \psi_n\psi^\ast_{n^\prime} \right)=\frac{1}{Z}\exp\left(- \sum_n \alpha_n I_n \right).
\end{equation}
We use the independent mode approximation so that in the second step, the off-diagonal elements are taken to be zero\sidenote{justify?}. The ${\alpha_n}$ coefficients are unknown and are determined by the condition of minimizing the grand potential. The density distribution function is normalized, so that the integration of $\sigma_{HF}$ over all of phase space is 1, by
\begin{equation}
 Z = \frac{1}{(2\pi\hbar)^N}\int d^N \theta \int d^N I e^{-\sum_n \alpha_n I_n} = \prod_{i=1}^N \frac{1}{\hbar\alpha_i}.
\end{equation}
The expectation value of each term in the Grand Potential is calculated, using the Hartree-Fock density distribution function $\sigma_{HF}$.  The expectation value of a generic observable is given by
\begin{align}
 \langle O \rangle &= \frac{1}{(2\pi\hbar)^N}\int d^N \theta \int d^N I O(\lbrace I_n, \theta_n \rbrace) \sigma_{HF}\\ 
&= \frac{1}{(2\pi)^N}\prod_{i=1}^N \alpha_i\int d^N \theta \int d^N I O e^{ -\sum_n \alpha_n I_n}
\end{align}
The expectation values of the relevant observables are listed below.

\paragraph{Norm}
\begin{equation}
\expv{N_a}=\expv{\frac{1}{\hbar}\sum_n  I_n} =\sum_{n=1}^N \frac{1}{\hbar\alpha_n}\\
\end{equation}
\paragraph{Hamiltonian - Kinetic Term}
\begin{equation}
\expv{H_0}=\expv{\sum_n I_n\omega_n } = \sum_{n=1}^N\frac{\omega_n}{\alpha_n}\\
\end{equation}
\paragraph{Hamiltonian - Interaction Term}
\begin{equation}
 \expv{H_I} =\expv{\frac{\mu_0}{2\hbar^2}\sum_{m,p,q,r} \left(I_m I_p I_q I_{r}\right)^{1/2}\delta_{m+p,q+r} e^{-i(\theta_m + \theta_p - \theta_q -\theta_{r})}}=\frac{\mu_0}{\hbar^2}\sum_{m,p} \frac{1}{\alpha_m}\frac{1}{\alpha_p} 
\end{equation}
\paragraph{Entropy}
\begin{equation}\begin{split}
 S=&-\frac{1}{(2\pi\hbar)^N}\int d^N \theta \int d^N I\sigma_{HF}\log\sigma_{HF} = N_s - \sum_j\log(\hbar \alpha_j)
\end{split}\end{equation}
\subsection{Minimization of the Grand Potential}
The thermal expectation value of the Grand Potential within Hartree-Fock is given by 
\begin{equation}
\langle F\rangle=\sum_m \frac{\omega_m}{\alpha_m} +\frac{\mu_0}{\hbar^2} \sum_{m,p} \frac{1}{\alpha_m}\frac{1}{\alpha_p} - T\left( N_s - \sum_m \log(\hbar\alpha_m) \right) -\mu \sum_m \frac{1}{\hbar\alpha_m}
\end{equation}
Taking the variation with respect to $\alpha_n$, and setting it equal to zero gives
\begin{equation}
 \frac{\delta \langle F\rangle}{\delta \alpha_n}= -\frac{\omega_n}{\alpha_n^2} -2\frac{\mu_0}{\hbar^2} \frac{1}{\alpha_n^2}\sum_m\frac{1}{\alpha_m} + \frac{T}{\alpha_n} + \frac{\mu}{\hbar\alpha_n^2} = 0.
\end{equation}
Using $\sum_m \alpha_m^{-1} = \hbar N_a $ and solving for $\alpha_n$,
\begin{equation}
\alpha_n = \frac{1}{\hbar T} \left[\hbar\omega_n+2\mu_0 N_a -\mu\right]
\end{equation}
The thermal expectation values of the occupation of momentum mode $n$ become
\begin{equation}
\langle I_n \rangle = \frac{1}{\alpha_n} = \frac{\hbar T}{\hbar\omega_n+2\mu_0 N_a -\mu}.
\end{equation}
In general, the coefficients $\mu$ and $T$ are unknown and are determined by imposing constraints on the norm and energy, which come from the dynamical code.  The constraints are 
\begin{equation}\begin{split}
N_a =& \langle N_a \rangle = \frac{1}{\hbar}\sum_n \langle I_n \rangle\\
E_T = & \langle H \rangle = \sum_n \omega_n\langle I_n \rangle + \frac{\mu_0}{\hbar^2}\sum_{m,n} \langle I_m \rangle \langle I_n \rangle = \sum_n \omega_n\langle I_n \rangle + \mu_0 N_a^2
\end{split}\end{equation}
Beginning with the expression for $\langle I_n \rangle$, we can solve for $T$ in terms of $\mu$, $N_a$ and energy,
\begin{equation}
T = \omega_n \langle I_n \rangle + 2\frac{\mu_0}{\hbar} N_a \langle I_n \rangle -\frac{\mu}{\hbar} \langle I_n \rangle.
\end{equation}
Summing over $n$,
\begin{equation}
T = \frac{1}{N_s}\left[E_k + 2\mu_0 N_a^2 -\mu N_a \right]
\end{equation}
where $E_k \equiv \sum_n \omega_n \langle I_n \rangle$. This expression for $T$ can be substituted back into the contraints to reduce the system to two equations with two unknowns.  Using the expression for temperature and normalization condition, a single constraint remains to be solved,
\begin{equation}
\frac{1}{N_s}\sum_n\frac{\left[E_k + 2\mu_0 N_a^2 -\mu N_a \right]}{\left[\hbar\omega_n + 2\mu_0 N_a -\mu \right]}- N_a  = 0.
\end{equation}

The Hartree-Fock approximation is known to overestimate the interaction energy in the regime of strong interactions. For sufficiently large $\mu_0$, the Hartree-Fock interaction energy, $\mu_0N_a^2$ becomes greater than the total energy resulting in negative kinetic energy, where the kinetic energy is $E_k = E_T-\mu_0N_a^2$. For this reason, we determine the temperature $T$ and the chemical potential $\mu$ using the time-averaged numerical kinetic energy (along with the norm) instead of the total energy. The quantity $E_k$ in the thermal distribution is fixed to the time-averaged kinetic energy of the final state from the dynamical code. We fix the norm to its numerical value, and subsequently solve iteratively for the norm to find all parameters.  In this way, the total energy is never used in the constraints. \sidenote{Add plot of KE vs. time and note that after initial redistribution between kinetic and interaction energies, KE does not relax further}

Additionally the solutions must satisfy the physical constraint that $I_n \geq 0$ for all $n$, which 
leads to bounds on $\mu$.  For $T>0$, the condition such that the denominator is greater than zero for all $n$, is $\hbar \omega_n + 2 \mu_0 N_a - \mu > 0$, which leads to an upper bound for $\mu$, $\mu < 2 \mu_0 N_a$. There is a critical kinetic energy that corresponds to infinite temperature, which leads to equal population of all the modes, $\expv{I_n} = \hbar N_a/N_s$.  The critical kinetic energy, which separates the positive and negative temperature regime can be calculated as $E_{k-cr}=\sum_n\frac{N_a}{N_s}\hbar\omega_n = \frac{N_a}{N_s}\sum_n 2J\left[1-\cos\left( \frac{2\pi n}{N}\right)\right]=2JN_a$. For $E_k >E_{k-cr}$ the temerature is negative, and the lower bound on $\mu$ is $\hbar\omega_n + 2 \mu_0 N_a -\mu < 0 $ for all $ n$ or $\mu > 4J + 2 \mu_0 N_a$.  Close to the critical kinetic energy, both the temperature and the chemical potential diverge.  By expanding the norm in powers of $\omega_n/(\mu - 2\mu_0N_a)$, an estimate for the chemical potential when $E_k=E_{k-cr} \pm \varepsilon$ is $\mu \approx 2 \mu_0 + E_{k-cr} \pm E_{k-cr}^2/(2\varepsilon)$. The temperature and the chemical potential were computed individually for each initial condition used. 

\begin{figure}
\begin{center}
\includegraphics[scale=0.6]{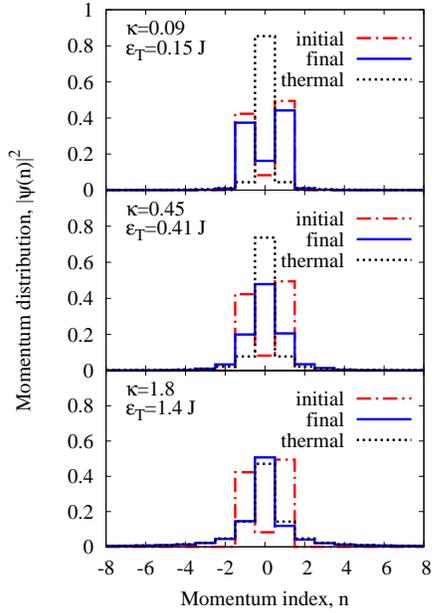}
\end{center}
\caption{\label{fig:avg_mom_dist} Initial, final and Hartree-Fock thermal momentum distributions for $\kappa=0.09,0.45,1.8,$ starting from the same initial state. $N_s=21$. The initial state is a representative state and the final state is time-averaged. $\epsilon_T$ is the total energy per particle.}
\end{figure}

In Fig.~\ref{fig:avg_mom_dist}, the initial and time-averaged momentum distributions of
a representative state are plotted for $\kappa=0.09,0.36$ and $0.9$, along with the thermal Hartree-Fock predictions,
$\langle |\psi_n|^2 \rangle = (T/N_{\mbox{\scriptsize a}})/(\hbar \omega_n + 2 \mu_0 N_{\mbox{\scriptsize a}} - \mu)$. 

\subsection{Spectral Entropy}\label{spec_entropy}
For coupled anharmonic oscillators, as in the FPU study, energy equipartition
among the normal momentum modes signified thermalization.
In the BHM, the additional conservation of the norm modifies the quantity that is equipartitioned.
To determine the best measure for the equipartition we use the variational Hartree-Fock Hamiltonian \citep{castin_coherent_2001, ohberg_hartree_1997},
$H^{\mbox{\scriptsize HF}} = \sum_{n} \hbar\omega^{\mbox{\scriptsize HF}}_n|\psi_n|^2$,
where the set of Hartree-Fock energies
$\{\hbar\omega^{\mbox{\scriptsize HF}}_n\}$ was regarded as the variational field. This procedure gives
$\hbar\omega^{\mbox{\scriptsize HF}}_n = \hbar\omega_n + 2 \mu_0 N_{\mbox{\scriptsize a}} - \mu$,
where $\mu$ is the chemical potential.

The new quantity to be equipartitioned is the distribution of the Hartree-Fock energy,
\begin{equation}
q_n(t) = \frac{|\psi_n(t)|^2 \hbar\omega^{\mbox{\scriptsize HF}}_n}
         {\sum_{n'} |\psi_{n'}(t)|^2 \hbar\omega^{\mbox{\scriptsize HF}}_{n'} }
\end{equation}

A quantitative measure of the distance from thermodynamic equilibrium is the spectral
entropy 
\begin{equation}
S(t) = -\sum_{n} q_n(t) \ln q_n(t) .
\end{equation}
In thermal equillibrium, $q_n$ is equipartioned, maximizing the spectral entropy at a value $S_{\text{max}}= \log (N_s)$. A more convenient quantity to study is the normalized spectral entropy \citep{livi_equipartition_1985},
\begin{equation}
\eta(t)=\frac{S_{\text{max}} - S(t)}{S_{\text{max}} - S(0)},
\end{equation}
which is unity at $t=0$ and vanishes as the system approaches thermal equilibrium.

\section{Results: N=21 Sites, Three-Mode Initial Conditions}
Initially, we study the FTMLE for a class of initial conditions where only the $k=0,\,\pm1$ modes are occupied. In this subspace we sample uniformly from the intersection of the 
microcanonical shells in energy and norm; the energy is chosen to be the infinite-temperature energy of the subsystem, and the norm is 1. For each value of $\kappa$, we sample 100 points, which we 
set as the initial points $\bm{x}_{0}$. To each initial point we add a small random vector, as little as machine precision allows, to obtain the corresponding $\widetilde{\bm{x}}_{0}$'s. Each pair we 
propagate for a time $t_{\mbox{\scriptsize fin}}$,  short enough to ensure linearity of the evolution of $\delta\bm{x}(t)$ but long enough 
to be able to clearly 
distinguish chaotic trajectories from regular ones on a plot of $\ln\delta\bm{x}(t)$ versus $t$:  the former
are straight lines of positive slope, while the latter are logarithm-like \citep{contopoulos_number_1978}. We also verify that the average of the FTMLE's over the ensemble of initial 
conditions does not depend on $t_{\mbox{\scriptsize fin}}$ as long as both criteria above are satisfied. 
\begin{figure}[ht]
\begin{center}
\includegraphics[scale=0.8]{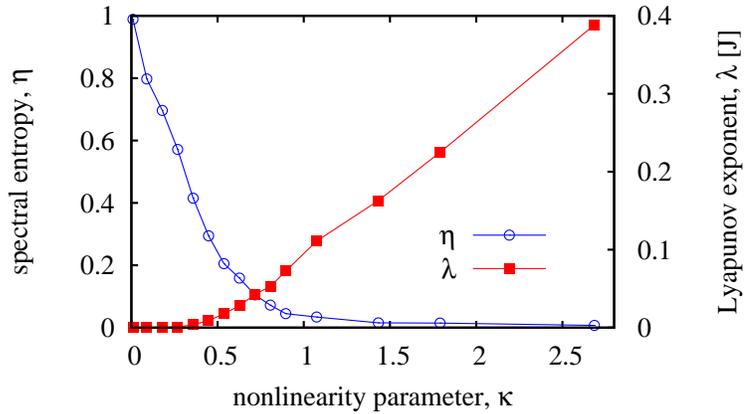}
\end{center}
\caption{
Ensemble-averaged finite-time maximal Lyapunov exponent, $\lambda$,
and normalized spectral entropy, $\eta$, as a function of the nonlinearity, $\kappa$. $N_s=21$.}
\label{fig:lyapunov_exp}
\end{figure}
%
In Fig.~\ref{fig:lyapunov_exp} the averaged FTMLEs
are plotted as a function of the interaction strength. 
There is a distinct regime with zero Lyapunov exponent for small $\kappa\lesssim 0.5$ and a strongly chaotic regime for $\kappa \gtrsim 1$ where all initial conditions have positive exponent.
Next we consider the relation between chaos and thermalization in the system.

In Fig.~\ref{fig:lyapunov_exp} the spectral entropy of the final time-averaged state, also averaged over 100 initial states
(drawn from the same ensemble that was used for the Lyapunov exponent calculation)
is plotted for each value of $\kappa$. For large nonlinearities, $\kappa \gtrsim 1$, the normalized spectral entropy goes to zero, indicating
remarkable agreement between the final state and the thermal predictions. Note that this corresponds to chaos threshold observed previously. 
For $\kappa \lesssim .5$ the normalized spectral entropy is above $.5$ signifying that 
during the time evolution the state of the system remains close to the initial state. 
\begin{figure}[ht]
\begin{center}
\includegraphics[scale=0.8]{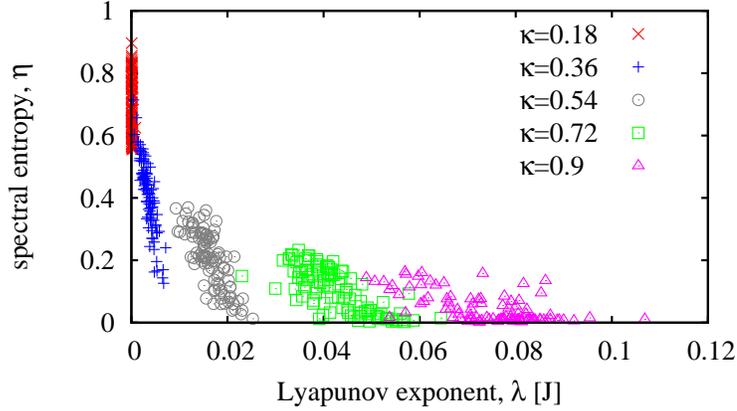}
\end{center}
\caption{\label{fig:entropy_vs_exp} Normalized spectral entropy of the final time-averaged state versus finite-time maximal Lyapunov exponent for each of the 100 initial condition used to compute the averaged value for $\kappa=0.36$, $0.54$, $0.72$, $0.9$. $N_s=21$.}
\end{figure}
In the Fig.~\ref{fig:entropy_vs_exp}, the normalized spectral entropy is plotted versus the FTMLE for each of the 100 individual runs for $\kappa=0.36$, $0.54$, $0.72$, $0.9$.  As seen in the plot, an individual initial state with larger FTMLE tends to have lower spectral entropy, i.e. to relax to a state which is closer to the thermal one. Beginning at $\kappa \approx 0.5$, where the averaged FTMLE is substantially non-zero, some of the initial states thermalize completely.

\paragraph{Method of Calculating Spectral Entropy}
For an individual realization, we calculate $\mu_0$ and $T$ given the norm and the final kinetic energy.  The spectral entropy is calculated for (1) the initial momentum distribution\sidenote{(check which energy)} and (2) the final mometum distribution, $\langle |\psi_n(t)|^2\rangle$, which is the time-averaged population of momentum mode $n$ from the dynamical code. 
\sidenote{(check time averaging process) - using ra4000 - last 4000 data point for all N=11, but I don't know about for this data. Additionally, I calculate $\eta(t)$ using both (1) the instantaneous momentum distribution and (2) the time-averaged momentum distribution. Note that for large enough nonlinearity where thermalization is expected, the spectral entropy of the time-averaged momentum distribution goes to zero, but for the instantaneous mometum distribution it does not, but rather fluctuates about a small value. (this is not surprising - expect fluctuations to decrease with increasing size of the system?)}

\subsection{Fluctuations}\label{sec:fluctuations}
In order to confirm that the system is thermal when $\eta\rightarrow 0$, we investigate the scaling of fluctuations of the kinetic energy for large $\kappa$.

For a system with $N$ particles and volume $V$ in thermal equillibrium, the fluctuations of extensive observables scale with the size of the system as $V^{-1/2}$, in the thermodynamic limit, $N \rightarrow\infty$, $V\rightarrow \infty$, $N/V=\text{ const}$.  
Consider an extensive quantity, $O$.  The temporal fluctuations of $O$ are given by
\begin{align*}
 \sigma_O^2 = \expv{O^2} - \expv{O}^2.
\end{align*}
where $\expv{O} = \frac{1}{T}\int_{t-\Delta T/2}^{t+\Delta T/2} dt \; O(t)$, and the relative fluctuations are $\sigma_O/\expv{O}$. Now consider two idential systems, $A$ and $B$, with corresponding extensive observables $O_A$ and $O_B$, such that $O=O_A+O_B$.  Then
\begin{align*}
 \sigma_O^2 &= \expv{O^2} - \expv{O}^2.\\
	    &= \expv{O_A^2 + O_B^2 + 2 O_A O_B} - \expv{O_A + O_B}^2\\
	    &= \expv{O_A^2} + \expv{O_B^2} + 2\expv{O_A O_B} - \expv{O_A}^2 - \expv{O_B}^2 - 2\expv{O_A}\expv{O_B}\\
	    &= \sigma^2_{O_A} + \sigma^2_{O_B} + 2\expv{O_A O_B} - 2\expv{O_A}\expv{O_B}
\end{align*}

For a system in thermal equilibrium, the two parts of the system are decorrelated, so that $\expv{O_AO_B} = \expv{O_A}\expv{O_B}$ and the last two terms cancel.  For a general rescaling of the system $N^\prime = \alpha N$ and the extensive observable $\expv{O}^\prime = \alpha\expv{O}$, the fluctuations scale as $\sigma_O^\prime = \sqrt{\alpha}\sigma_O$ and the relative fluctuations scale as $\sigma_O^\prime/\expv{O} = {\alpha}^{-1/2}\sigma_O/\expv{O}$.  In contrast, consider the case where two identical systems in identical initial states are concatenated in a regime where the behavior is regular.  In this case, there will be strong correlations between the two parts of the system\sidenote{justify, or explain better or just justify for the case of kinetic energy.}. In the extreme case of $O_A(t) = O_B(t)$, then $\sigma_O = 2\sigma_{O_A} = 2\sigma_{O_B}$ and the relative fluctuations $\sigma_O^\prime/\expv{O}$ will be constant, independent of the size of the system.

For the system under consideration, the thermodynamic limit is taken by scaling the number of atoms and length as $N_{a}^\prime = \alpha N_{a}$, $N_s^\prime = \alpha N_s$, while the interaction parameter $\mu_0$ and the lattice spacing, $a$, remain constant ($U$, $J$ = constant as well). In order to simulate the thermodynamic limit, the initial conditions are generated by concatenating $\alpha$ duplicates of the real-space wavefunction of the reference lattice. This is equivalent to generating an initial state with momentum modes ${\psi_k^\prime}$, given by $\psi_0^\prime=\psi_0$, $\psi_{\pm\alpha}^\prime=\psi_{\pm1}$, from the initial state in the $N_{s0}$ lattice, ${\psi_k}$.  Generating the initial conditions in this way preserves the average energy per particle, in units of $J=\hbar^2/2ma^2$. A small perturbation is added to the initial wavefunction to break the symmetry associated with the translational invariance. 

For the thermodynamic limit, we want to take the case where $N_{a} \rightarrow \infty$,$L \rightarrow \infty$, with $N_{a}/L = \text{const}$, with the lattice spacing, $a$, and interaction strength, $g$, remaining constant.  Consider the case where $L^{\prime}=\alpha L$, $N_{a}^{\prime} = \alpha N_{a}$. The scaling of relevant parameters is given in Table \ref{tab:TL}.

\begin{table}
\caption{\label{tab:TL}Thermodynamic Limit and Scaling}
\centering
\begin{empheq}[box=\fbox]{align*} 
\hbar\tilde{\omega}_1^{\prime}&=\frac{\hbar^2}{2m}\left(\frac{2\pi}{L^{\prime}}\right)^2=\frac{\hbar\tilde{\omega}_1}{\alpha^2} \\
 \cline{1-2}\tau_{\text{talbot}}^{\prime}&= \frac{2\pi}{\tilde{\omega}_1}=\alpha^2\tau_{\text{talbot}}\\
 \cline{1-2} \mu_0^{\prime}&=\frac{gN_{a}}{L}=\mu_0\\
 \cline{1-2} J^{\prime}&=\frac{\hbar^2}{2ma^2}=J\\
 \cline{1-2} U^{\prime}&=\frac{g}{a} = U\\
 \cline{1-2} U^{\prime}/J^{\prime}&=U/J\\
 \cline{1-2} \xi^{\prime}&=\frac{\mu_0}{\hbar\tilde{\omega}_1} =\frac{N_{a}N_s}{4\pi^2}\frac{U}{J} = \alpha^2 \xi\\
 \cline{1-2} \hbar\omega_n^{\prime}&=2J^{\prime}\left\lbrace1-\cos\left(\frac{2\pi n}{N_s^{\prime}}\right)\right\rbrace=2J\left\lbrace 1-\cos\left(\frac{2\pi n}{\alpha N_s}\right)\right\rbrace
\end{empheq}
\end{table}

We study the standard deviation of the fluctuations for systems with $N_s=21$ sites and $\alpha$=2,3,4. To compare fluctuations for different lengths, we calculate:
\begin{equation}
\bar{\sigma}(N_s,N_{s0}) \equiv \frac{\sigma_{E_k}(N_s)/\,\overline{E_k(N_s)}}{\sigma_{E_k}(N_{s0})/\,\overline{E_k(N_{s0})} }
\end{equation}
where $\sigma_E(N_s)$ and $\overline{E_k(N_s)}$ are the standard deviation and time average of the kinetic energy for a chain with $N_s$ lattice sites. The reference lattice size is $N_{s0}=21$ and the calculation begins after the system has already thermalized. \sidenote{propagation times:  averaging time $\approx 30 \lambda^{-1}$ and starting time for averaging is $\approx 30 \lambda^{-1}$, corresponding to the time it takes the divergence of the trajectories to reach one, starting from an initial separation of $10e-10$.}
\begin{figure}
\begin{center}
\includegraphics[scale=0.8]{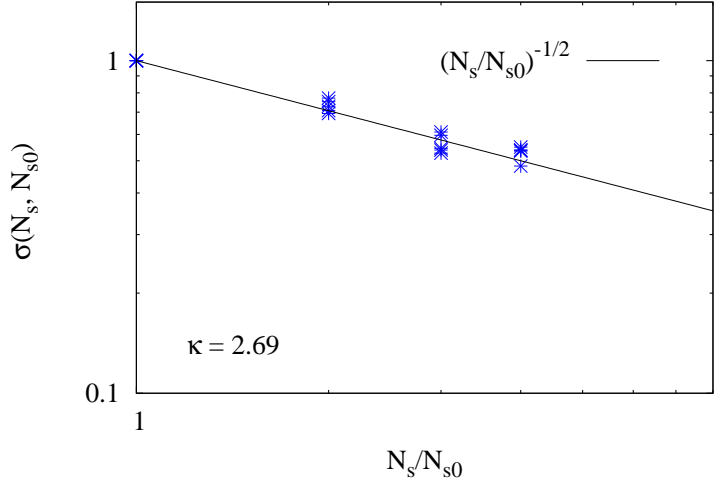}
\end{center}
\caption{\label{fig:fluctuations} Reltative fluctuations in kinetic energy. Normalized standard deviation/mean for $N_s$=21, 42, 63, 84 lattice sites and $N_0=21$. Data points are for five sample runs with equivalent initial conditions for different lattice sizes and $\kappa=2.69$.} 
\end{figure}
In Fig.~\ref{fig:fluctuations} we plot $ \bar{\sigma}(N_s,N_{s0})$ as a function of $\alpha = N_s/N_{s0}$ for various lattice sizes on a log-log scale for $\kappa=2.69$.  As can be seen clearly from the plot, the fluctuations scale as $N_s^{-1/2}$, indicating that the fluctuations are indeed thermal. \sidenote{For comparision, the scaling of the relative fluctuations of kinetic energy are plotted for $\kappa = [1] $ and equivalent initial conditions used in \ref{fig:fluctuations}. }

\section{Chaos Threshold for Different Lattice Sizes}
Let us start from the notion that the parameter $\kappa$ introduced in (\ref{kappa}) is the only dimensionless combination of the parameters 
of the problem that remains finite in the thermodynamic limit, $N_{\mbox{\scriptsize s}} \to \infty$, $N_{\mbox{\scriptsize a}}/N_{\mbox{\scriptsize s}} = \mbox{const}, J = \mbox{const},
U = \mbox{const}$. Curiously, the chaos threshold for $N_{\mbox{\scriptsize s}} = 21$  is at $\kappa \approx .5$, {\it i.e.} $\kappa \sim 1$.
Another observation comes from a related work \citep{villain_fermi-pasta-ulam_2000} on chaos threshold in NLSE with hard-wall boundary conditions. The authors find that the boundary between regular and chaotic motions of momentum mode, $n$, is given by $(\mu_{0}|\psi_n|^2)/(\hbar\omega_{1} n) \sim 1$, where $\hbar\omega_{1}$ is the lowest
excitation energy, {\it e.g.} the energy of the first excited mode in the case of the Hamiltonian (\ref{eqn:hamiltonian}). Assuming that the shape of the 
momentum distribution $|\psi_n|^2$ as a function of $n/N_{\mbox{\scriptsize s}}$ should be fixed in the thermodynamic limit, the left-hand side of the above relationship also remains finite.
These observations lead to a conjecture that the chaos criterion involves only intensive parameters and observables, {\it i.e. those that are finite in the thermodynamic limit}. 
\begin{figure}[ht!]
\begin{center}
\includegraphics[scale=0.8]{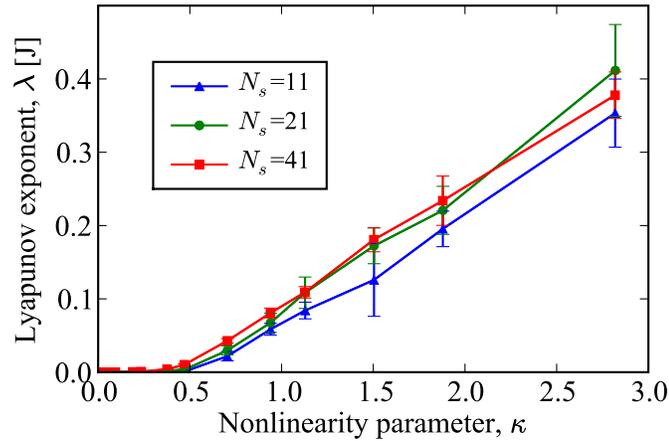}
\end{center}
\caption{\label{fig:lambda_size_scaling} Averaged Finite Time Lyapunov exponent, $\lambda [J]$, for three different system sizes, $N_{\mbox{\scriptsize s}}=11,21,41$.  
For each $\kappa$, the same energy-per-particle was used for each lattice size. The error bars represent one standard deviation. }
\end{figure}
Our test for the above conjecture is based on the fact that for a chaotic motion the majority of the trajectories cover the whole available phase space, and as a result the 
LE becomes (for a given set of parameters) a function of the energy only. This implies that for the same energy-per-particle and the same nonlinearity 
parameter, $\kappa$, Lyapunov exponents for different lattice sizes should be similar. In Fig.~\ref{fig:lambda_size_scaling} the time-averaged finite-time maximal Lypunov exponent is plotted for three different 
lattices, $N_{\mbox{\scriptsize s}}=11$, $21$, and $41$. For each $\kappa$, the same energy-per-particle (in units of J) is used for all three lattices. 

From the plot it is indeed evident that the LE is universal with respect to the size of the lattice and that the values of the LE for $N_{\mbox{\scriptsize s}}=11$ already give a very good estimate of both the value of the LE and the threshold.

\section{Two Parametric Theory of the Chaos Threshold}
The universality observed above suggests the most relevant pair of variables for mapping the chaos threshold, namely $\kappa$ and the total energy-per-particle, $\epsilon_{T}/J$.  In order to test these parameters, we independently vary the nonlinearity, $\kappa$ and the energy-per-particle, $\epsilon_T$. For each $\kappa$ and $\epsilon_T$, we generate ten initial states with microcanonical weight in the reduced phase space of three ($n=0,\pm 1$ ) or five ($n=0,\pm 1,\pm 2$ ) momentum modes.  It is necessary to generate initial states with five momentum modes $n=0,\pm 1,\pm 2$ because there is an upper limit on the energy of an initial state with only three modes occupied. The finite-time maximal Lyapunov exponent and normalized spectral entropy are calculated for each of the ten realizations and averaged over this ensemble.  For the rest of this work we will use the term Lyapunov exponent (LE) to denote the ensemble-averaged finite time maximal Lyapunov exponent and normalized spectral entropy (NSE) to denote the ensemble-averaged normalized spectral entropy, unless otherwise specified. The total data represents over 3200 runs. 

\begin{figure}[ht!]
\begin{center}
\includegraphics[scale=0.7]{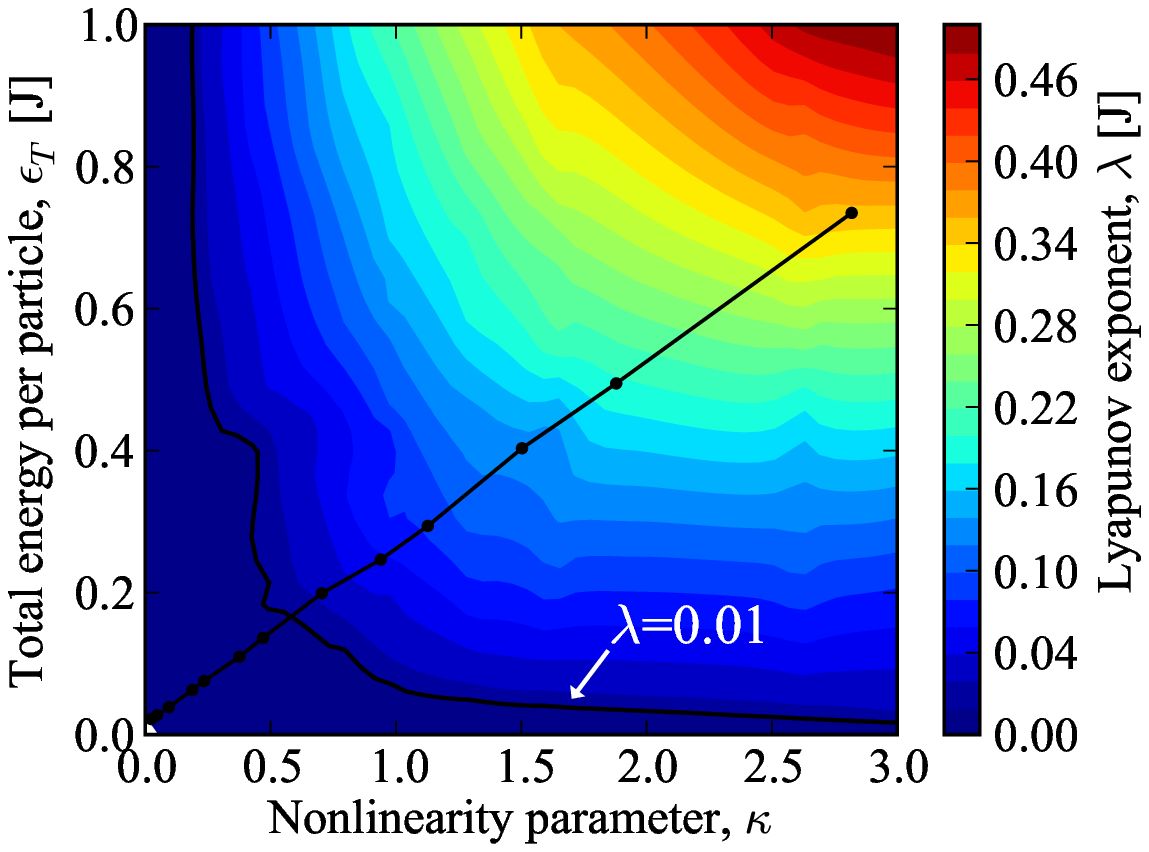}
\includegraphics[scale=0.7]{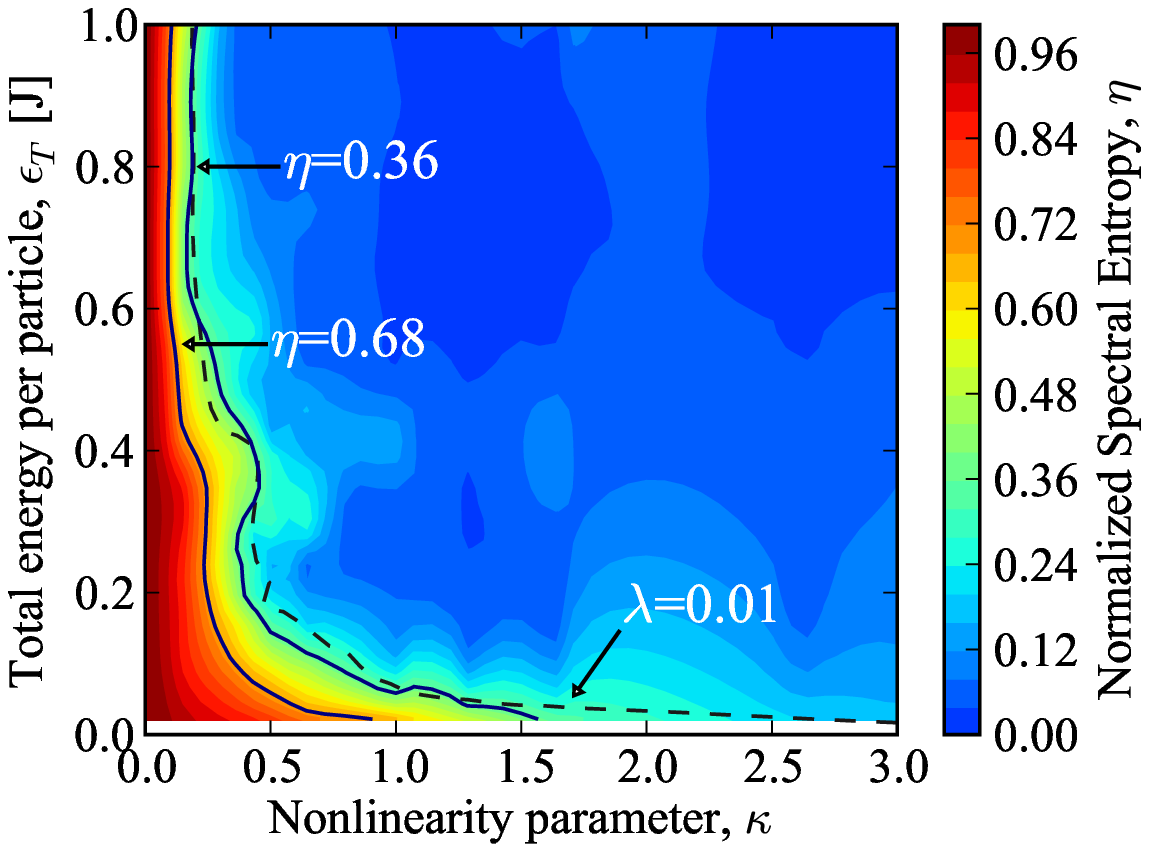}
\end{center}
\caption{\label{fig:lambda_contour}  (a) Contour lines of the averaged FTMLE versus the nonlinearity, $\kappa$, and energy-per-particle, $\epsilon_T = (H-H_{0})/N_{\mbox{\scriptsize a}}$, where $H$ is the Hamiltonian (\ref{eqn:hamiltonian}), and $H_{0} = -2J + (1/2)\mu_{0}$ is the ground state value of $H$. The solid contour line corresponds to $\lambda_c=0.01$. The diagonal solid line represents the set of energies and nonlinearities used in Fig.~\ref{fig:lambda_size_scaling}. (b) Contour lines of the averaged normalized spectral entropy versus the nonlinearity, $\kappa$ and energy-per-particle. Solid contour lines correspond to $\eta=0.68$ and $\eta=0.36$. For reference, the threshold line from the FTMLE in (a) is plotted (dashed line). $N_{\mbox{\scriptsize s}}=11$.
        }
\end{figure}

In Fig.~\ref{fig:lambda_contour}(a) contour lines of the LE for $N_{\mbox{\scriptsize s}}=11$ are plotted versus the nonlinearity parameter and total energy-per-particle. One can observe an initial plateau in the LE for $\lambda \lesssim\lambda_c=0.02$, given by the solid line. The threshold depends on both the nonlinearity and the total energy-per-particle.  Based on the parameter regime investigated, it appears that the threshold persists for small $\kappa$ no matter how much energy is present. For small energies it is unclear whether the threshold will persist or vanish for $\kappa \gg 1$.

After crossing the critical line the LE increases with uniform slope. The critical line resembles 
a hyperbola with the point of closest approach to the 
origin at $(\kappa,\,\epsilon_{T}) \sim (0.5,\,0.2 J)$, so that the hopping
parameter $J$ appears to be a relevant energy scale. This is probably not an
accident: for $\epsilon_{T} \gg J$ the dispersion law $\omega_{n}$
begins to deviate from the (quadratic) dispersion law of the continuous NLSE with periodic boundary conditions, which is integrable. 

The normalized spectral entropy was calculated for the same set of data runs and is plotted in Fig.~\ref{fig:lambda_contour}(b). There are two solid contour lines, at $\eta=0.68$ and $\eta=0.36$. The second contour line at $\eta=0.36$ follows closely the dotted line, which is the threshold from the Lyapunov exponent.  It is apparent that the two plots have the same general features, and that there is a strong correspondence between the presense of chaos and thermalization in the BHM. 

\begin{figure}[ht!]
\begin{center}
\includegraphics[scale=0.7]{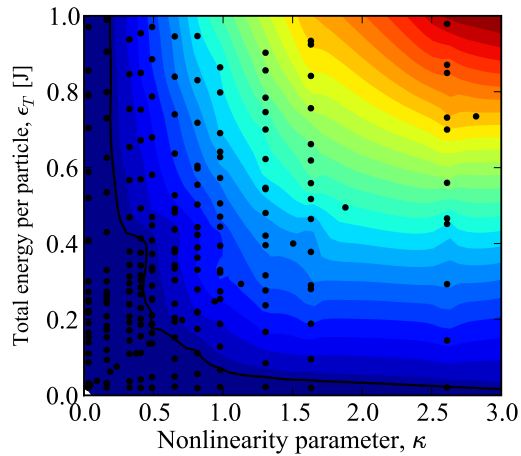}
\end{center}
\caption{\label{fig:contour_data_points} Data points used for interpolation for contour plots in Fig.~\ref{fig:lambda_contour} superimposed on data for Lyapunov exponent.
        }
\end{figure}

A few features deserve discussion. 

In the region of $\eta \gtrsim 0.68, \lambda=0$, which is enclosed by the first contour line in the NSE and the x- and y-axes, the system relaxes to a state that is closer to the thermal state even though the Lyapunov exponent is zero.  The time dependence of the spectral entropy reveals that this relaxation takes place very quickly, after which the spectral entropy remains flat for many fundamental cycles, indicating that no further relaxation will take place (See Fig.~\ref{fig:spect_ent_time_Eta1}). \sidenote{in the same time scale of the redistribution of kinetic and internal energies; dephasing; (1 talbot time?)}
This raises a few questions: To what state does the system relax in this region? It is possible to describe it by a constrained ensemble, where the constrained quantities are the conserved quantities of near-by intergrable systems? We will return to these questions later.

For $ \lambda \gg \lambda_c$, the region of strong chaos, the majority of initial states relax to the thermal state and all final states are close to the thermal. It is likely in this region, that full relaxation is not seen due to slow relaxation times and the spectral entropy would vanish for longer propagation times. 

We consider the two limits, $\kappa \rightarrow 0, \epsilon_T \sim J$ and $\epsilon_T/J \rightarrow 0, \kappa\approx 3$. In the limit of small $\kappa$, the chaos threshold and NSE contour line $\eta=0.36$ overlap and for $\kappa \rightarrow 0,\; \epsilon_T \gtrsim 0.6 J$ converge to a value that is independent of the total energy-per-particle. For the parameter region explored there is no indication that the threshold will vanish, even for very large energies.
This suggests a dependence on the ratio of the nonlinear to linear terms, similar to the critical Reynolds number found in FPU \citep{livi_equipartition_1985}. In the opposite limit of $\epsilon_T/J \rightarrow 0, \kappa \gtrsim 1.5 $ the behavior is quite different.  While the Lypunov exponent is zero, there is significant relaxation in the momentum distribution.  It is important to note that in the limit that $\epsilon_T \rightarrow 0 $ the initial state approaches the state where only the $n=0$ mode is populated, which is also the thermal state and thus the normalized spectral entropy is not well-defined in that limit.  For this reason, data is plotted for $\epsilon_T > 0.02 J$, the lowest energies simulated. However even for $\epsilon_T \sim 0.02 J$, relaxation is visible in the momentum distributions. 

\begin{figure}[ht!]
\begin{center}
\includegraphics[scale=0.7]{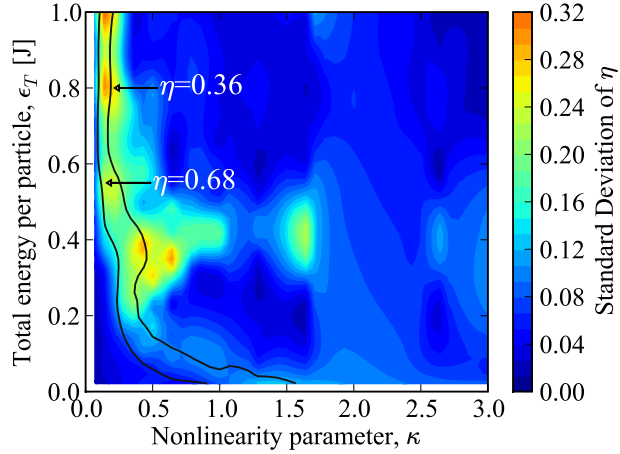}
\end{center}
\caption{\label{fig:nse_stdev} Contour lines of the standared deviation of the NSE versus the nonlinearity, $\kappa$ and energy-per-particle. Solid contour line corresponds to at $\eta=0.36$ and $\eta=0.68$  from Fig.~\ref{fig:lambda_contour}(b). $N_{\mbox{\scriptsize s}}=11$. 
        }
\end{figure}

In Fig.~\ref{fig:nse_stdev} the standard deviation of the normalized spectral entropy is plotted along with the contour lines at $\eta=0.36$ and $\eta=0.68$ from Fig.~\ref{fig:lambda_contour}(b). The contour line at $\eta=0.36$ follows closely the chaos threshold from Fig.~\ref{fig:lambda_contour}(a). Far above the threshold, where the $\eta \rightarrow 0$, the standard deviation is also small indicating that most of the states themalize, as expected.  Below $\eta=0.68$, (in the region bounded by the axes), relaxation is minimal and  the standard deviation is small, indicating that most initial states will not thermalize. In the vacinity of the threshold ($\lambda=\lambda_c$, which is close to $\eta=0.36$), the standard deviation is larger. We conjecture that this is because there is a large spread in the amount of relaxation expected for different initial states with the same parameters and/or that some states have not fully relaxed due to insufficient propagation times as a result of multiple relaxation time scales.

\begin{figure}
\begin{center}
\includegraphics[scale=0.8]{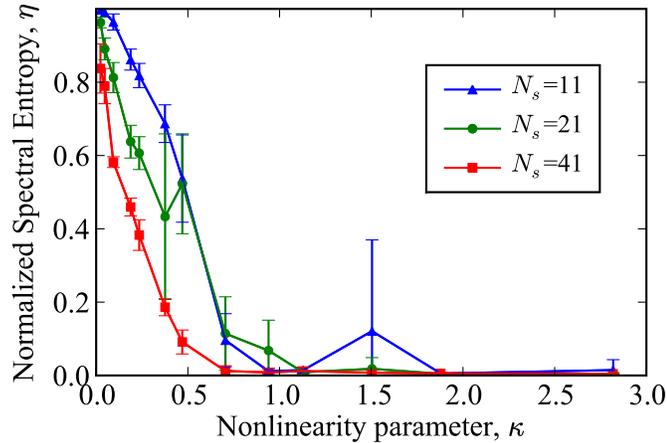} 
\end{center}
\caption{\label{fig:eta_size_scaling} Averaged normalized spectral entropy, $\eta$, for three different system sizes, $N_{\mbox{\scriptsize s}}=11,21,41$.  For each $\kappa$, the same energy-per-particle was used for each lattice size.
        }
\end{figure}

In Fig.~\ref{fig:eta_size_scaling} the normalized spectral entropy is plotted for three different lattice lengths, $N_{\mbox{\scriptsize s}}=11,21,41$ for the same energy-per-particle at each $\kappa$, using the same energy values as in Fig.~\ref{fig:lambda_size_scaling}.  While the main features are similar for the different lattice sizes, the size scaling of the NSE is not as universal as the scaling of Lyapuonv exponent. For small $\kappa$'s with the same energy-per-particle, more relaxation is seen in larger lattices. This suggests that the number of modes in involved in the dynamics may plays a role in relaxation. In addition the standard deviation of the NSE is larger than for the Lyapunov exponents. In contrast to the size scaling of the Lyapunov exponents, the variance of the NSE increases with smaller chains. We conjecture that the large variance can be attributed to multiple relaxation scales.  For example, for  $N_s=11 , \kappa=1.5$, individual states reveal that while most relax fully to the thermal state there is a single state that remains very far from thermal, which is the cause of the large variance.  

The large variance of the spectral entropy is the reason that there are more features in the contour plot of the normalized spectral entropy compared with the contour plot of the Lyapunov exponent.  Repeating the simulations for longer times would likely smooth some of the features of the NSE contour plot, and decrease the variance in regions where it is currently large.

\subsection{Thermalization Times and Slow Relaxation}
\begin{figure}[ht!]
\begin{center}
\includegraphics[scale=0.6]{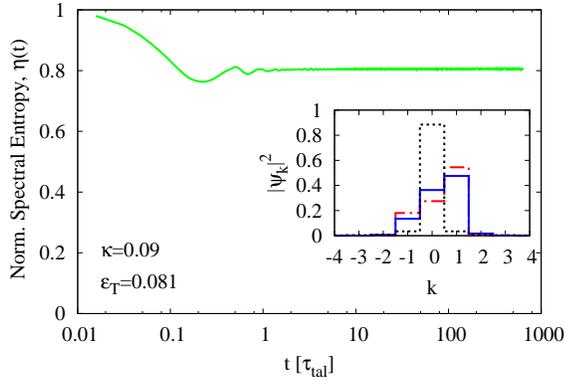} 
\end{center}
\caption{\label{fig:spect_ent_time_Eta1} Sample time dependence of normalized spectral entropy, $\eta$. $\kappa=0.09, \epsilon_T=0.081 J$, $N_{\mbox{\scriptsize s}}=21$. Inset: Initial (dashed red), final (solid blue) and thermal (dotted black) momentum distributions.
        }
\end{figure}

The time dependence of the normalized spectral entropy is plotted in Fig.~\ref{fig:spect_ent_time_Eta1} for $\kappa=0.09$ and $\epsilon_T=0.081 J$, which is in the non-chaotic region. The initial, thermal and final momentum distributions are plotted in the inset. The time-dependent spectral entropy is calculated from a running average over the momentum distribution and plotted in units of $\tau_{\text{tal}}$, the talbot time, which is the period associated with the lowest frequency of the non-interacting system with quadratic dispersion. After an initial relaxation during the first talbot time, the spectral entropy saturates and remains flat for close to $1000 \tau_{\text{tal}}$. The momentum distributions confirm that there is some relaxation, but that the state remains far from the thermal state.

\begin{figure}[ht!]
\begin{center}
\includegraphics[scale=0.6]{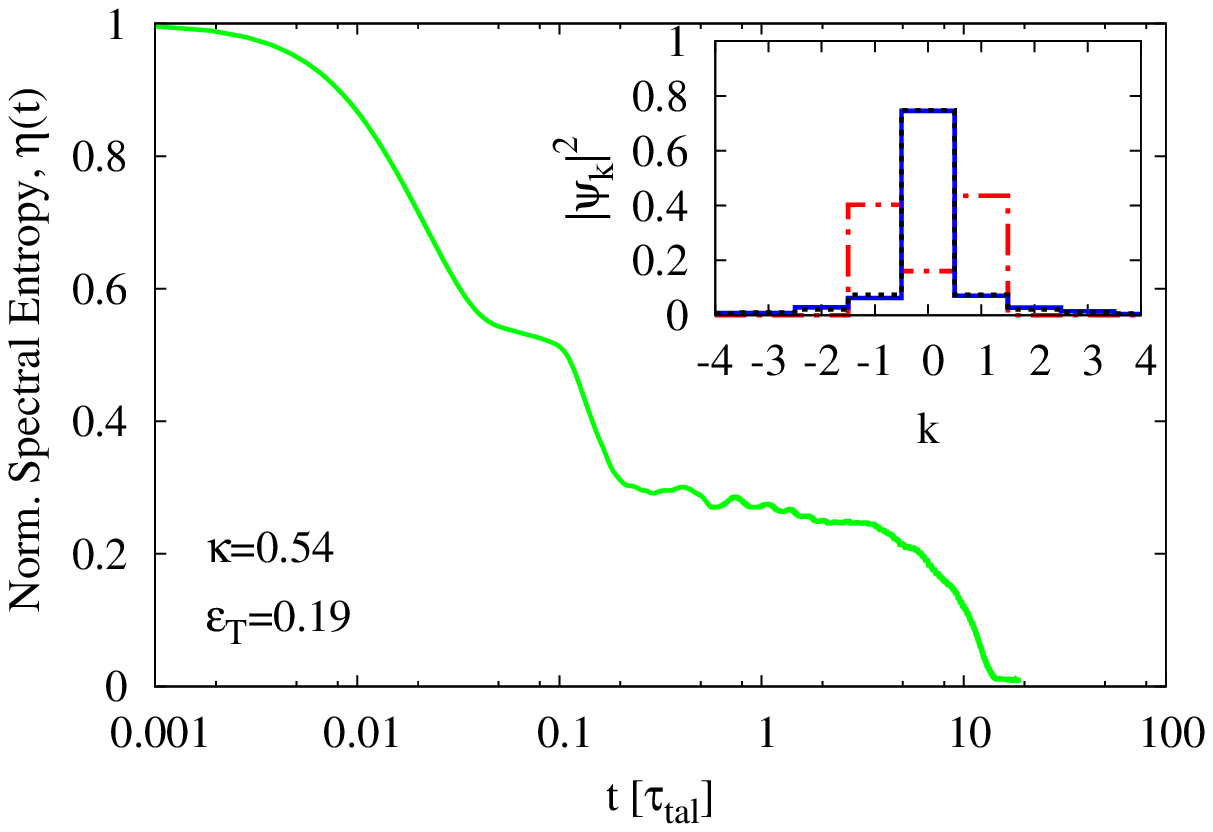}  
\includegraphics[scale=0.6]{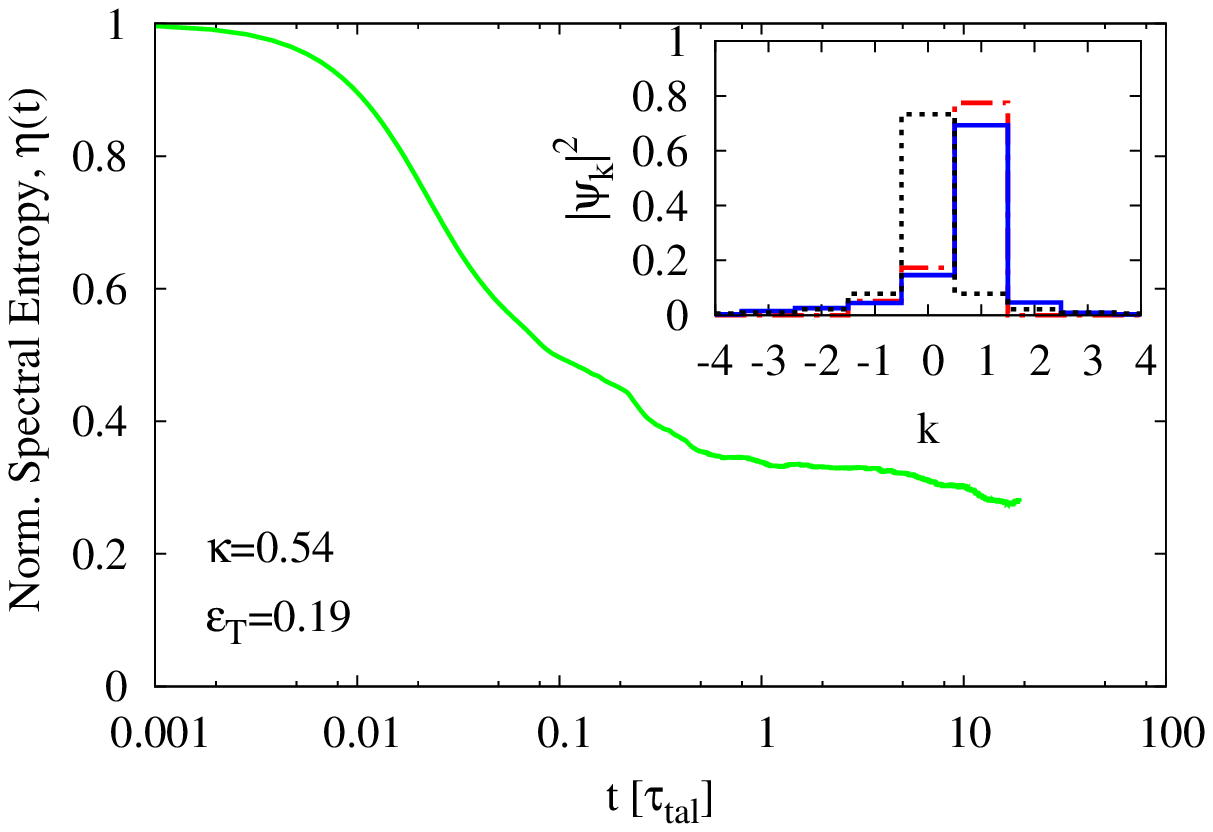}
\end{center}
\caption{\label{fig:spect_ent_time_Eta6} Sample time dependence of normalized spectral entropy, $\eta$, for two different initial states. $\kappa=0.54, \epsilon_T=0.19 J$ for both. $N_{\mbox{\scriptsize s}}=21$. Inset: Initial (dashed red), final (solid blue) and thermal (dotted black) momentum distributions.
        }
\end{figure}

In Fig.~\ref{fig:spect_ent_time_Eta6} the time dependence of the normalized spectral entropy is plotted for two different initial states that both have $\kappa=0.54$ and total energy $\epsilon_T=0.19 J$. In Fig.~\ref{fig:spect_ent_time_Eta6}(a) the normalized spectral entropy vanishes indicating that the state relaxes to the thermal state, which is also seen in the final momentum distribution. The normalized spectral entropy drops in several stages suggesting that there are multiple relaxations time scales.
In Fig.~\ref{fig:spect_ent_time_Eta6}(b), the state does not fully relax during the observed propagation time. After the initial relaxation, which is very similar to the previous case, the normalized spectral entropy slowly relaxes further but does not vanish in the observed time.  Both states have a positive finite-time maximal Lyapunov exponent and thus are in the chaotic regime.
These observations brings up several questions.  Given states in the chaotic sea with the same total energy and strength of nonlinearity, why do some fully relax while other do not? Will these states fully thermalize for longer propagation times?  What are the relevant time scales? What governs the slow relaxation times?

Comparing the momentum distributions for both plots, the initial momentum distribution is almost symmetric in Fig.~\ref{fig:spect_ent_time_Eta6}(a) so that the total quasi-momentum of the initial state is close to zero. Total quasi-momentum is not a conserved quantity of the BHM, although it is a conserved quantity of the noninteracting model, the continous model (in which case is becomes the true momentum) and the Ablowitz-Ladik discretization of the NLS. The total quasi-momentum is zero in the thermal state. For small $\kappa$'s, there is very little redistribution among the momentum modes and thus the total quasi-mementum is well conserved.  We call the total quasi-momentum a \ldq quasi-conserved quantity'' because it is not actually conserved in the BHM, but it conserved in the nearby integrable models and thus is expected to be conserved in the BHM when is \ldq close'' to one of the integrable limits. Proposed future work includes the investigation of the role of the conserved quantities of the nearby integrable models in the dynamics of the BHM. 

While these plots are sample runs, the pattern just described is observed in other individual runs for different values of $\kappa$ and $\epsilon_T$ in the chaotic region. Relaxation occurs on multiple time scales and the propagation times used in the simulations are long enough for the fast relaxation, but are not always long enough for the slow relaxation. For a given set of parameters, $(\kappa,\epsilon_T) $ there are different slow relaxation times for different initial states.  Insufficient propagation times are one possible reason for large variation of the individual NSE's observed in Fig.~\ref{fig:nse_stdev}. In the strongly chaotic regime, it is expected that the the normalized spectral entropy will converge to zero for longer propagation times. 
However it is also possible that for $\lambda \approx \lambda_c$, ($\eta \approx 0.36$) some initial states will not fully relax, even for very long times.  Furthermore, for $0.68 \gtrsim \eta \gtrsim 0.36$ it is likely that the variance will remain large. It is clear from Fig.~\ref{fig:spect_ent_time_Eta1} that some states do not relax, even for very long times.  

In summary, we have observed a threshold for chaos in the BHM, which depends on two parameters, the strength of the nonlinearity, $\kappa$ and the total energy-per-particle, $\epsilon_T/J$. Far above the threshold, the state relaxes to the one predicted by statistical mechanics.  Below the chaos threshold, we observe relaxation to a non-thermal steady-state.  For small nonlinearities, $\kappa$'s the chaos threshold and absense of thermalization persist even for large energy-per-particle, $\epsilon_T \sim J$. For regions just above the threshold, there are multiple relaxation times, with different intitial states relaxing on different time scales.
These observations bring up several questions: What is the origin of the chaos threshold?  What governs the long relaxation times? Is the nonthermal steady-state affected by the conserved quantities of the nearby integrable systems?

%
%
%

\chapter{Resonance Model and Failure of Chirikov's Criterion}
\label{chap:resonances_1DMFBH}
%
%
%
%
In this chapter we study a Hamiltonian coupling a single set of modes for initial states with few modes occupied.  First we use a perturbation theory expansion that is valid for small nonlinearities. Second we study nonlinear resonances of the one-dimensional mean-field Bose-Hubbard model to predict the chaos criterion by Chirikov's method of overlapping resonances.
\section{Perturbation Theory Study of BHM}
To study a perturbation theory expansion we introduce the a prefactor $\epsilon$ in the driving term, which will be set to unity in the end.

The full Hamiltonian in the momentum-space wavefunction representation is:
\begin{equation}
H = \sum_m \left( \hbar\omega_m|\psi_m|^2 - \frac{\mu_0}{2} |\psi_m|^4 \right)
  + \epsilon \frac{\mu_0}{2}\sum_{m^\prime,l^\prime, i^\prime, j^\prime}
    \psi_{m^\prime}^\ast \psi^\ast_{l^\prime} \psi_{i^\prime} \psi_{j^\prime}
\end{equation}
where the sum carries the restrictions: $m^\prime + l^\prime = i^\prime + 
j^\prime$; $m^\prime \neq i^\prime, j^\prime$;  $l^\prime \neq i^\prime, 
j^\prime$. 
We define the unperturbed Hamiltonian as 
\begin{equation}\label{eqn:H_0}
H_0 = \sum_m \left( \hbar\omega_m|\psi_m|^2 - \frac{\mu_0}{2} |\psi_m|^4 \right), 
\end{equation}
which consists of decoupled nonlinear oscillators. The frequencies of the unperturbed Hamiltonian are 
\[ \Omega_n \equiv 
\omega_n -(\mu_0/\hbar)|\psi_n|^2=\tilde{\omega}_1 n^2 -(\mu_0/\hbar)|\psi_n|^2.
\]
 The equations of motion of the full Hamiltonian are given by:
\begin{align}
\frac{\partial}{\partial t}\psi_n &= -\frac{i}{\hbar}\frac{\partial H}{\partial 
\psi_n^\ast}
= -i\left(\omega_n-\frac{\mu_0}{\hbar}|\psi_n|^2 \right)\psi_n - i\epsilon \frac{\mu_0}
{\hbar}\sum_{l,i,j; i\neq l,n}\psi_l^\ast\psi_i\psi_{n+l-i}\\
&= -i\Omega_n\psi_n - i\epsilon\frac{\mu_0}{\hbar}
\sum_{l,i;i\neq n,l}\psi_l^\ast\psi_i\psi_{n+l-i}
\end{align}

Now we make the following assumption:
(1) All modes except $n$ are treated as independent oscillators with equations of motion given by the unperturbed Hamiltonian,
$\dot{\psi_l} = -i\Omega\psi_l$ which has solutions of 
$\psi_l(t) = \bar{\psi}_le^{-i\Omega_l t}$
and (2) the nonlinear frequency of mode $n$ associated with the unperturbed Hamiltonian is fixed, $\Omega_n=$ const. With these restrictions, the equations of motion of mode $n$ become
\begin{equation}
\frac{\partial}{\partial t}\psi_n
= -i\Omega_n\psi_n - i \epsilon\frac{\mu_0}{\hbar}
\sum_{l,j; j\neq n,l}\bar{\psi}_l\bar{\psi}_j\bar{\psi}_{n+l-j}
e^{-i(\Omega_j+\Omega_{n+l-j}-\Omega_l)t}.
\end{equation}

\subsection{Dynamics of an Initially Unpopulated Mode }

First we study the maximum value of initially unpopulated mode in the perturbation theory expansion. We consider the initial state where a block of modes from $ [-N_{max}, N_{max}]$ are equally populated and study the time dynamics of mode $Q=N_{max}+1$. In order to do this, we seek a solution of
\begin{equation}
\frac{d}{dt}\psi_n + i\Omega_n\psi_n = f(t)
\end{equation}
We want to find an integrating factor $h(t)$, satisfying $\dot{h}(t)=i\Omega_nh(t)$ such that
\begin{equation}
\frac{d}{dt}[\psi_n(t)h(t)] = h(t)\frac{d}{dt}\psi_n + i\Omega_n h(t)\psi_n 
=f(t) h(t).
\end{equation}  
The solution is $h(t)= \exp(i\Omega_n t)$. Let $f(t)=\sum_\alpha 
g(\alpha)\exp(i\Delta_\alpha t)$ where
\begin{equation}
\begin{split}
\sum_\alpha &= \sum_{l,j; j\neq n,l} \\
g(\alpha) &= -\frac{\mu_0}{\hbar}\bar{\psi}_l\bar{\psi}_j\bar{\psi}_{n+l-j} \\
\Delta_\alpha &=  \Omega_l -\Omega_j - \Omega_{n+l-j}.
\end{split}\end{equation}

Next we integrate over time
\begin{equation}\begin{split}
\int_0^\tau dt \frac{d}{dt}\left(\psi_n(t)e^{i\Omega_n t} \right) &= 
\int_0^\tau dt \sum_\alpha g(\alpha)e^{i\left(\Omega_n+\Delta_\alpha\right)t}\\
\psi_n(\tau)e^{i\Omega_n \tau}-\psi_n(0) &=\sum_\alpha \frac{g(\alpha)}
{i\left(\Omega_n+\Delta_\alpha\right)}\left( e^{i\left(\Omega_n + \Delta_\alpha
\right)\tau}-1\right)\\
\psi_n(t) &=\sum_\alpha \frac{g(\alpha)}{i\left(\Omega_n+
\Delta_\alpha\right)}\left( e^{i \Delta_\alpha t}-e^{-i\Omega_n t}\right)
\end{split}\end{equation}
The time-averaged value of $\psi_n(t)$ is given by
\begin{equation}
\begin{split}
\overline{|\psi_n|^2} = &\lim_{T\rightarrow\infty}\frac{1}{T}\int_{0}^{T} dt |\psi_n(t)|^2\\
=&\lim_{T\rightarrow\infty}\frac{1}{T}\int_{0}^{T} dt \sum_{\alpha,\beta}\frac{g(\alpha)
g(\beta)\left(e^{i(\Delta_\alpha - \Delta_\beta)t} - e^{i(\Delta_\alpha-
\Omega_n)t} - e^{-i(\Delta_\beta-\Omega_n)t} - 1 \right)}{\left(\Omega_n+\Delta_\alpha\right)\left(\Omega_n + \Delta_\beta 
\right)}\\
=&\sum_{\alpha,\beta}\frac{g(\alpha)g(\beta)}{\left(\Omega_n+
\Delta_\alpha\right)\left(\Omega_n + \Delta_\beta 
\right)} \lbrack\delta(\Delta_\alpha-\Delta_\beta)
- \delta(\Delta_\alpha-\Omega_n) - \delta(\Delta_\beta-\Omega_n)-1 \rbrack\\
=&\sum_{\alpha,\beta}\frac{g(\alpha)g(\beta)}{\left(\Omega_n+
\Delta_\alpha\right)\left(\Omega_n + \Delta_\beta 
\right)} \lbrack\delta(\Delta_\alpha-\Delta_\beta)
-1 \rbrack,
\end{split}
\end{equation}

We introduce the dimesionless variables, 
\[ \boxed{\xi=\frac{\mu_0}{\hbar\tilde{\omega}_1}} , \qquad \qquad \left( \xi = \kappa\left(\frac{N_s}{2\pi}\right)^2\right) \]
which is directly proportional to the nonlinearity parameter, $\kappa$, but scales with the size of the system in the thermodynamic limit. We also introduce a new time scale, $\tau=\tilde{\omega}_1 t$.

\paragraph{Occupation of mode $N_{max}+1$ for Quadratic Dispersion}

We are interested in the population of mode $Q \equiv N_{max}+1$  when 
initially modes $(-N_{max},N_{max})$ are equally occupied with population $\bar{\psi}$.

When only low momentum modes are excited, the true cosine dispersion laws can be approximated by a quadratic dispersion relation, $\omega_n = \tilde{\omega}_1 n^2$ and the unperturbed frequencies can be written as $\Omega_l = \omega_l(l^2 - \xi|\bar{\psi}|^2)$.  Dropping terms $O(\xi)$ in the denominator, the equation of motion for
$\psi_Q$ can be written as
\begin{equation}\begin{split}
\psi_Q(t) = &-\epsilon\xi\tilde{\omega}_1e^{-i\Omega_Q t}
\bar{\psi}^3\sum_{|j|,|l|<Q; j\neq l} \frac{e^{i(\Omega_Q+\Omega_l-\Omega_j-
\Omega_{Q+l-j})t}-1}{\Omega_Q + \Omega_l -\Omega_j - \Omega_{Q+l-j}}\\
\psi_Q(\tau) = &-\epsilon\xi\bar{\psi}^3\sum_{|j|,|l|<Q\ j\neq l}\frac{e^{i(l^2-j^2-(Q+l-j)^2)\tau}-e^{-iQ^2\tau}}
{Q^2 + l^2-j^2-(Q+l-j)^2}\\
\psi_Q(\tau) = &-\epsilon\xi\bar{\psi}^3\sum_{|j|,|i|<Q}\bar{\psi}_j\bar{\psi}_{Q+l-j}
\frac{e^{i(2(i-Q)(j-Q)-Q^2)\tau}-e^{-iQ^2\tau}}{2(i-Q)(j-Q)}
\end{split}\end{equation}
where we make the substitution $l=j+i-Q$ in the last expression.
Consider the case where modes $0,\pm 1$ are initially occupied with equal occupation, $\bar{\psi}$ and $Q=2$. The nonzero terms in the sum correspond to modes $(2,0\leftrightarrow 1,1), \; (2,-1 \leftrightarrow 1,0),\; (2,-1 \leftrightarrow 0,1)$.
The time evolution becomes:
\begin{equation}\begin{split}
\psi_2(t) =& -e^{-i\Omega_2t}\frac{\mu_0\bar{\psi}^3}{\hbar}\left[
\frac{e^{i(\Omega_2+\Omega_0-2\Omega_1)t} - 1}{\Omega_2+\Omega_0-2\Omega_1}
+2\frac{e^{i(\Omega_2+\Omega_{-1}-\Omega_0-\Omega_1)t} - 1}
{\Omega_2 + \Omega_{-1} - \Omega_0 -\Omega_1}\right]\\
\psi_2(\tau) =& \xi\bar{\psi}^3e^{-i4\tau}\left(
\frac{e^{i2\tau}-1}{2}+ 2\frac{e^{i4\tau} - 1}{4}\right)=
\xi\bar{\psi}^3\left(
\frac{1 + e^{-i2\tau} - 2e^{-i4\tau}}{2}\right)
\end{split}\end{equation}
So that for $N_{max}=1$,the time evolution of $|\psi_2(\tau)|^2$ is:
\begin{equation}
|\psi_2(\tau)|^2 = \xi^2\bar{\psi}^6\left(\frac{
3 - \cos(2\tau)- 2\cos(4\tau)}{2}
\right)
\end{equation}
\paragraph{Occupation of mode $N_{max}+1$ for Cosine Dispersion}
In the previous case, we assumed a quadratic dispersion, which corresponds to the free space.  The true dispersion of the BHM, which is a lattice model, is cosine, 
\begin{equation}
\omega_n=-\tilde{\omega}_1\frac{N_s^2}{2\pi^2}\left(1-N_s^2\cos\left(\frac{2\pi n}{N_s}\right).
\right)
\end{equation}

and the time evolution of mode population $|\psi_2(\tau)|^2$ becomes
\begin{align*}
\psi_2(t) = -\xi\bar{\psi}^3 e^{iN_s^2\cos(4\pi/N_s)t}\frac{\tilde{\omega}_1}{N_s^2}\Bigg(
&\frac{e^{it N_s^2(\cos(4\pi/N_s)+1-2\cos(2\pi/N_s))} - 1}
{-\cos(4\pi/N_s)-1+2\cos(2\pi/N_s)}\\
& +2\frac{e^{it N_s^2(\cos(4\pi/N_s)-1)} - 1}
{-\cos(4\pi/N_s)+1}\Bigg) 
\end{align*}

Defining $\Delta_1\equiv N_s^2(\cos(4\pi/N_s)+1-2\cos(2\pi/N_s))/\tilde{\omega}_1$, $\Delta_2\equiv N_s^2(\cos(4\pi/N_s)-1)/\tilde{\omega}_1$, the

\begin{equation}
|\psi_2(\tau)|^2 = \xi^2\bar{\psi}^6\left(
\frac{e^{i\tau \Delta_1} - 1}{\Delta_1}
+2\frac{e^{i\tau \Delta_2} - 1}{\Delta_2}\right)
\left(\frac{e^{-i\tau \Delta_1} - 1}{\Delta_1}
+2\frac{e^{-i\tau \Delta_2} - 1}{\Delta_2}\right)
\end{equation}
\begin{equation}
\begin{split}
|\psi_2(\tau)|^2 = \xi^2\bar{\psi}^6\Bigg(
&\frac{e^{i\tau \Delta_1} - 1}{\Delta_1}
+2\frac{e^{i\tau \Delta_2} - 1}{\Delta_2}\Bigg)
\Bigg(\frac{e^{-i\tau \Delta_1} - 1}{\Delta_1}
+2\frac{e^{-i\tau \Delta_2} - 1}{\Delta_2}\Bigg)\\
= \xi^2\bar{\psi}^6\Bigg(
&\frac{2(1-\cos(\tau\Delta_1))}{\Delta_1^2} + 8\frac{1-\cos(\tau \Delta_2)}{\Delta_2^2}\\
&+\frac{4\left(\cos(\tau(\Delta_1-\Delta_2)) - \cos(\tau\Delta_1) - \cos(\tau\Delta_2) + 1 \right)}{\Delta_1\Delta_2}\Bigg)
\end{split}\end{equation}

 \sidenote{Note that in the units used, $m=L=\hbar=1$, $\tilde{\omega}_1=2\pi^2$.- Fix this!}
 This expression for the time evolution of the modes is expected to be accurate for small values of $\xi$.  Comparison of these predictions with numerics shows very good agreement for small $\xi$. 

\subsection{Nonlinear Frequencies in the Real-Space Dynamics}
\begin{center}
\begin{figure}
\includegraphics[scale=0.8]{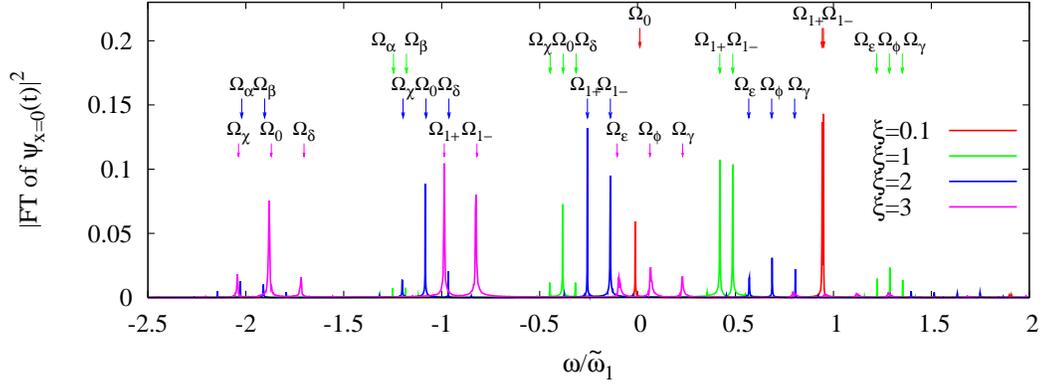}
\caption{\label{fig:fft_pos0}Frequencies of the real-space wavefunction at the center of the box, $|FT[\psi_{x=0}(t)]^2|$ for $\xi=0.1,1,2,3$.}
\end{figure}
\end{center}

For small values of $\xi$, the perturbation theory expansion accurately predicts the time-dynamics of the real-space wavefunctions. Next we consider the same model as the nonlinear coupling increases. In Fig.~\ref{fig:fft_pos0} we plot the modulus-squared of the Fourier transform of the real-space wavefunction at the center of the box, $\psi_{x=0}(t)$, is plotted for $N_s=21$ and $\xi=0.1,1,2,3$ for an initial state with momentum modes $\psi_{n=0}, \psi_{n=\pm1}$ occupied.  Using a non-interacting model (no coupling between modes, but nonlinear terms associated with each mode), the predicted value of each of the resonances is $\Omega_n = \omega_n - \frac{\mu_0}{\hbar^2} I_n$.  Taking $I_n$ as the time-averaged value of $I_n(t) $.   The driving term is of the form $e^{i(\Omega_m+\Omega_n-\Omega_l)}$, so additional frequencies are expected to be linear combinations of three other frequencies. 
\begin{center}
\begin{tabular}{ll}
  $ \Omega_{\alpha}=2\Omega_0-\Omega_{1\text{-}}$ &  \\ 
  $ \Omega_{\beta}=2\Omega_0-\Omega_{1\text{+}}$ &  \\ 
  $ \Omega_{\chi}=\Omega_0+\Omega_{1\text{+}}-\Omega_{1\text{-}}$ &  \\ 
  $ \Omega_{\delta}=\Omega_0+\Omega_{1\text{-}}  - \Omega_{1\text{+}}$ &  \\ 
  $ \Omega_{\epsilon}=2\Omega_{1\text{+}} - \Omega_0 $ &  \\ 
  $ \Omega_{\phi}=\Omega_0-\Omega_{1\text{+}}  - \Omega_{1\text{-}}$ &  \\ 
  $ \Omega_{\gamma}=2\Omega_{1\text{-}}  - \Omega_0 $ & 
\end{tabular}
\end{center}
\section{Chirikov's Criterion}
Next, we study the Bose-Hubbard Model in the Born-Oppenheimer approximation (BOA).  In the BOA, the populations of all but one mode are fixed.  The fixed modes rotate in phase-space with constant frequency. The motion of the free mode is coupled to the the fixed modes and motion is governed by the equations of motion.

In the next section, we deduce the Chirikov chaos criterion by studying single resonances within the Born Oppenheimer approximation.  Within the resonant approximation it is assumed that near a resonance, the resonance dominates the motion, so that driving terms can be studies independently.  Within the BOA, we assume that the action variable of the single mode under study is small compared to the other modes, $p,q,r$, and that the other modes can be well-described by the integrable Hamiltonian, $H_0$ that describes independent modes,
\[
 H_0 = \sum_m \left(\omega_m - \frac{\mu_0}{2\hbar^2}I_m \right)I_m.
\]
 Solving the equations of motion for the action variable, $\dot{I_m} = -\pd{H_0}{\theta_m}$ gives $I_m = \text{const} \equiv \bar{I}_m$, while the equation of motion for the angle variable, $\dot{\theta}_m=\frac{\partial H_0}{\partial I_m} $  gives  
\begin{equation}
\theta_m = \left(\omega_m - \frac{\mu_0}{\hbar^2}\bar{I}_m\right) t \equiv\Omega_m(\bar{I}_m) t. 
\end{equation}
Thus the action variable of modes $p,q,r$ is fixed and the angle rotates with constant frequency.

\paragraph{Summary of assumptions:} Throughout this section, we assume that 
\begin{enumerate}
 \item Single resonance approximation: we isolate individual driving terms of the Hamiltonian and study the resonances of these models.
 \item Born-Oppenheimer approximation: The population of all modes, except the one under study are fixed.  The fixed modes have frequencies $\Omega_m = \left(\omega_m - \frac{\mu_0}{\hbar^2}\bar{I}_m\right)$.
\item Quadratic Dispersion: For low-energy modes the dispersion $\omega_n = 2\tilde{\omega}_1(N_s/2\pi)^2[1-\cos (2 \pi n/N_s)] $ can be approximated by a quadratic dispersion, $\omega_n = \tilde{\omega}_1 n^2$.
\item Equal population of fixed modes:  The populations of the fixed modes are taken to be equal in order to reduce the number of parameters.
\end{enumerate}
\subsection{Classification of Resonances}
We classify the single resonances into three categories. We use the notation $(p,q)\rightarrow (n,r)$, where the first two modes are the ``feeding'' modes and the second two modes are the ``filling modes''.  Mode $n$ is the mode of interest, whose dynamics are governed by the resonant Hamiltonian, while $p,q$ and $r$ are three \textit{different} modes with fixed action variables.  Additionally, the momentum indices must satisfy $p+q=r+n$. The resonances are classified according to the order of the exponent of mode $n$ in the driving term. The three classes of Hamiltonians are:
\begin{enumerate}
    \item  \textbf{First-order Resonance} $(p,q) \rightarrow (r,n)$\\
	The two feeding modes $p,q$ fill two different modes, $r,n$.  The first-order resonant Hamiltonian is
	\begin{equation}
	H_n = \omega_n I_n - \frac{\mu_0}{2\hbar^2}I_n^2
	+\frac{4\mu_0}{\hbar^2}\left(I_n I_p I_q I_r\right)
	^{1/2}\cos\left(\theta_n +\theta_r - \theta_p - \theta_q\right).
	\end{equation}
  \item  \textbf{1-R Resonance} $(p,p) \rightarrow (r,n)$\\
	A single feeding mode $p$ fills two different modes, $r,n$. The 1-R resonant Hamiltonian is
	\begin{equation}  
	H_n = \omega_n I_n - \frac{\mu_0}{2\hbar^2}I_n^2
	+\frac{2\mu_0}{\hbar^2}\left(I_n I_r I_p^2\right)
	^{1/2}\cos\left(\theta_n +\theta_r - 2\theta_p\right).
	\end{equation}
	The 1-R resonance differs from the first-order resonance only in the prefactor in front of the driving term.
   \item  \textbf{Second-order Resonance} $(p,q) \rightarrow (n,n)$\\
	Two different feeding modes, $p,q$ fill the same mode, $n$. The second order resonant Hamiltonian is
	\begin{equation}
	H_n = \omega_n I_n - \frac{\mu_0}{2\hbar^2}I_n^2 
	+\frac{2\mu_0}{\hbar^2}I_n \left(I_p I_q\right)^{1/2}\cos\left(2\theta_n - \theta_p - \theta_q\right)
	\end{equation}

\end{enumerate}

\begin{table}
\caption{\label{tab:resonances} Resonance Parameters }
\centering
\renewcommand{\arraystretch}{0.4}
\begin{tabular}{|c|c|l|l|}
\hline  & & & \\Order & Modes & Bare Detuning & \begin{tabular}{c} Mean Occupation \\ of Fixed Modes\end{tabular}\\
\hline  & & & \\First 
	& $\begin{array}{c} p,q \rightarrow r,n \\ \\ p+q = r+n\end{array}$ 
	& $\Delta_{01} \equiv p^2+q^2 - r^2-n^2$  
	& $\bar{I}_{1} \equiv (\bar{I_p}\bar{I_q}\bar{I_r})^{3/2} $\\
\hline  & & &\\1R 
	& $\begin{array}{c} p,p \rightarrow r,n \\ \\ 2p = r+n\end{array}$ 
	& $\Delta_{01r} \equiv 2p^2 - r^2-n^2$ 
	& $\bar{I}_{1r} \equiv  (\bar{I_p}^2\bar{I_r})^{3/2}$ \\
\hline  & & &\\Second 
	& $\begin{array}{c} p,q \rightarrow n,n \\ \\ p+q = 2n\end{array}$ 
	& $\Delta_{02} \equiv  p^2 + q^2 - 2n^2$
	& $ \bar{I}_{2} \equiv  (\bar{I_p}\bar{I_q})^{1/2}$ \\
\hline
\end{tabular}
\\[1em]
\begin{tabular}{|c|l|l|}
\hline  & & \\Order & Drive Frequency & Nonlinear Detuning \\
\hline  & & \\First 
	& $\begin{aligned}\nu_1  &\equiv  \Omega_p+\Omega_q - \Omega_r \\ 
		&= \omega_p + \omega_q - \omega_r -\frac{2 \mu_0}{\hbar}(\bar{I}_p + \bar{I}_p - \bar{I}_r) \end{aligned}$
	& $\begin{aligned} \Delta_{1}  &\equiv (\nu_1 - \omega_n )/\tilde{\omega}_1\\ & = \Delta_{01} - \xi \bar{I}_{1} \end{aligned}$\\
\hline  & & \\1R 
	& $\begin{aligned} \nu_{1r}  &\equiv 2\Omega_p - \Omega_r \\
		&= 2 \omega_p - \omega_r - \frac{2 \mu_0}{\hbar}(2 \bar{I}_p - \bar{I}_r)\end{aligned}$
	& $\begin{aligned}\Delta_{1r} & \equiv (\nu_{1r} - \omega_n )/\tilde{\omega}_1 \\ 
		&= \Delta_{01r}-\xi \bar{I}_{1r}\end{aligned}$\\
\hline  & & \\Second 
	& $ \begin{aligned} \nu_2  &\equiv  \Omega_p + \Omega_q \\
		&= \omega_p + \omega_q -\frac{2 \mu_0}{\hbar}(\bar{I}_p + \bar{I}_q) \end{aligned}$
	& $\begin{aligned} \Delta_{2} & \equiv (\nu_{2} - 2\omega_n )/(2\tilde{\omega}_1) \\
		&= \Delta_{02}/2 - \xi \bar{I}_{2}  \end{aligned}$\\ 
\hline
\end{tabular}

\end{table}

In Table \ref{tab:resonances} the important parameters and definitions are listed for the three classes of resonances. The parameters include: the geometric mean of the fixed modes, $\bar{I}$, the bare detuning, $\Delta_0$, and the frequency of the drive $\nu$.  For the nonlinear detuning, the second equality holds when all the fixed modes have the same population. For a generic driving term of each type, the bare detuning, $\Delta_0$ is always negative for 1-R terms, always positive for 2nd order terms and can be positive or negative for 1st order terms (see Table \ref{tab:bare_detuning}).  Next we will study the resonances in each class of Hamiltonians.

\begin{table}
\caption{\label{tab:bare_detuning}Bare Detuning: Generic Values}
\centering
\renewcommand{\arraystretch}{0.4}
\begin{tabular}{|c|c|l|}
\hline & & \\Order & Modes & Bare Detuning\\ 
\hline & & \\
First & $n+m,n+l \rightarrow n,n+l+m$  & $\begin{aligned}\Delta_{01} \equiv & (n+m)^2+(n+l)^2 \\ &- (n+m+l)^2-n^2 \\ = & -2ml \end{aligned}$  \\
\hline & & \\1R &   $ n+m,n+m \rightarrow n,n+2m$  & $\begin{aligned}\Delta_{01r} \equiv& 2(n+m)^2 \\&- (n+2m)^2-n^2 \\  =&  - 2m^2 \end{aligned}$ \\
\hline & & \\Second & $n+m,n-m \rightarrow n,n $ & $\begin{aligned}\Delta_{02} \equiv& (n+m)^2+(n-m)^2\\& - 2n^2\\ =& 2 m^2\end{aligned}$\\ 
\hline
\end{tabular}
\end{table}
\subsection{First-Order Resonances in BOA}
Consider a first-order resonance where $n+r=p+q$; $p,q\neq n,r$. Fix the populations 
of the feeding modes, $p,q$ and the filling mode $r$, so that the only
variables are the population and angle of mode $n$.  The frequencies of modes $p,q,r$ are fixed to their values in the unperturbed Hamiltonian (\ref{eqn:H_0}).  Solving $\dot{\theta}_n=\frac{\partial H_0}{\partial I_n} $ for the angle gives 
\begin{equation}
\theta_n = \left(\omega_n - \frac{\mu_0}{\hbar^2}\bar{I}_n\right) t \equiv\Omega_n(\bar{I}_n) t. 
\end{equation}
In this approximation, which we call the Born-Oppenheimer approximation (BOA), we keep the eight terms corresponding to $(p,q) \rightarrow (r,n)$
from the full Hamiltonian. The Hamiltonian for first-order resonances in the Born-Oppenheimer approximation is:
\begin{equation}
\begin{split}
H_n = & \omega_n I_n - \frac{\mu_0}{2\hbar^2}I_n^2
+\frac{4\mu_0}{\hbar^2}\left(I_n I_p I_q I_r\right)
^{1/2}\cos\left(\theta_n +\theta_r - \theta_p - \theta_q\right)\\
= & \omega_n I_n - \frac{\mu_0}{2\hbar^2}I_n^2
+\frac{4\mu_0}{\hbar^2}\left(I_n I_p I_q I_r\right)
^{1/2}\cos\left(\theta_n -\nu t + \phi_1\right)
\end{split}
\end{equation}
where $\nu \equiv \Omega_p+\Omega_q - \Omega_r$.
Now we let $\bar{I}_1 \equiv (\bar{I}_p\bar{I}_q\bar{I}_r)^{1/3}$, divide by $\tilde{\omega}_1$, set $\hbar=1$ and use $\xi=\mu_0/(\hbar\tilde{\omega}_1)$ to get
\begin{equation}
h_n \equiv \frac{H_n}{\tilde{\omega}_1} = \frac{\omega_n}{\tilde{\omega}_1} I_n - \frac{\xi}{2} I_n^2
+4 \xi \left(\bar{I}_1\right)^{1/2}\cos\left(\theta_n -\nu t + \phi_1\right).
\end{equation}

We seek a canonical transformation to a rotating reference frame ($I_n, \theta_n \rightarrow \tilde{I}_n,\tilde{\theta}_n)$, where the new angle, $\tilde{\theta}_n$ is slowly varying. Introducing the type 2 generating function
\begin{equation}
 \Phi=(\theta_n - \nu t+\phi_1)\tilde{I}_n,
\end{equation}
the new canonical variables are given by
\begin{align}
  I_n &\equiv \pd{\Phi}{\theta_n}=\tilde{I}_n \\
  \tilde{\theta}_n &\equiv \pd{\Phi}{\tilde{I}_n}=\theta_n - \nu t+\phi_1.\\
\end{align}
The Hamiltonian, which transforms according to $\tilde{h}_n = h_n + \pd{\Phi}{t}$, becomes
\begin{equation}\label{eqn:1st_order_h}
\boxed{\widetilde{h}_n = (\omega_n-\nu_1)/\tilde{\omega}_1\tilde{I}_n - \frac{\xi}{2}\tilde{I}_n^2
+4\xi \bar{I}_1^{3/2}\tilde{I}_n^{1/2}\cos\tilde{\theta}_n.}
\end{equation}
The corresponding equations of motion are
\begin{align}
\dot{\tilde{I}}_n  &= -\pd{\tilde{h}_n}{\tilde{\theta}_n}
= 4\xi\widetilde{I}_n^{1/2}\left(\bar{I}_1\right)^{3/2}
\sin\tilde{\theta}_n\\
\dot{\tilde{\theta}}_n  &= \phantom{-}\pd{\tilde{h}_n}{\tilde{I}_n}
= -\Delta_1 - \xi \tilde{I}_n
+2\xi \frac{\bar{I}_1^{3/2}}{\tilde{I}_n^{1/2}}\cos\tilde{\theta}_n,
\end{align}
where $\Delta_1\equiv \nu_1 - \omega_n$. 
\subsubsection{Resonance Condition}
\begin{figure} 
\begin{center}
\includegraphics[scale=0.8]{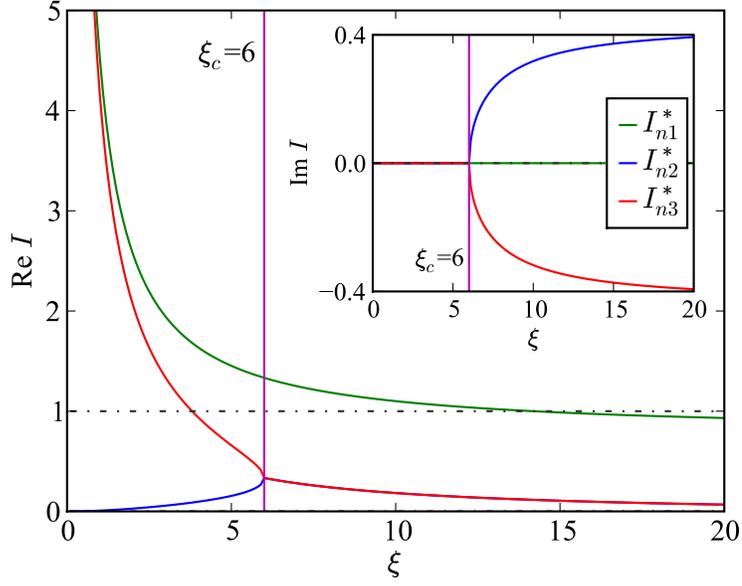}
\end{center}
\caption{\label{fig:1st_order_roots} Real and imaginary parts of the fixed points of first-order Hamiltonian (\ref{eqn:1st_order_h}).  $\Delta_{01} = -4, \; \bar{I}_1=1/3.$  }
\end{figure}

Resonance occurs at the stationary points, that is
\begin{align}
 \dot{\tilde{I}}_n\Big|_{(\tilde{I}_n^\ast,\tilde{\theta}_n^\ast )}  &= 0 \qquad \Rightarrow \qquad  4\xi(\tilde{I}_n^\ast)^{1/2}\bar{I}_1^{3/2}
\sin\tilde{\theta}_n^\ast = 0\\
\dot{\tilde{\theta}}_n\Big|_{(\tilde{I}_n^\ast,\tilde{\theta}_n^\ast )} &= 0 \qquad \Rightarrow \qquad
   -\Delta_1 - \xi \tilde{I}_n
+2\xi \frac{\bar{I}_1^{3/2}}{(\tilde{I}_n^\ast)^{1/2}}\cos\tilde{\theta}_n^\ast = 0. \label{eqn:1st_order_dtheta}
\end{align}
The first condition is satisfied by (a) $\tilde{I}_n^\ast = 0 $ or (b) $ \sin\tilde{\theta}_n^\ast = 0$.  For $\tilde{I}_n =0$ the phase is not well-defined and corresponds to a stationary point, even though (\ref{eqn:1st_order_dtheta}) is not satisfied. The resonances of this model will correspond to taking $ \sin\tilde{\theta}_n^\ast = 0$ and solving (\ref{eqn:1st_order_dtheta}) for $\tilde{I}_n^\ast$,
\begin{equation}
-\Delta_1\tilde{I}_n^{1/2} - \xi \tilde{I}_n^{3/2}=  \mp 2\xi \bar{I}_1^{3/2}.
\end{equation}
Squaring both sides and rearranging gives,
\begin{equation}
\tilde{I}_n^3 +2 \frac{\Delta_1}{\xi} \tilde{I}_n^2 + \frac{\Delta_1^2}{\xi^2} \tilde{I}_n - 4\bar{I}_1^3 =0.
\end{equation}
The solution to this cubic equation gives the fixed points,
\begin{align}
 \tilde{I}_{n1}^\ast &= - \frac{2\Delta_1}{3\xi} + A + B\\
 \tilde{I}_{n2}^\ast &= - \frac{2\Delta_1}{3\xi} -\frac{1}{2}\left( A + B\right) + 
			i\frac{\sqrt{3}}{2}\left(A-B\right)\\
 \tilde{I}_{n3}^\ast &= - \frac{2\Delta_1}{3\xi} -\frac{1}{2}\left( A + B\right) -
			i\frac{\sqrt{3}}{2}\left(A-B\right),
\end{align}
where
\begin{align}
 A &= \left[ \left(\frac{\Delta_1}{3\xi}\right)^3 + 2\bar{I}_1^3 
	+ 2 \bar{I}_1^{3/2}\left(\left(\frac{\Delta_1}{3\xi}\right)^3 + \bar{I}_1^3 \right)^{1/2} \right]^{1/3}\\
 B &= \left[ \left(\frac{\Delta_1}{3\xi}\right)^3 + 2\bar{I}_1^3 
	- 2 \bar{I}_1^{3/2}\left(\left(\frac{\Delta_1}{3\xi}\right)^3 + \bar{I}_1^3 \right)^{1/2} \right]^{1/3}.
\end{align}
The real and imaginary parts of the roots are plotted in Fig.~\ref{fig:1st_order_roots} for $\bar{I}_1 =1/3$ and $\Delta_{01} = -4$, which corresponds to $(n+2,n+1)\rightarrow(n,n+3)$, which is the negative detuning closest to zero. For the initial condition studied by the previous perturbation expansion, where a block of consecutive modes are initially occupied, the bare detuning is always negative.  Thus, we still study the case of negative detuning although it is possible to have a first-order resonance with positive or negative values. The value of $\xi$ the separates regions with three real solutions to one real solution is given by $\left(\frac{\Delta_1}{3\xi_c}\right)^3 + \bar{I}_1^3= 0$ or 
\begin{equation}
   \frac{\Delta_1}{3\xi_c}  = \frac{\Delta_{01} - \xi_c \bar{I}_1}{3\xi_c} =- \bar{I}_1 \qquad \Rightarrow \qquad \boxed{ \xi_{c2} \equiv - \frac{\Delta_{01}}{2\bar{I}_1}}
\end{equation}
As will be shown below there is another relevant critical value of $\xi$ at lower values, so this critical value is defined as $\xi_{c2}$. There are three physical fixed points $(\tilde{I}_n^\ast \geq 0$ for $\xi \leq \xi_{c2} $ and one physical fixed point for $\xi > \xi_{c2}$.  

\begin{figure} 
\begin{center}
\includegraphics[scale=0.55]{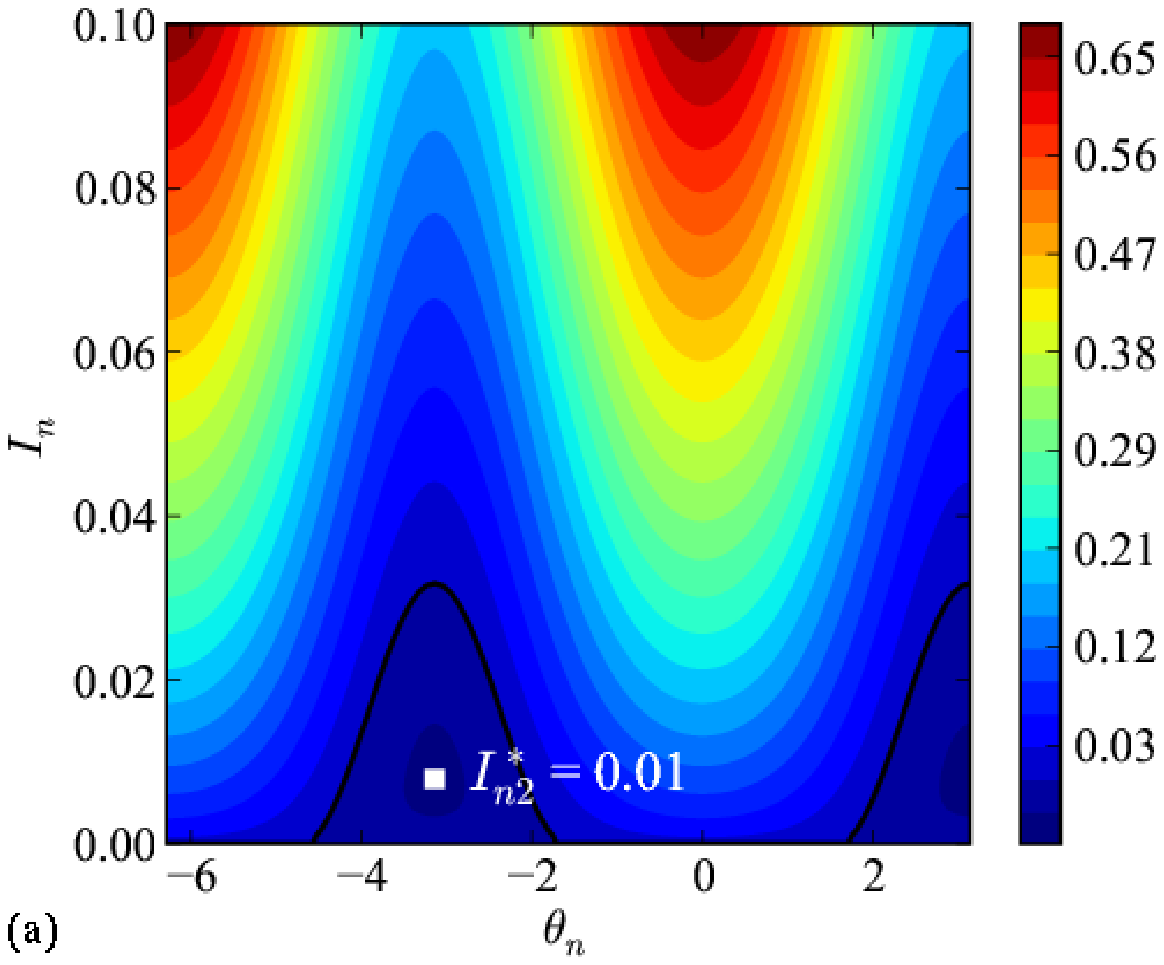}
\includegraphics[scale=0.55]{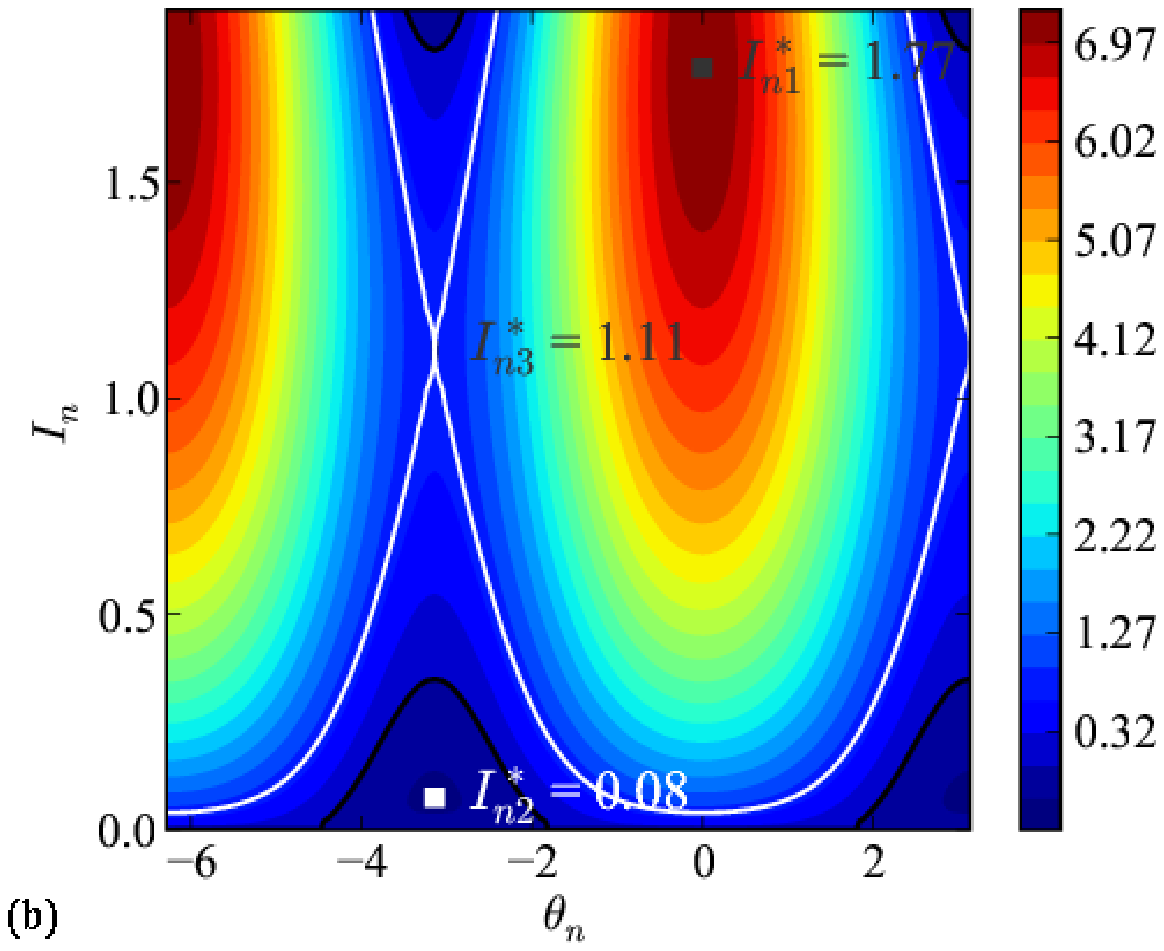}
\includegraphics[scale=0.55]{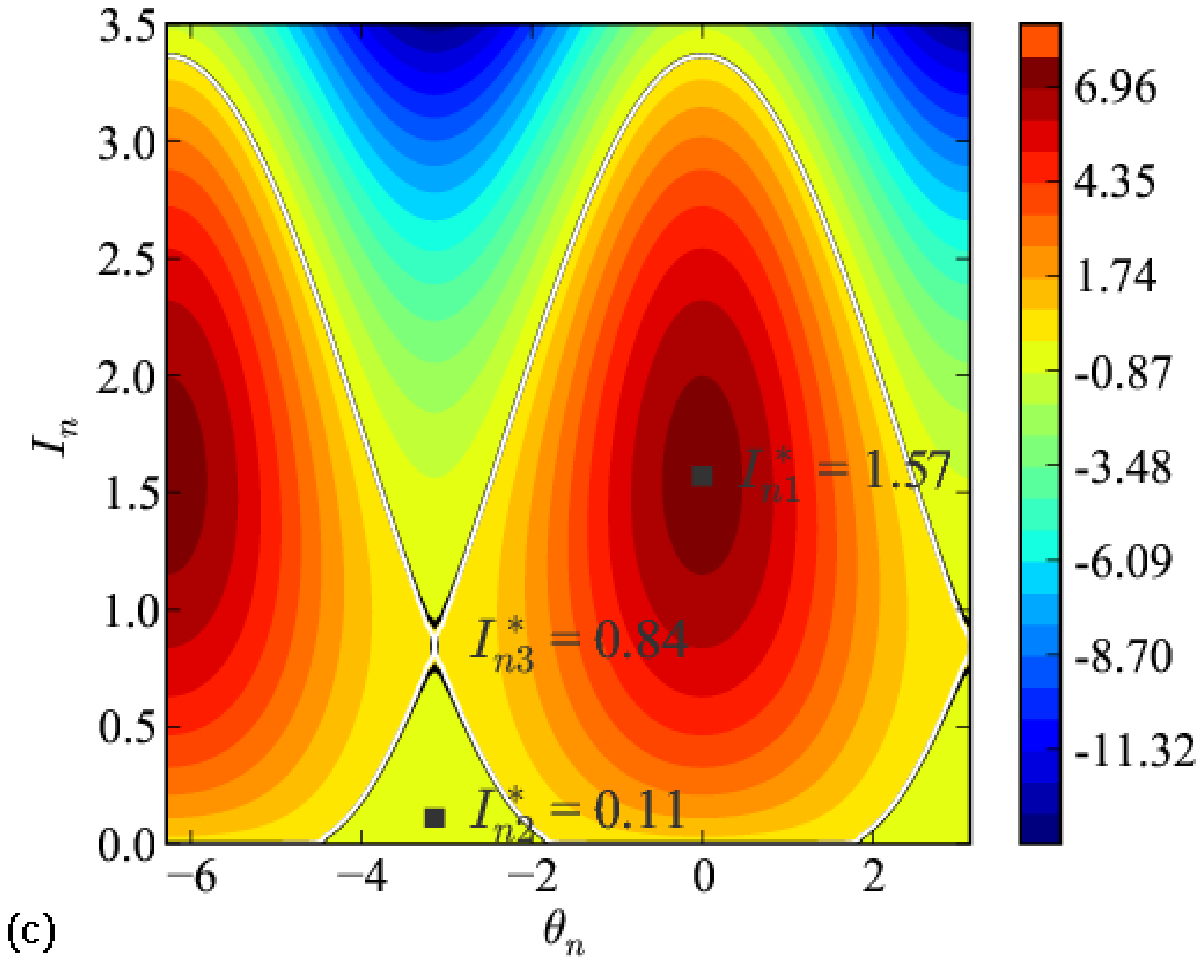}
\includegraphics[scale=0.55]{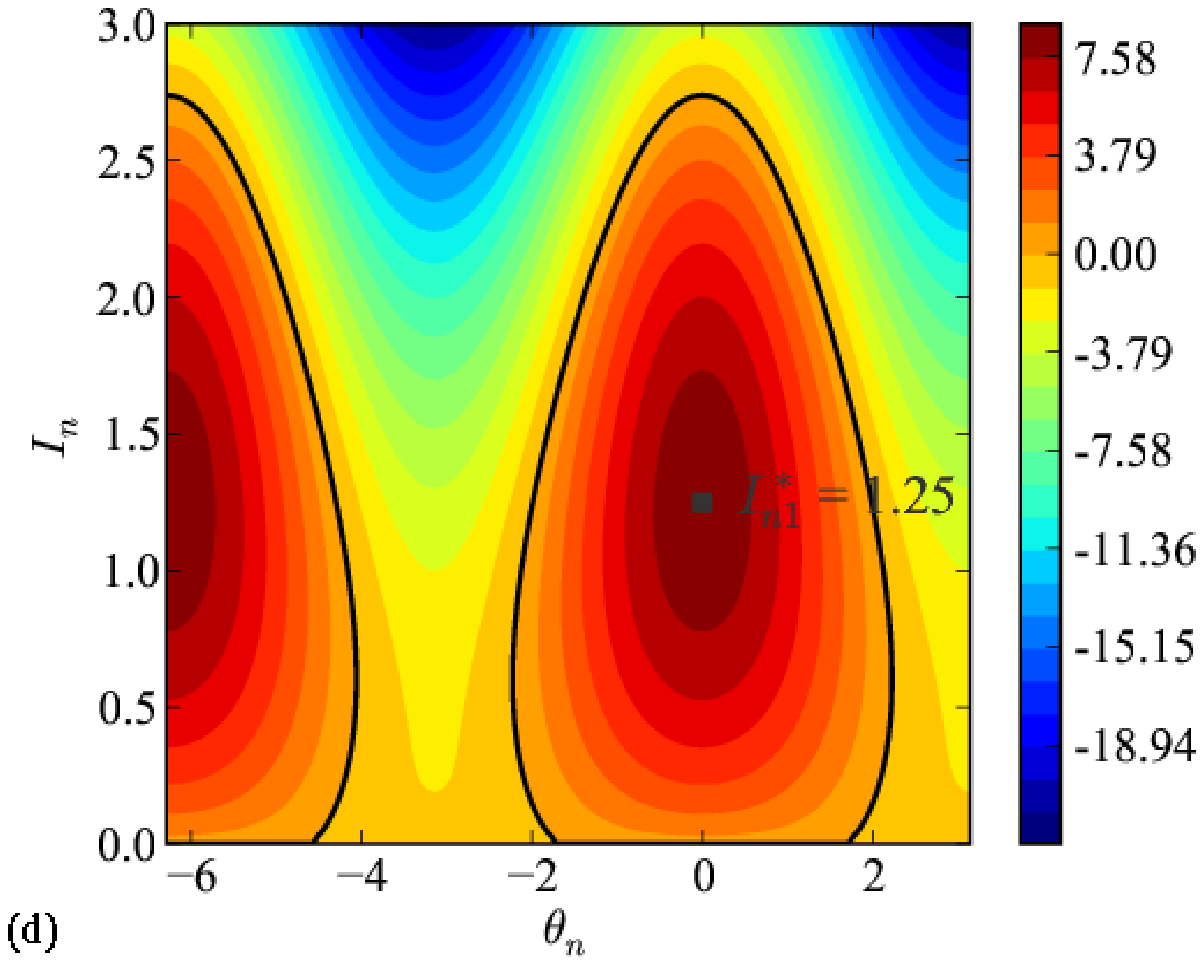}
\end{center}
\caption{\label{fig:1st_order_contours} Action-angle phase-space plots for variables $\tilde{I}_n$, $\tilde{\theta}_n$ for the first-order resonant Hamiltonian (\ref{eqn:1st_order_h}). $(n+2,n+1) \rightarrow (n,n+3)$ $(\Delta_{01} = -4)$. $\bar{I}_1=1/3$. (a) $\xi=1$ (b) $\xi=3.5$ (c)  $\xi=\xi_{c1}=4.3 $ (d) $\xi=7 > \xi_{c2}$. The black contour line corresponds to $\tilde{h}_n = 0$ and the white contour line to $\tilde{h}_n(\tilde{I}_{n}, \tilde{\theta}_n) = \tilde{h}_n(\tilde{I}_{n3}^\ast, -\pi)$. }
\end{figure}

\subsubsection{Separatrix Width}  
There are two separatricies:  one is associated with $\tilde{h}_n = 0$ another passes through the fixed point $(\tilde{I}_{n3}^\ast,\tilde{\theta}_n^\ast \pm \pi)$ and is defined by the contour $\tilde{h}_n = \tilde{h}_n (\tilde{I}_{n3}^\ast, \pm \pi)$.

\begin{figure}
\begin{center}
\includegraphics[scale=0.8]{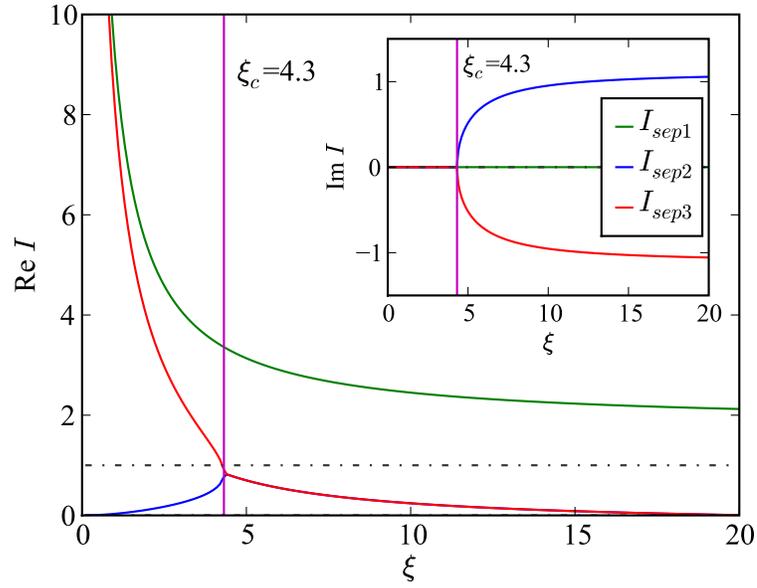}
\end{center}
\caption{\label{fig:1st_order_sep} Real and imaginary parts of the solutions of $\tilde{h}_n = 0$ for the first-order Hamiltonian (\ref{eqn:1st_order_h}).  $\Delta_{01} = -4 \; (n+1,n+2)\rightarrow (n,n+3)$, $\bar{I}_1=1/3$. For $\xi \leq \xi_{c1}$, $\tilde{I}_{sep2}$ gives the separatrix for the the resonance at $(\tilde{I}_{n2}^\ast,\tilde{\theta}_n =\pm \pi) $.  For $\xi \geq \xi_{c1}$, $\tilde{I}_{sep1}$ gives the separatrix for the the resonance at $(\tilde{I}_{n1}^\ast,\tilde{\theta}_n =0) $.  }
\end{figure}

In this section we calculate the maximum value of $\tilde{I}_n$ for the separatrix defined by $\tilde{h}_n = 0$. The maximum value of the separatrix occurs at $ \tilde{\theta}_n = \pm m\pi$ for integer $m$. We thus solve $\tilde{h}_n=0$ for $ \cos(\tilde{\theta}_n)=\pm 1$.
\begin{equation}
-\Delta_1\tilde{I}_n - \frac{1}{2}\xi \tilde{I}_n^2 =  \mp 2\xi \bar{I}_1^{3/2}\tilde{I}_n^{1/2}.
\end{equation}
Squaring both sides and rearranging gives,
\begin{equation}
\tilde{I}_n^3 + 4 \frac{\Delta_1}{\xi} \tilde{I}_n^2 + 4 \frac{\Delta_1^2}{\xi^2} \tilde{I}_n - 64\bar{I}_1^3 =0.
\end{equation}
Solving this cubic equation gives the roots, 
\begin{align}
 \tilde{I}_{sep1} &= - \frac{4\Delta_1}{3\xi} + A + B\\
 \tilde{I}_{sep2} &= - \frac{4\Delta_1}{3\xi} -\frac{1}{2}\left( A + B\right) + 
			i\frac{\sqrt{3}}{2}\left(A-B\right)\\
 \tilde{I}_{sep3} &= - \frac{4\Delta_1}{3\xi} -\frac{1}{2}\left( A + B\right) -
			i\frac{\sqrt{3}}{2}\left(A-B\right)
\end{align}
where
\begin{align}
 A &= \left[ 8\left(\frac{\Delta_1}{3\xi}\right)^3 + 4\bar{I}_1^3 
	+ 16\sqrt{2} \bar{I}_1^{3/2}\left(\left(\frac{\Delta_1}{3\xi}\right)^3 + 2\bar{I}_1^3 \right)^{1/2} \right]^{1/3}\\
 B &= \left[ 8\left(\frac{\Delta_1}{3\xi}\right)^3 + 4\bar{I}_1^3 
	- 16\sqrt{2} \bar{I}_1^{3/2}\left(\left(\frac{\Delta_1}{3\xi}\right)^3 + 2\bar{I}_1^3 \right)^{1/2} \right]^{1/3}.
\end{align}
The second critical value of $\xi$ is given by $\left(\frac{\Delta_1}{3\xi_c}\right)^3 + 2\bar{I}_1^3= 0$ or 
\begin{equation}
   \frac{\Delta_1}{3\xi_c}  = \frac{\Delta_{01} - \xi_c \bar{I}_1}{3\xi_c} =- \sqrt[3]{2}\bar{I}_1 \qquad \Rightarrow \qquad \boxed{ \xi_{c1} \equiv - \frac{\Delta_{01}}{(1+3\sqrt[3]{2})\bar{I}_1}}
\end{equation}
The maximum value of the separatrix occurs at $ \tilde{\theta}_n = \pm \pi, \; \left(\cos(\tilde{\theta}_n\right)=-1) $  for $\xi \leq \xi_{c1}$ and at $ \tilde{\theta}_n=0 \pm 2\pi, \; \left(\cos(\tilde{\theta}_n\right)=+1)$ for $ \xi \geq \xi_{c1}$.  In Fig.~{\ref{fig:1st_order_sep}} the real and imaginary parts of these solutions are plotted. 

From the fixed points of the Hamiltonian, solutions of $\tilde{h}_n = 0$ and sample contour plots, the phase-space of the first-order Hamiltonian can be characterized as follows:
\begin{description}
 \item[$\xi < \xi_{c1}$] There are two resonances.  One is at $(\tilde{I}_{n}, \tilde{\theta}_n=)= (\tilde{I}_{n2}^\ast,\pm \pi)$ with the separatrix defined by the contour at $\tilde{h}_n = 0$. This resonance can be seen in the phase-space diagrams of Fig.~\ref{fig:1st_order_contours}(a)-(b).
The height of the separatrix is given by $I_{sep2}$ which increases from zero at $\xi=0$ to a maximum value $\tilde{I}_{n2-\text{max}} = 2 \sqrt[3]{2}\bar{I}_1 $ at $\xi=\xi_{c2}$ as seen in Fig.~\ref{fig:1st_order_sep}. For $\bar{I}_1=1/3$ the maximum value is  $\tilde{I}_{n2-\text{max}} \approx 0.84$. The second resonance is given by  $(\tilde{I}_{n1}^\ast, \tilde{\theta}_n=0)$. The separatrix is defined by the contour that passes through the saddle point $( \tilde{I}_{n3}^\ast,\pm \pi$).  This resonance diverges as $\xi \rightarrow 0$ and remains above 1 in this parameter range is thus not physically accessible. However, the lower separatrix comes into the accessible phase space as seen in Fig.~\ref{fig:1st_order_contours}(b).Due to the normalization, the occupation of $ \tilde{I}_{n} $ is bounded by one.  Larger values of $ \tilde{I}_{n} $ are plotted in Fig.~\ref{fig:1st_order_contours} to gain a of deeper understanding of the phase space.
 \item[$\xi = \xi_{c1}$]  At the first critical point, the separatricies of the two resonances overlap as show in Fig.~\ref{fig:1st_order_contours}(c).  At this critical point, the lower bound of the separatrix defined by $ \tilde{h}_n= \tilde{h}_n(\tilde{I}_{n3}^\ast, -\pi)$ touches zero. Starting with an arbitrarily small occupation, the entire physical phase space becomes accessible to $\tilde{I}_{n}$ once the two resonances touch.
 \item[$\xi_{c1} < \xi < \xi_{c2}$] The two resonances at $(\tilde{I}_{n1}^\ast,0) $ and $(\tilde{I}_{n2}^\ast,\pm \pi) $ remain, with the separatrix of the first defined by $\tilde{h}_n=0$ and the separatrix of the second resonance defined by $ \tilde{h}_n= \tilde{h}_n(\tilde{I}_{n3}^\ast, -\pi) $. The association of the separatricies have switched from the case where $\xi < \xi_{c1}$.
  \item[$\xi \geq \xi_{c2}$]  At the second critical point the fixed points $(\tilde{I}_{n2}^\ast, \pm \pi)$ and $(\tilde{I}_{n3}^\ast, \pm \pi)$ merge and annihilate and a single resonance at $(\tilde{I}_{n},\tilde{\theta}_{n} ) =(\tilde{I}_{n1}^\ast,0) $ persists. Above $\xi_c$, there is a single resonance at $(\tilde{I}_{n},\tilde{\theta}_{n} ) =(\tilde{I}_{n1}^\ast,0) $, as show in Fig.~\ref{fig:1st_order_contours}(d) and the separatrix is defined by the contour $\tilde{h}_n=0$.  The height of the separatrix is given by $\tilde{I}_{sep1}$ in Fig.~\ref{fig:1st_order_sep}.
 \end{description}

\subsection{1-R Resonances in BOA. $\Omega_p=\Omega_q$}
Consider a first-order resonance where the feeding modes are the same, 
$n+r=2p$. In this case, we keep four terms from the full Hamiltonian, and the Hamiltonian for the 
first-order resonances (1-R) is
\begin{equation}
H_n =\omega_n I_n - \frac{\mu_0}{2\hbar^2}I_n^2
+\frac{2\mu_0}{\hbar^2}\left(I_n I_p^2 I_r\right)
^{1/2}\cos\left(\theta_n +\theta_r - 2\theta_p\right)
\end{equation}
After making the transformation to a rotating reference frame, 
introducing $\nu_{1r}= 2\Omega_p - \Omega_r$, fixing the populations of 
modes $I_p, I_r$, and dividing by $\omega_n$, the Hamiltonian, $\tilde{h}_n\equiv\widetilde{H}_n/\omega_n$ becomes
\begin{equation}
\boxed{\tilde{h}_n = -\Delta_{1r}\tilde{I}_n - \frac{\xi}{2}\tilde{I}_n^2
+ 2\xi\tilde{I}_n^{1/2}\bar{I}_{1r}^{3/2}\cos\tilde{\theta}_n}
\end{equation}
where $\Delta_{1r}\equiv \nu_{1r} - \omega_n$. The corresponding equations of motion are
\begin{align}
\dot{\tilde{I}}_n  &= -\frac{\partial \tilde{h}_n}{\partial \tilde{\theta}_n}
= 2\xi\tilde{I}_n^{1/2}\bar{I}_{1r}^{3/2}\sin\tilde{\theta}_n\\
\dot{\tilde{\theta}}_n & = \frac{\partial \tilde{h}_n}{\partial \tilde{I}_n}
= -\Delta - \xi \tilde{I}_n
+2 \xi \left(\frac{\bar{I}_{1r}^{3/2}}
{\tilde{I}_n}\right)^{1/2}\cos\tilde{\theta}_n
\end{align}

The Hamiltonian for the 1R resonances differs from the first-order Hamiltonian only by prefactors.  The resonances of the 1-R Hamiltonian will thus be very similar in form to those of the first-order resonances, and can be determined from the previous analysis by a proper rescaling of $\bar{I}_1$.

\subsection{Second Order Resonances in BOA}
For the second order case, we consider a single filling mode $n$ and restrict the resonance to modes that satisfy $p+q=2n$.  The Hamiltonian for a single second order resonance in the Born Oppenheimer approximation is 
\begin{equation}\begin{split}
H_n =& \omega_n I_n - \frac{\mu_0}{2\hbar^2}I_n^2
+\frac{2\mu_0}{\hbar^2}I_n \left(I_p I_q\right)
^{1/2}\cos\left(2\theta_n - \theta_p - \theta_q\right)\\
= & \omega_n I_n - \frac{\mu_0}{2\hbar^2}I_n^2
+\frac{2\mu_0}{\hbar^2}I_n\left(\bar{I}_p \bar{I}_q\right)
^{1/2}\cos\left(2\theta_n -\nu_2 t\right)
\end{split}
\end{equation}
where in the second line we set $I_p$ and $I_q$ to their unperturbed values and \[\nu_2 \equiv \Omega_p(\bar{I}_p)+\Omega_q(\bar{I}_q) = \omega_p + \omega_q - \frac{\mu_0}{\hbar^2} \left(\bar{I}_p + \bar{I}_q\right). \]
Next we make a canonical transformation to a rotating reference frame ($I_n, \theta_n \rightarrow \tilde{I}_n,\tilde{\theta}_n)$, through the type 2 generating function
\begin{equation}
  \Phi=\frac{1}{2}(2\theta_n - \nu_2 t+\phi_1)\tilde{I}_n.
\end{equation}
The new canonical variables are determined by
\begin{align}
  I_n &= \pd{\Phi}{\theta_n} = \tilde{I}_n \\
  \tilde{\theta}_n&= \pd{\Phi}{\tilde{I}_n}= \frac{1}{2}\left(2\theta_n - \nu_2 t+\phi_1\right),
\end{align}
and the Hamiltonian transforms according to $\tilde{H}_n(\tilde{I}_n\tilde{\theta}_n) = H_n (I, \theta_n) + \pd{\Phi}{t}$
becomes
\begin{equation}
\boxed{\widetilde{H}_n = \left(\omega_n-\frac{1}{2}\nu_2\right)\tilde{I}_n - \frac{\mu_0}{2\hbar^2}\tilde{I}_n^2
+\frac{2\mu_0}{\hbar^2}\tilde{I}_n\left(\bar{I}_p\bar{I}_q\right)
^{1/2}\cos2\tilde{\theta}_n}
\end{equation}
Next we divide by $\tilde{\omega}_1$, which is equivalent to rescaling time by a factor $\tau = \tilde{\omega}_1 t$, introduce $\xi\equiv \mu_0/\hbar\tilde{\omega}_1$ and set $\hbar = 1$.  Furthermore, we define $\bar{I}_2 \equiv (\bar{I}_p\bar{I}_q)^{1/2}$ as the geometric mean of the filling modes $\bar{I}_p, \bar{I}_q$ and $\Delta_2 \equiv (\frac{1}{2}\nu_2 - \omega_n)/\tilde{\omega}_1$.

\begin{equation}\label{eqn:2nd_order_h}
\boxed{\widetilde{h}_n\equiv \frac{\widetilde{H}_n}{\tilde{\omega}_1} = -\Delta_2 \tilde{I}_n - \frac{\xi}{2}\tilde{I}_n^2 + 2\xi \tilde{I}_n\bar{I}_2\cos2\tilde{\theta}_n}
\end{equation}

The corresponding equations of motion are
\begin{align}
\dot{\tilde{I}}_n &= -\pd{\widetilde{h}_n}{\tilde{\theta}_n} = 
4\xi \tilde{I}_n\bar{I}_2\sin2\tilde{\theta}_n\\
\dot{\tilde{\theta}}_n &= \pd{\widetilde{h}_n}{\tilde{I}_n} = -\Delta_2 - \xi\tilde{I}_n
+ 2\xi \bar{I}_2\cos2\tilde{\theta}_n.
\end{align}
If we furthermore assume that the filling modes have equal populations, $\bar{I}_p=\bar{I}_q = \bar{I}_2$ and that the dispersion is quadratic, $\omega_n = n^2\tilde{\omega}_1 $ then the detuning reduces to 
\begin{align}\Delta_2 &= \frac{1}{2\tilde{\omega}_1}\left(\nu - 2\omega_n \right) = \frac{1}{2\tilde{\omega}_1}\left[(p^2+q^2 - 2n^2)\tilde{\omega}_1 - \xi\left(\bar{I}_p+\bar{I}_q\right) \right]\\
 &= \frac{\Delta_{02}}{2} - \xi\bar{I}_2.
\end{align}

\subsubsection{Resonances Conditions}
Unlike the first-order resonances there is a simple expression for the fixed points of the second order resonances. The conditions for the fixed points are $\dot{\tilde{I}}_n|_{(\tilde{I}_n^{\ast}, \tilde{\theta}_n^{\ast})} = 0, \dot{\tilde{\theta}}_n|_{(\tilde{I}_n^{\ast}, \tilde{\theta}_n^{\ast})}=0$.
\begin{align} 
 \dot{\tilde{I}}_n\Big|_{(\tilde{I}_n^\ast,\tilde{\theta}_n^\ast )}  & = 0 \qquad \Rightarrow \qquad \boxed{4\xi \tilde{I}_n^{\ast}\bar{I}_2\sin2\tilde{\theta}_n^{\ast} = 0}\\
\dot{\tilde{\theta}}_n\Big|_{(\tilde{I}_n^\ast,\tilde{\theta}_n^\ast )}  &= 0 \qquad \Rightarrow \qquad  \boxed{-\Delta_2 - \xi\tilde{I}_n^{\ast}
+ 2\xi \bar{I}_2\cos2\tilde{\theta}_n^{\ast}=0}
\end{align}

\begin{enumerate}
 \item \textbf{Case 1:} $\tilde{I}_n^{\ast} = 0$.\\
 The condition for $\dot{\tilde{I}}_n  = 0$ is satisfied by $\tilde{I}_n^{\ast} = 0$ and the condition on $\tilde{\theta}_n^{\ast}$ such that $\dot{\tilde{\theta}}_n=0$ is 
    \[ \cos2\tilde{\theta}_n^{\ast} =  \frac{\Delta_2}{2\xi \bar{I}_2} =  \frac{\Delta_{20}}{4\xi \bar{I}_2} - \frac{1}{2}.\]
Note that if $I_n=0$, the phase of the mode $n$ is not well-defined. In this case, the motion of the action variable is always stationary.  Thus a second-order resonant Hamiltonian can never populate an initially unoccupied mode.  However, once a small seed is present, the mode can grow and enter the dynamics.
\item \textbf{Case 2:} $\sin 2 \tilde{\theta}_n^{\ast} = 0$ ($\cos 2 \tilde{\theta}_n^{\ast} = \pm 1$)\\
   The condition for $\dot{\tilde{I}}_n  = 0$ is satisfied by $\sin 2 \tilde{\theta}_n^{\ast} = 0$ and the condition on $\tilde{I}_n^{\ast}$ such that $\dot{\tilde{\theta}}_n=0$ is 
    \[\tilde{I}_n^{\ast} =  -\frac{\Delta_2}{\xi}\pm 2 \bar{I}_2 
=  -\frac{\Delta_{02}}{2\xi} + \bar{I}_2 \pm 2 \bar{I}_2 \qquad (\text{for }\cos2\tilde{\theta}_n^{\ast} = \pm 1).\]
\end{enumerate}

\subsubsection{Classification of Stationary Points}
Following \cite{tabor_chaos_1989} we outline a method for determining the stability of the stationary points by linearizing about the fixed points.  Consider a general second order differential equation, written as a pair of first-order differential equations,
\begin{align}
 \dot{x}=f(x,y)\\
 \dot{y}=g(x,y)
\end{align}
that has stationary points $\{(x^\ast, y^\ast)|
f(x^\ast,y^\ast)=0,\;  g(x^\ast,y^\ast)=0\}$.  The stability of these points can be determined by
linearizing about the stationary points.
\begin{equation}
\left(\begin{array}{c} \Delta \dot{x} \\ \Delta \dot{y} \end{array}\right)
=\left(\begin{array}{cc} 
\pd{f}{x}|_{x^\ast, y^\ast} &
\pd{f}{y}|_{x^\ast, y^\ast}\\
\pd{g}{x}|_{x^\ast, y^\ast} &
\pd{g}{y}|_{x^\ast, y^\ast}
\end{array}\right)
\left(\begin{array}{c} \Delta x \\ \Delta y \end{array}\right).
\end{equation}
The solution to these equations of motion, $\mathbf{\dot{x}} = \mathbf{Ax}$, is
$\mathbf{x} = c_1 e^{\lambda_1 t}\mathbf{v_1} +c_2 e^{\lambda_2 t}\mathbf{v_2}$
where $\lambda_j$ is an eigenvalue of $\mathbf{A}$ with corresponding 
eigenvector $\mathbf{v_j}$.  The stability of the fixed points is determined 
from the eigenvalues, $\lambda_j = a_j + ib_j,\; a_j,b_j \in \mathbf{R}$. The fixed points are classified as follows,
\begin{enumerate}
\item center:  $a_{1,2}=0$ 
\item spiral:  $b_{1,2}\neq 0$. Stable for $a_{1,2} < 0$.  Unstable for $a_{1,2}>0$.
\item node: $b_{1,2}=0$. Stable for $a_{1,2} < 0$. Unstable for $a_{1,2}>0$ 
\item saddle point:  $b_{1,2}=0,\; a_1<0,\; a_2>0$
\end{enumerate}
\sidenote{Note that we need to be careful in the marginal cases (centers) where the 
equations have been linearized.}

The linearized equations of motion for the second order Hamiltonian (\ref{eqn:2nd_order_h}) are
\begin{equation}
\v{A}=\left(\begin{array}{cc} 
-\pdpd{\tilde{h}_n }{\tilde{\theta}_n}{\tilde{I}_n}|_{\tilde{I}_n^\ast, \tilde{\theta}_n^\ast} &
-\pdd{\tilde{h}_n }{\tilde{\theta}_n}|_{\tilde{I}_n^\ast, \tilde{\theta}_n^\ast}\\
\pdd{\tilde{h}_n }{\tilde{I}_n}|_{\tilde{I}_n^\ast, \tilde{\theta}_n^\ast} &
\pdpd{\tilde{h}_n }{\tilde{I}_n}{\tilde{\theta}_n}|_{\tilde{I}_n^\ast, \tilde{\theta}_n^\ast}
\end{array}\right)
=\left(\begin{array}{cc} 
4\xi \bar{I}_2\sin2\tilde{\theta}_n^{\ast} &
8\xi \tilde{I}_n\bar{I}_2\cos2\tilde{\theta}_n^{\ast}\\
-\xi&
- 4\xi \bar{I}_2\sin2\tilde{\theta}_n^{\ast}
\end{array}\right).
\end{equation}
For each set of fixed points, we calculate the eigenvalues of the matrix, given by $\det\left|\v{A}- \lambda\v{I} \right| = 0$ to determine the type of fixed point.
\begin{enumerate}
 \item  \textbf{Fixed Point 1:} $(\tilde{I}_n^{\ast}, \tilde{\theta}_n^{\ast})= (0,\frac{1}{2}\cos^{-1}\left(\Delta_2/(2\xi\bar{I}_2)\right))$ \\
The eigenvalues are given by
\begin{equation}
 	\lambda = \pm 4 \xi\bar{I}_2 \left[ 1 - \left(\frac{\Delta_2}{2\xi\bar{I}_2}\right)^2\right]^{1/2}
		= \pm 2 \xi \left[ \left(2\bar{I}_2\right)^2 - \left(\frac{\Delta_2}{\xi}\right)^2\right]^{1/2}.
\end{equation}
These eigenvalues are real for $|\Delta_2| < 2\xi\bar{I}_2$.  Given that the bare detuning, $\Delta_{02}$, is always positive for the second-order Hamiltonian (see Table \ref{tab:bare_detuning}) and the typical population of the fixed modes, $\bar{I}_2$ is also positive, this condition leads to a critical value of the nonlinearity parameter, $\xi$,
\begin{equation}\boxed{ \xi_c = \frac{\Delta_{02}}{6 \bar{I}_2}}\end{equation}
 For $ \xi \geq \xi_c$ this fixed point is a saddle point and the separatrix passes through this point.  
\item \textbf{Fixed Point 2:} $(\tilde{I}_n^{\ast}, \tilde{\theta}_n^{\ast})= (-\Delta_2/\xi + 2\bar{I}_2 ,m\pi), \; m=\text{integer}$\\
\begin{equation}
 	\lambda= \pm 2\sqrt{2} \xi \bar{I}_2^{1/2} \left[\frac{\Delta_2}{\xi} - 2 \bar{I}_2\right]^{1/2}.
\end{equation}
For $\tilde{I}_n^\ast = -\Delta_2/\xi + 2\bar{I}  > 0$ the fixed point is a center, and for $\tilde{I}_n^\ast <  0$ it is a saddle point. We exclude the unphysical values $\tilde{I}_n^\ast < 0$. The stationary points for $\tilde{I}_n > 0$ is at $ \cos(2\tilde{\theta}_n) = + 1$, are the physical resonances for the second order driving terms, with resonant value given by

\begin{equation}\label{eqn:2nd_order_res}\boxed{I_{\text{res}} = -\frac{\Delta_{02}}{2\xi} + 3 \bar{I}_2.}\end{equation}

Furthermore the condition for the existence of this type of resonance is $\xi > \xi_c$.
\item \textbf{Fixed Point 3:} $(\tilde{I}_n^{\ast}, \tilde{\theta}_n^{\ast})= (-\Delta_2/\xi + 2\bar{I} , (2m+1)\pi/2),\; m=\text{integer}$\\
For $\cos(2\tilde{\theta}_n^\ast) = -1$, the value of $\tilde{I}_n$ is always negative and thus not relevant to the current analysis.
\end{enumerate}

\begin{figure}[ht!]
\begin{center}
\includegraphics[scale=0.7]{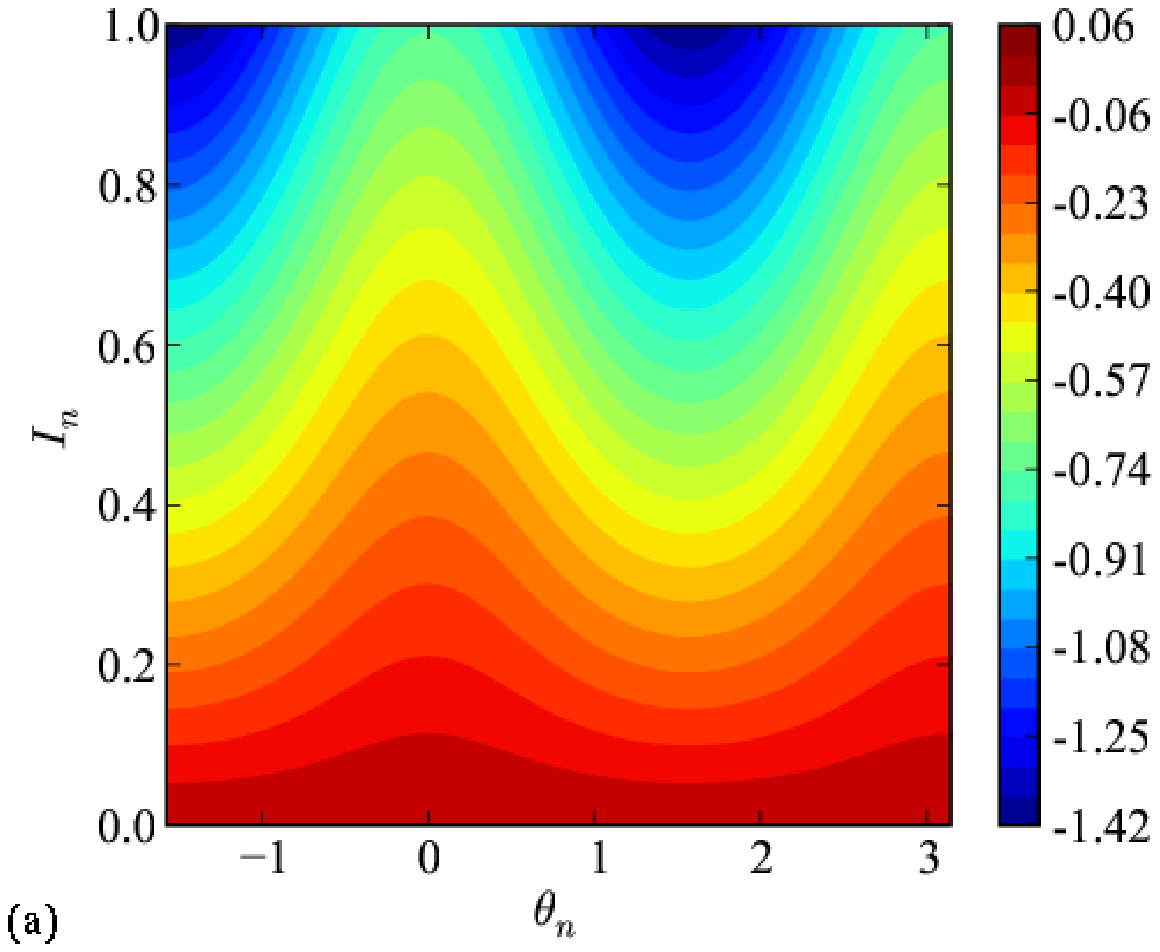}
\includegraphics[scale=0.7]{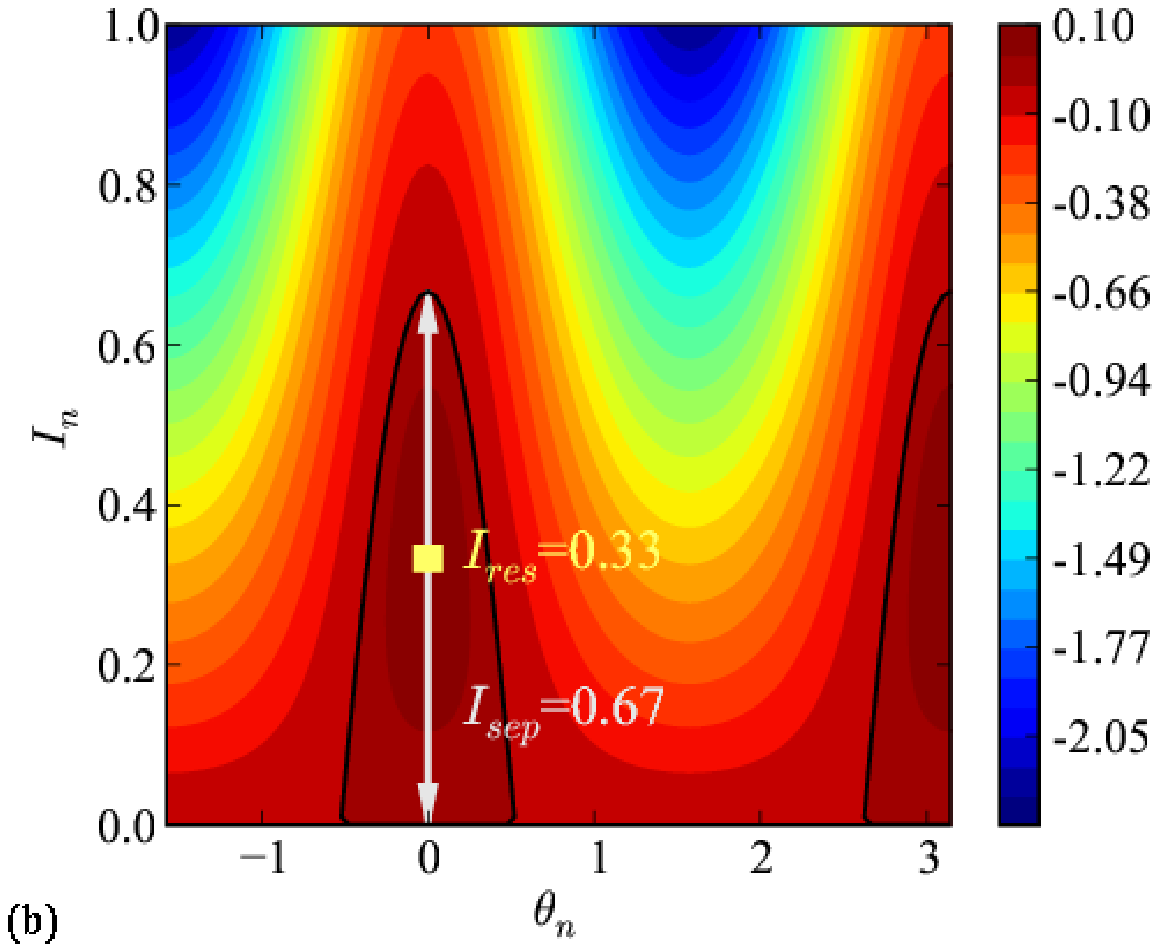}
\end{center}
\caption{\label{fig:2nd_order_contours} Phase-space plots for the second order Hamiltonian.  (a) $\xi =0.5 \; (\kappa=0.16 \text{ for } N_s=11).$ There is no resonance in the range $0 \leq \tilde{I}_n \leq 1$ for $\xi < \xi_c=1 $. (b)  $\xi =1.5 \; (\kappa=0.49 \text{ for } N_s=11).$  $\bar{I}_2 = 1/3, \; \Delta_{02} = 2$. For $\xi > \xi_c$ the resonant value is given by (\ref{eqn:2nd_order_res}) and the separatrix with is given by (\ref{eqn:2nd_order_sep}).}
\end{figure}

\subsubsection{Separatrix Height}
The separatrix passes through the fixed point at $\tilde{I}_n^\ast = 0, \tilde{\theta}_n^\ast= \frac{1}{2}\cos^{-1}\left(\Delta_2/(2\xi\bar{I}_2)\right)$ when $\tilde{\theta}_n^\ast$  is real.    The contour plots of $\tilde{h}_n$ confirm that the separatrix passes through this fixed point.  The maximum value of $\tilde{I}_n$ along the separatrix can be found by solving for $\tilde{I}_n$ when $\tilde{h}_n=0$. Given
\begin{align*}
\tilde{h}_n &= -\Delta_2 \tilde{I}_n - \frac{\xi}{2}\tilde{I}_n^2 + 2\xi \tilde{I}_n\bar{I}_2\cos2\tilde{\theta}_n = 0,\\
\tilde{I}_n &=-\frac{2\Delta_2}{\xi} + 4\bar{I}_2\cos2\tilde{\theta}_n  = 
-\frac{\Delta_{02}}{\xi} + 2\bar{I}_2(1+2\cos2\tilde{\theta}_n).  \\
\end{align*}
The maximum height of the separatrix occurs at $\tilde{\theta}_n=0$ and is given by
\begin{equation}\label{eqn:2nd_order_sep}
\boxed{\tilde{I}_{\text{sep}} =-\frac{\Delta_{02}}{\xi} + 6\bar{I}_2.}
\end{equation}

In Fig.~\ref{fig:2nd_order_contours}(a)-(b) the phase plots for the second order Hamiltonian is plotted for nonlinearities above and below the critical value. For $\xi\leq \xi_c $, there is no resonance in the phase-space that corresponds to physical values of $\tilde{I}_n$.  In Fig.~\ref{fig:2nd_order_contours}(b), where $\xi > \xi_c$, there is a resonance and the separatrix is defined by the contour $\tilde{h}_n=0$  which is plotted in black. The resonance values and separatrix height are labeled by the values given by (\ref{eqn:2nd_order_res}) and (\ref{eqn:2nd_order_sep}), respectively.

\begin{figure}[ht!]
\begin{center}
\includegraphics[scale=0.7]{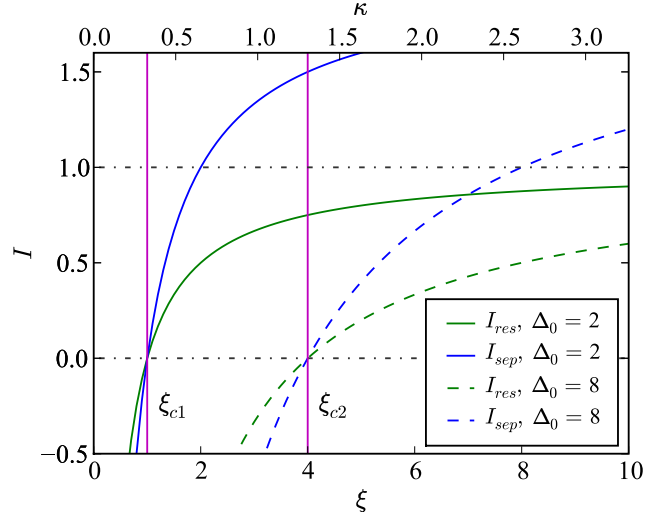}
\end{center}
\caption{\label{fig:2ndorder_Ires} Resonant values and separatrix width for second-order resonance for $\Delta_{02}=2$ $[(n+1,n-1)\rightarrow(n,n)]$ and for $\Delta_{02}=8$ $[(n+2,n-2)\rightarrow(n,n)]$. $\bar{I}=0.33$. Vertical lines: $\Delta_{20}=2$: $\xi_{c1} = 1/3\bar{I}_2 = 1$, $\Delta_{02}=8$:  $\xi_{c2} = 4/3\bar{I}_2 = 4$. The upper x-axis gives the corresponding values of $\kappa$ for $N_s=11$. }
\end{figure}

The values of the action variable at resonance and the separatrix height are plotted in Fig.~\ref{fig:2ndorder_Ires} for the two second-order Hamiltonians as a function of the nonlinearity, $\xi$ for the two lowest bare detunings. The two detunings are $\Delta_{02}=2$, corresponding to $(n+1,n-1)\rightarrow(n,n)$ and $\xi_c = 2\bar{I}_2/3$, and $\Delta_{02}=8$ for $(n+2,n-2)\rightarrow (n,n)$ and $\xi_c = 4\bar{I}_2/3$.  It is clear from the plot that for the second order resonant Hamiltonian there is a critical value of the nonlinearity such that below that value there are no physical resonances present. 

\section{\label{sec:chirikov}Failure of Chirikov's Criterion}
The Chirikov criterion for the onset of global chaos is governed by the ratio of the width of the separatrix and the distance between the resonances for individual resonances.  When the width of the separatricies becomes comparable to the distance between the resonances, the Chirikov parameter $\mathcal{K} $ satisfies the inequality,
\begin{equation}
  \mathcal{K} \equiv \frac{\text{separatrix width}}{\text{distance between resonances}} = \frac{\Delta I }{I_{\text{res}1} - \bar{I}_{\text{res}2}} \gtrsim 1,
\end{equation}
and the system is predicted to be chaotic.

We should note that the study of resonances in the BHM is quite different from the case of the kicked rotor.  In particular, it is not possible to independently vary the frequency of the drive and the strength of the drive in the BHM.  In addition, in the canonical transformation to the rotating reference frame the action variable is unchanged, $\tilde{I}_n=I_n$, so that the resonant values of $\tilde{I}_n$ are not well-spaced in the original action coordinates.  Furthermore in the BHM, for the first order resonances, a single driving term has resonances at multiple action values.

Note that for a generic driving term, $(n+p,n+m-p) \rightarrow (n,n+m) $, the bare detuning is given by
\begin{equation}\begin{split}
\Delta_0 &= (n+p)^2 + (n+m-p)^2 - n^2 - (n+m)^2 = \\
	 &= n^2 + 2np + p^2 + n^2 + m^2 + p^2 - 2np - 2mp + 2nm\\
	 &\phantom{=} - \left[ n^2 + n^2 + 2nm + m^2\right]\\
	 &= 2p^2 - 2mp,
                \end{split}
\end{equation}
which is always constant.

We present several ways to deduce the criterion governing the threshold for the mean-field Bose-Hubbard model and see that the various methods are in agreement.

\paragraph{Onset of Second-Order Resonances}
First we consider the case of the second-order resonance, where the analytic expression is simple. 

From the previous analysis for the second order resonance, the separatrix width is given by,
\begin{equation}
 I_{\text{sep}} = \frac{-\Delta_0}{\xi} + 6 \bar{I}.
\end{equation}
The distance between resonances is proportional to $\xi^{-1}$,
\begin{equation}
 \Delta I_{\text{res}} = I_{\text{res}}(\Delta_b) - I_{\text{res}}(\Delta_a) = \frac{\Delta_{0a}-\Delta_{0b}}{\xi}.
\end{equation}
For the lowest two resonances, a naive comparison of the separatrix height and distance between the resonances gives,
\begin{equation}
\mathcal{K} \equiv \frac{I_{\text{sep}}}{\Delta I_{\text{res}}}  = \frac{-2/\xi + 6\bar{I}}{6/\xi}
\end{equation}
and chaos exists for 
\begin{equation}
\mathcal{K} \gtrsim 1 \Rightarrow \boxed{\bar{I} \xi\gtrsim \frac{5}{3}}
\end{equation}
However, but looking at Fig.~\ref{fig:2ndorder_Ires} , we see that once the second resonance appears, the two resonances overlap.  Thus we take the appearance of the second resonance as the criterion for the second-order case, which gives 
\begin{equation}
\xi > \xi_{c2} = \frac{4}{3\bar{I}}  \text{  or  }  \boxed{\bar{I}\xi \gtrsim \frac{4}{3}}.
\end{equation}

\paragraph{Overlap of resonances within first-order resonant Hamiltonian}  
For the first-order resonance, the resonances of $\tilde{I}_{n2}^\ast$ grow from zero and exist for any $\xi > 0$.  However, the separatrix of the resonance is small and thus even though the resonances may overlap, the population is still expected to be confined in a narrow region of phase-space.  However, there are additional resonances at larger values of $\tilde{I}_n$, which are initially inaccessible and move down from above.  For the first-order resonance we use the criterion that the two separatricies of the same first-order Hamiltonian overlap - i.e. all of phase space is accessible to a single mode.  
The condition for being able to explore the entire phase-space for a given $I_n$ is thus 
\begin{equation}
  \xi \geq \xi_{c1} \equiv - \frac{\Delta_{01}}{(1+3\sqrt[3]{2})\bar{I}_1}
\end{equation}
which is equivalent to 
\begin{equation}
  \bar{I}\xi \geq - \frac{\Delta_{01}}{(1+3\sqrt[3]{2})}.
\end{equation}
For the lowest bare detuning, $\Delta_{01} = -4$, the criterion is
\begin{equation}
\boxed{\bar{I}\xi \geq \frac{4}{(1+3\sqrt[3]{2})} \approx 0.84}
\end{equation}
which is the comparable to the results for the second order resonance.

\paragraph{Dimensional Analysis}
An alternate way of coming to this conclusion is to consider dimensional grounds.  
The starting point is to assume:
\begin{enumerate}
 \item Typical mode occupation:  $\bar{I} \sim \frac{1}{\Delta n}$
 \item Resonant approximation:  only include resonant terms in equations of motion.
 \item Quadratic dispersion instead of cosine: $\omega_n = \tilde{\omega}_1 n^2 $
\end{enumerate}
These are the same ingredients for the previous analytic analysis.  A quadratic dispersion leads to translational invariance.  The momentum distribution can be shifted with no effect. The resonant equations are invariant under a shift in 'n', thus the frequency scale is set by $\tilde{\omega}_1$. As a consequence of the resonant approximation, the equations of motion only couple neighboring modes. Thus, $\mu_0$ and $\Delta n$ can only enter as $\frac{\mu_0}{\Delta n}$.  Thus the only parameters are $\mu_0/\Delta n$ and $\tilde{\omega}_1$.  The only dimensionless combination of these quantities is $\mu_0/(\hbar \tilde{\omega}_1\Delta n) = \xi/\Delta n$, so this parameter must be what governs the Chirikov threshold. Signficantly, all of these methods give the same functional form of the criterion and are identical up to numerical factors.

\subsection{Thermodynamic Limit}

Next we consider how $\mathcal{K}$ scales in the thermodynamic limit, ($N_s \rightarrow \infty$ while $U=\text{const, } J=\text{const, } N_a/N_s=\text{const}$).  Additionally, the width of the momentum distribution remains fixed, $\Delta n/N_s=\text{const}$.  In this limit, the nonlinearity parameter $\kappa$ is constant while $\xi \sim N_s^2$.  The typical occupation of the fixed modes is given by $\bar{I} \sim (\Delta n)^{-1} \sim (N_s)^{-1}$
We find that the Chirikov parameter,
\begin{equation}
\mathcal{K} \approx \xi \bar{I} \sim \frac{N_s^2}{N_s} \sim N_s
\end{equation}
diverges in the thermodynamic limit, predicting that there is no threshold in that limit and $\mathcal{K}$ is always greater than one indicating that the system is always chaotic.  However, this is not what we have observed numerically. At a minimum Chirikov predicts the threshold to scales linearly with the size of the system and our numerics indicate that the threshold depends on parameters that are independent of the system size.

\subsection{Continuous Limit}
To check this threshold, we also consider the continuum limit, in which the length of the system, normalization and interaction parameter are fixed, while the distance between lattice sites goes to zero:  $L,N_a,\mu_0=\text{const}, \, a \rightarrow 0, N_s \rightarrow \infty$.
\begin{equation}
 \mathcal{K} \sim  \frac{U}{J}\left(\frac{N_a}{\Delta n}\right) N_s = \frac{\mu_0N_s}{N_a}\frac{2mL^2}{\hbar^2N_s^2} \left(\frac{N_a}{\Delta n}\right) N_s
= \frac{2\mu_0mL^2}{\hbar^2}\left(\frac{1}{\Delta n}\right) \sim \frac{1}{N_s} \rightarrow 0
\end{equation}
In the continuum limit, the number of modes in the initial momentum distribution, $\Delta n$, must diverge, so that $N_a = \Delta n |\bar{\psi}_n|^2 $ remains constant.  Alternatively we can say that $\Delta n / N_s$ is fixed.  From either perspective, it is clear that $\mathcal{K}$ vanishes in the continuum limit, predicting regular motion, as expected due to the known integrability of the continuous NLS equation.

This leads to the question: Why does the Chirikov analysis fail in this case?  Several possible explanations are:
\begin{enumerate}
 \item non-quadraticity of the spectrum 
 \item break down of the resonant approximation
 \item interferences between resonances.
\end{enumerate}

First there are possible corrections due to the lattice and that we have assumed a quadratic dispersion instead of a cosine dispersion. We conjecture that the Chirikov analysis is incorrect because the the resonant approximation is wrong and off-resonant terms are significant. That is the motion is not dominated by single driving terms.  It is also possible that the single resonance approximation is incorrect and a multi-resonance model is necessary for recovering the correct scaling.



\chapter{Conserved Quantities of the Ablowitz-Ladik Lattice}
\label{chap:IDNLS}
%
%
From the numerical work on the Bose-Hubbard Model, we have seen regions where although individual trajectories are chaotic the system does not relax to the expected thermal state.  Additionally there are regions in the parameter space of nonlinearity, $\kappa$, and energy-per-particle, $\epsilon_T$, where the system relaxes even though chaos is not present.  Additionally it has been found that the slow relaxation times for individual states with the same nonlinearity and energy-per-particle vary widely. What is the origin of these phenomena? 

There are at least three nearby integrable systems of the BHM.  These are the noninteracting case $(\kappa = 0)$, the continuum limit, which is the continuous nonlinear Schr\"odinger equation and the integrable discrete nonlinear Schr\"odinger equation (IDNLS) also know as the Ablowitz-Ladik (AL) lattice, which we discuss in this section.   Looking at the relaxation of individual realizations, it appears that initial states which have a higher quasi-momentum also have a higher final spectral entropy.  The total quasi-momentum is not a conserved quantity of the BHM.  It is however, a conserved quantity of all three nearby integrable systems.  In some cases the normalized spectral entropy many not vanish because the system has a very slow relaxation time.  In other cases it may be that it will never fully relax because of the nearby conserved quantities.  Are the slow relaxation times governed by the conserved quantities of the nearby integrable systems?  In the regions where it is clear that the steady-state is not the thermal state, what governs the steady-state? This evidence suggests that the integrals of motion of the nearby conserved quantities play a role in the relaxation dynamics of the BHM.  In this section, we derive the conserved quantities of the AL.

\section{Integrable Discrete Nonlinear Schr\"odinger (IDNLS) Equation }
The continuous Nonlinear Schr\"odinger (NLS) Equation is given by,
\begin{equation}
 i \dot{q} + q_{xx} + \sigma |q|^2q = 0
\end{equation}
where $\dot{q} = dq/dt$ and $q_x=dq/dx$, is know to be completely integrable. The corresponding Hamiltonian is
\begin{equation}
 H(q,q^\ast)= - \int_0^L \left[ q_x q^\ast_x + \frac{1}{2}\sigma(qq^\ast)^2\right] dx  ,
\end{equation}
with canonical pairs $q,q^\ast $. Periodic boundary conditions are assumed. 
The mean-field Bose-Hubbard model is a discretization of the NLS equation. In real space, the Hamiltonian can be written as:
\begin{equation}
 H = -\sum_n  \left(q_n q^\ast_{n+1} + q^\ast_n q_{n+1} - 2 |q_n|^2\right) + \frac{\sigma}{2} \sum_n |q_n|^4
\end{equation}
with the Poisson brackets
\begin{equation}
 \{ q_m, q_n^\ast \} = i \delta_{m,n}, \qquad \{q_m, q_n\} = \{q_m^\ast, q_n^\ast\} = 0.
\end{equation}
The equations of motion, which are given by $\dot{q}_n = \{ H, q_n\}$, are
\begin{equation}
 i \dot{q}_n = - \left( q_{n+1} + q_{n-1} - 2 q_n  \right) + \sigma  |q_n|^2 q_n.
\end{equation}
This equation, also know as the Diagonal Discrete Nonlinear Schr\"odinger (DDNLS) Equation, does not preserve the integrability of the continuous case.  An alternate discretization, which is integrable \citep{ablowitz_nonlinear_1976}, has equations of motion:
\begin{equation}
 i\dot{q}_n = -\left(q_{n+1} + q_{n-1} - 2q_{n}\right) + \sigma |q_n|^2(q_{n+1} + q_{n-1})/2
\end{equation}
and is suitably called the Integrable Discrete Nonlinear Schr\"odinger (IDNLS) Equation, also know as the Ablowitz-Ladik lattice (AL). This equation can be derived from the Hamiltonian \citep{scharf_properties_1991,herbst_symplectic_1994}
\begin{equation}\label{eqn:H_AL}
 H = -\sum_n  \left(q_n q^\ast_{n+1} + q^\ast_n q_{n+1} \right) - \frac{4}{\sigma} \sum_n \ln\left(1-\frac{\sigma}{2}|q_n|^2\right)
\end{equation}
with the nonstandard Poisson brackets
\begin{equation}
 \{ q_m, q_n^\ast \} = i(1- \frac{\sigma}{2}|q_n|^2) \delta_{m,n}, \qquad \{q_m, q_n\} = \{q_m^\ast, q_n^\ast\} = 0.
\end{equation}
The equations of motion are derived in the usual way, $ \dot{q}_m = \{H,q_m \}$. Using the following properties of Poisson brackets,
\begin{align}
 \{ a,bc \} = b\{ a,c \} + \{ a,b \}c \\
 \{f(a),b \} = \frac{\partial f}{\partial a} \{a,b\},
\end{align}
the AL equation can be derived from the equations of motion of Hamiltonian (\ref{eqn:H_AL}),
\begin{align*}
 \dot{q}_m = &\{H,q_m \}\\ 
 =& -\sum_n  \left(\{q_n q^\ast_{n+1},q_m \} + \{q^\ast_n q_{n+1},q_m \} \right) - \frac{4}{\sigma} \sum_n \{\ln\left(1-\frac{\sigma}{2}|q_n|^2\right),q_m \}\\
= &-\sum_n  \left(q_n \{q^\ast_{n+1},q_m \} + \{q^\ast_n ,q_m \} q_{n+1} \right) - \frac{4}{\sigma} \sum_n 
\left[\frac{\partial}{\partial q_n^\ast}\ln \left(1-\frac{\sigma}{2}|q_n|^2\right)\right]\{q_n^\ast,q_m \}\\
  =& -\sum_n  \left[ q_n \cdot (-i)(1- \frac{\sigma}{2}|q_{n+1}|^2) \delta_{m,n+1} + q_{n+1} \cdot(-i)(1- \frac{\sigma}{2}|q_n|^2) \delta_{m,n} \right]\\
& \phantom{-\sum_n} + \frac{4}{\sigma} \sum_n \frac{(\sigma/2) q_n}{1-\frac{\sigma}{2}|q_n|^2}  \cdot(-i)(1- \frac{\sigma}{2}|q_n|^2) \delta_{m,n}\\
=& i \left[q_{m-1}(1 - \frac{\sigma}{2}|q_{m}|^2) + q_{m+1}(1- \frac{\sigma}{2}|q_m|^2) \right] - 2i q_m\\
=& i \left(q_{m-1} + q_{m+1} - 2 q_m \right) -i  \frac{\sigma}{2}|q_{m}|^2\left(q_{m-1} + q_{m+1} \right).
\end{align*}

This equation is completely integrable and has an infinite number of conserved quantities (for the infinite lattice) and can be solved by the method of Inverse Scattering Transform (IST) developed by Gardner, Greene, Kruskal and Mira \citep{gardner_method_1967}.
%
\section{Conserved Quantities of the AL equation}
In this section, we outline the approach of inverse scattering and the work of Ablowitz and Ladik for calculating the conserved quantities of the AL equation \citep{ablowitz_nonlinear_1976, ablowitz_solitons_1981}. 

The method of inverse scattering is analogous to the Discrete Fourier Transform.  In the direct scattering problem, scattering data is derived from initial data, the potential. The time evolutions of the scattering data is simple.  From the scattering data at some time 't', the potential can be calculated via the (non-trivial) inverse scattering transform. In order to calculate the conserved quantities is it only necessary to calculate the scattering data.  From the scattering data it is clear which quantities are time independent.

The following is an outline of \citet{ablowitz_nonlinear_1976} with details filled in using \citep{ablowitz_solitons_1981}.  
For the AL equation, we taking $S_n(t)=T_n(t)=0$ in \citep{ablowitz_nonlinear_1976}.  The canonical pairs are $R_n$ and $Q_n$.  In the end we set $R_n = \pm \alpha Q_n^\ast$.

To begin, consider the generalized eigenvalue problem,
\begin{equation}\label{eqn:gev}
 \begin{split}
  V_{1,n+1} = &\, z V_{1,n} + Q_n(t) V_{2,n}\\
  V_{2,n+1} = &\, \frac{1}{z} V_{2,n} + R_n(t) V_{1,n},
 \end{split}
\end{equation}
which can be equivalently expressed as
\begin{equation}
 \begin{pmatrix} \hat{E} & -Q_n(t)\\ - R_n(t) & \hat{E} \end{pmatrix} 
 \begin{pmatrix} V_1\\ V_2 \end{pmatrix}  =
\begin{pmatrix} z & 0 \\ 0 & 1/z \end{pmatrix} 
\begin{pmatrix} V_1\\ V_2 \end{pmatrix}  
\end{equation}
where $\hat{E}$ is the shift operator: $\hat{E}X_n = X_{n+1}$.  In this form, one can see the analogy with the Schr\"odinger equation in quantum mechanics with $Q_n$ and $R_n$ playing the role of the potential.  The potentials correspond to the real-space classical fields $\psi_n,\psi_n^\ast$ in the AL. The time-dependence is postulated to have the form
\begin{equation}\label{eqn:time_evol}
 \begin{split}
  \dot{V}_{1,n} = &\, A_n(t) V_{1,n} + B_n(t) V_{2,n}\\
  \dot{V}_{2,n} = &\, C_n(t) V_{1,n} + D_n(t) V_{2,n}.
 \end{split}
\end{equation}
The associated equations of motion for $A_n, B_n, C_n, D_n$ are generated by cross-differentiating (\ref{eqn:gev}) and (\ref{eqn:time_evol})
\begin{equation}
 \begin{split}
  \pd{}{t}\left(V_{1,n+1}\right) 
	=&\, z \dot{V}_{1,n} + \dot{Q}_n V_{2,n}+ Q_n\dot{V}_{2,n} \\
	=&\, z \left(A_n V_{1,n} + B_n V_{2,n}\right) + \dot{Q}_n V_{2,n}+ Q_n\left(C_n V_{1,n} + D_n
	 V_{2,n}\right)\\
  \pd{}{t}\left(V_{2,n+1}\right) 
	= &\, \frac{1}{z} \dot{V}_{2,n} + \dot{R}_n V_{1,n} + R_n \dot{V}_{1,n}\\
	= &\, \frac{1}{z} \left(C_n V_{1,n} + D_n V_{2,n}\right) + \dot{R}_n V_{1,n} + R_n \left(A_n V_{1,n}
	 + B_n V_{2,n}\right)\\
  \left(\pd{V_{1,n^\prime}}{t}\right)_{n^\prime=n+1} 
	= &\, A_{n+1} V_{1,n+1} + B_{n+1} V_{2,n+1}\\
	= &\, A_{n+1} \left( z V_{1,n} + Q_n V_{2,n}\right) + B_{n+1} \left(\frac{1}{z} V_{2,n} + R_n 
	V_{1,n}\right)\\
  \left(\pd{V_{2,n^\prime}}{t}\right)_{n^\prime=n+1} 
  	= &\, C_{n+1} V_{1,n+1} + D_{n+1} V_{2,n+1}\\
	= &\, C_{n+1} \left(z V_{1,n} + Q_n V_{2,n}\right)  + D_{n+1} \left(\frac{1}{z} V_{2,n} + R_n V_{1,n}\right)
 \end{split}
\end{equation}
where the eigenvalue, $z$ is time-invariant and the explicit time-dependence has been dropped.  Requiring $\frac{\partial}{\partial t} V_{i,n+1} =  \left(\frac{\partial}{\partial t} V_{i,n^\prime}\right)_{n^\prime = n+1}$, 
\begin{equation}
 \begin{split}
	z \left(A_n V_{1,n} + B_n V_{2,n}\right) + &\dot{Q}_n V_{2,n}+ Q_n\left(C_n V_{1,n} + D_n
	 V_{2,n}\right)\\
	&= \, A_{n+1} \left( z V_{1,n} + Q_n V_{2,n}\right) + B_{n+1} \left(\frac{1}{z} V_{2,n} + R_n 
	V_{1,n}\right)\\
	\frac{1}{z} \left(C_n V_{1,n} + D_n V_{2,n}\right) + &\dot{R}_n V_{1,n} + R_n \left(A_n V_{1,n}
	 + B_n V_{2,n}\right)\\
	&= \, C_{n+1} \left(z V_{1,n} + Q_n V_{2,n}\right)  + D_{n+1} \left(\frac{1}{z} V_{2,n} + R_n
	 V_{1,n}\right)
\end{split}
\end{equation}
and equating the coefficients of $V_{i,n}$ for each equation results in the following time-evolution equations,
\begin{equation}\label{eqn:time_evol_An}
 \begin{split}
  z\Delta_n A_n + R_n B_{n+1} - Q_nC_n &\,= 0\\
  (1/z)B_{n+1} - z B_n + Q_n(A_{n+1} - D_{n}) &\,= \dot{Q}_n\\
  zC_{n+1} - (1/z) C_n - R_n(A_n - D_{n+1}) &\,= \dot{R}_n\\
  (1/z)\Delta_n D_n + Q_n C_{n+1} - R_nB_n &\,= 0,
 \end{split}
\end{equation}
where $\Delta_n X_n\equiv X_{n+1}-X_n$. 

The linear dispersion relation and the above equations suggest the following expansions for the functions of (\ref{eqn:time_evol}):
\begin{equation}
 \begin{split}
  A_n =&\; z^2A_n^{(2)} + A_n^{(0)}\\
  B_n =&\; zB_n^{(1)} + \frac{1}{z}B_n^{(\text{-}1)}\\
  C_n =&\; zC_n^{(1)} + \frac{1}{z}C_n^{(\text{-}1)}\\
  D_n =&\; D_n^{(0)} + \frac{1}{z^2}D_n^{(\text{-}2)} 
 \end{split}
\end{equation}
The coefficients of these expansions are determined by subsituting back into (\ref{eqn:time_evol_An}),
\begin{equation}
 \begin{split}
  z^3\Delta_n A_n^{(2)} + z\Delta_n A_n^{(0)}+ zR_n B_{n+1}^{(1)} + (1/z)R_n B_{n+1}^{(\text{-}1)} &\\
	 - zQ_nC_n^{(1)} - (1/z)Q_nC_n^{(\text{-}1)} &\,= 0\\
  B_{n+1}^{(1)}+ (1/z^2)B_{n+1}^{(\text{-}1)} - z^2 B_n^{(1)} - B_n^{(\text{-}1)} + Q_n( z^2A_{n+1}^{(2)} + 		A_{n+1}^{(0)}  &\\
 	- D_n^{(0)} - (1/z^2)D_n^{(\text{-}2)} ) &\,= \dot{Q}_n\\
  z^2C_{n+1}^{(1)} + C_{n+1}^{(\text{-}1)}- C_n^{(1)}-(1/z^2) C_n^{(\text{-}1)} - R_n(z^2A_{n}^{(2)} + A_{n}^{(0)} & \\
	  - D_{n+1}^{(0)}- (1/z^2)D_{n+1}^{(\text{-}2)}) &\,= \dot{R}_n\\
  (1/z)\Delta_n D_n^{(0)} + (1/z^3)\Delta_nD_n^{(\text{-}2)} + zQ_n C_{n+1}^{(1)} - (1/z)Q_nC_{n+1}^{(\text{-}1)}& \\
	  - zR_n B_n^{(1)} - (1/z)R_nB_n^{(\text{-}1)} &\,= 0,
 \end{split}
\end{equation}
and solving in powers of 'z':
\begin{equation}
\begin{array}{ll}
	\mathcal{O} (z^3):& \Delta_n A_n^{(2)} = 0 \qquad \Rightarrow \; A_{n+1}^{(2)} = A_n^{(2)} \equiv A\_^{(2)}\\
	\mathcal{O} \left(\frac{1}{z^3}\right):& \Delta_n D_n^{(\text{-}2)} = 0 \qquad \Rightarrow \;  D_{n+1}^{(\text{-}2)} = D_n^{(\text{-}2)} \equiv D\_^{(\text{-}2)}\\
	\mathcal{O} \left(z^2\right):& B_n^{(1)} - Q_n A_{n}^{(2)} = 0 \qquad \Rightarrow \; B_n^{(1)} = Q_n A\_^{(2)}\\
	& C_{n+1}^{(1)} - R_n A_{n}^{(2)} = 0 \qquad \Rightarrow \; C_{n}^{(1)} = R_{n-1} A\_^{(2)}\\
	\mathcal{O} \left(\frac{1}{z^2}\right):& B_{n+1}^{(\text{-}1)} - Q_n D_{n}^{(\text{-}2)} = \qquad \Rightarrow \; B_n^{(\text{-}1)} = Q_{n-1} D\_^{(\text{-}2)}\\
	& C_{n}^{(\text{-}1)} - R_n D_{n}^{(\text{-}2)} = 0 \qquad \Rightarrow \; C_{n}^{(1)} = R_{n} D\_^{(\text{-}2)}\\
	\mathcal{O} \left(z \right): & \Delta_n A_n^{(0)} = Q_nC_{n}^{(1)} - R_n B_{n+1}^{(1)}   \\
	&\phantom{\Delta_n A_n^{(0)}}  = - A\_^{(2)} \left( Q_{n+1} R_{n} - Q_n R_{n-1} \right)\\
	 & \Rightarrow  \; A_n^{(0)} = - A\_^{(2)} Q_{n} R_{n-1} + A\_^{(0)} \\
	\mathcal{O}\left(\frac{1}{z}\right):&  \Delta_n D_n^{(0)} =  R_n B_{n}^{(\text{-}1)}- Q_nC_{n+1}^{(\text{-}1)}  \\
	&\phantom{\Delta_n D_n^{(0)}} = - D\_^{(\text{-}2)} \left( Q_{n} R_{n+1} - Q_{n-1} R_{n} \right)  \\
	& \Rightarrow \;  D_n^{(0)} = - D\_^{(2)} Q_{n-1} R_{n} + D\_^{(0)}.
\end{array}
\end{equation}
The zeroeth order yields the time-dependence of the potentials,
\begin{equation}
 \begin{split}
	\mathcal{O}\left(z^0\right):& \\
	\qquad \dot{Q}_n &= B_{n+1}^{(1)} - B_{1}^{(\text{-}1)} +  Q_n\left( A_{n+1}^{(0)}- D_{n}^{(0)}\right) \\
	\qquad \dot{Q}_n &= \left( A\_^{(2)} Q_{n+1} - D\_^{(\text{-}2)} Q_{n-1}  \right) \left(1 - Q_n R_n \right) + Q_n \left(A\_^{(0)} - D\_^{(0)} \right)\\
	 \qquad \dot{R}_n &= C_{n+1}^{(\text{-}1)} - C_{n}^{(1)} - R_n\left( A_{n}^{(0)}- D_{n+1}^{(0)}\right) \\
	\qquad \dot{R}_n &= \left( D\_^{(2)} R_{n+1} - A\_^{(\text{-}2)} R_{n-1}  \right) \left(1 - Q_n R_n \right) - R_n \left(A\_^{(0)} - D\_^{(0)} \right).
 \end{split}
\end{equation}

In summary, the coefficients of the time-evolution equations are given by
\begin{equation}
 \begin{split}
  A_n &= A\_^{(2)}(z^2 - Q_n R_{n-1}) + A\_^{(0)}\\
  B_n &= A\_^{(2)}Q_nz - D\_^{(\text{-}2)}Q_{n-1} (1/z)\\
  C_n &= A\_^{(2)}R_{n-1}z - D\_^{(\text{-}2)}R_n(1/z)\\
  D_n &= D\_^{(\text{-}2)}((1/z^2) - Q_{n-1}R_n) + D\_^{(0)},
 \end{split}
\end{equation}
while the time evolutions of $Q_n$ and $R_n$ are governed by 
\begin{subequations}
\begin{align}
\dot{Q}_n = &\;\left( A\_^{(2)} Q_{n+1} - D\_^{(\text{-}2)} Q_{n-1}  \right) \left(1 - Q_n R_n \right) + Q_n \left(A\_^{(0)} - D\_^{(0)} \right) \label{eqn:time_evol_QR}\\
\dot{R}_n = &\;\left( D\_^{(2)} R_{n+1} - A\_^{(\text{-}2)} R_{n-1}  \right) \left(1 - Q_n R_n \right) - R_n \left(A\_^{(0)} - D\_^{(0)} \right)
\end{align}
\end{subequations}
\subsection{AL equation}
To obtain the AL equation, we let $R_n = \pm \alpha Q_n^\ast$, $A\_^{(2)} = -D\_^{(\text{-}2)} = i$, and $A\_^{(0)} = -D\_^{(0)} = -i$.  In this case equation (\ref{eqn:time_evol_QR}) becomes
\begin{equation}\label{eqn:AL}
 \dot{Q}_n = i\left[ Q_{n+1} + Q_{n-1} - 2Q_{n} \mp \alpha |Q_n|^2(Q_{n+1} + Q_{n-1}) \right],
\end{equation}
which is the AL equation. With these substitutions, the time-dependent function $A_n,...,D_n$ obey
\begin{equation}
 \begin{split}
  A_n &= i(z^2 \mp \alpha Q_nQ^\ast_{n-1} -1)\\
  B_n &= i(zQ_n - (1/z)Q_{n-1})\\
  C_n &= \pm i \alpha(zQ^\ast_{n-1} - (1/z)Q^\ast_n)\\
  D_n &= i (1 - (1/z^2) \pm \alpha Q_{n-1}Q^\ast_n).
 \end{split}
\end{equation}
%
\subsection{Direct Scattering Problem}
\paragraph{Asymptotic Solution of Generalized Eigenvalue Problem} 
First of all we assume that $Q_n$ and $R_n$ are on compact support, that is that as $|n|\rightarrow\infty$, $Q_n, R_n \rightarrow 0$.  In the limit $|n|\rightarrow\infty$, the generalized eigenvalue problem becomes,
\begin{equation}
|n|\rightarrow\infty: \qquad \left\{
\begin{array}{rl}
  V_{1,n+1} = &\, z V_{1,n} \\
  V_{2,n+1} = &\, \frac{1}{z} V_{2,n}.   
\end{array}
\right. 
\end{equation}
To solve this direct scattering problem, define the time-independent eigenfunctions, $\phi_n, \bar{\phi}_n, \psi_n, \bar{\psi}_n $ which have the asymptotic forms:
\begin{equation}
 \begin{split}
   n\rightarrow - \infty:\qquad  &\phi_n \sim \begin{pmatrix} 1 \\ 0 \end{pmatrix} z^n,\qquad
    			   \bar{\phi}_n \sim \begin{pmatrix} 0 \\ -1 \end{pmatrix} z^{-n}\\
   n\rightarrow + \infty:\qquad  &\psi_n \sim \begin{pmatrix} 0 \\ 1 \end{pmatrix} z^{-n}, \qquad
   			   \bar{\psi}_n \sim \begin{pmatrix} 1 \\ 0 \end{pmatrix} z^{n}.
 \end{split}
\end{equation}
These eigenfunctions satisfy the generalized eigenvalue problem at a fixed time, which will be taken to be t=0.  They do not solve the time-evolution equations, which we will return to later. 
One can establish by induction, that 
\begin{equation}
 \begin{split}
  & z^{-n}\phi_n, \: z^n\psi_n \text{ are polynomial in powers of }\frac{1}{z} \text{ and are analytic for } |z| > 1 \\
  & z^{n}\bar{\phi}_n, \: z^{-n}\bar{\psi}_n \text{ are polynomial in powers of } z \text{ and are analytic for } |z| < 1 \; \left( |z| < \infty\right)
 \end{split}
\end{equation}
given that $Q_n$ and $R_n$ are on compact support.  
%
\paragraph{Wronskian and Linear Independence}
The Wronskian of a set of functions $\{\phi_1, ... \phi_n\}$ is defined by:
\begin{equation}
 W(\phi_1, ... \phi_n) \equiv \begin{vmatrix} \phi_1 & \phi_2 & \cdots& \phi_n \\
                               		\phi^\prime_1 & \phi^\prime_2 & \cdots& \phi^\prime_n\\
					\vdots & \vdots & \ddots & \vdots \\
                               		\phi^{(n-1)}_1 & \phi^{(n-1)}_2 & \cdots & \phi^{(n-1)}_n
                              \end{vmatrix}
\end{equation}
If the Wronskian is non-zero in some interval, then the functions are linearly independent in that interval.

For this problem, the Wronskian is defined as $W_n(u,v) = u_{1,n}v_{2,n} - u_{2,n}v_{1,n}$. The Wronskian of the functions that obey (\ref{eqn:gev}) can be found by
\begin{equation}
 \begin{split}
	W_{n+1}(u,v) 	= &\; u_{1,n+1}v_{2,n+1} - u_{2,n+1}v_{1,n+1}  \\
 			= &\; \left(z u_{1,n} + Q_n u_{2,n}\right)\left( (1/z) v_{2,n} + R_n v_{1,n}\right) \\
			  & \;- \left( (1/z) u_{2,n} + R_n u_{1,n}\right)\left(z v_{1,n} + Q_n v_{2,n}\right)\\
			= &\; u_{1,n}v_{2,n} + z R_nu_{1,n}v_{1,n} + 
			 (1/z) Q_n  u_{2,n}v_{2,n} + Q_nR_n  u_{2,n}v_{1,n}\\
			  &\; - \left( u_{2,n}v_{2,n} + (1/z)Q_n u_{2,n}v_{2,n} 
			+ zR_n  u_{1,n}v_{1,n} + Q_nR_n  u_{2,n}v_{1,n}  \right)\\
			= &\; \left( 1 - R_nQ_n\right) \left(u_{1,n}v_{2,n} -
			 u_{2,n}v_{1,n} \right)\\
			=& \; \left( 1 - R_nQ_n\right)W_n(u,v)
 \end{split}
\end{equation}
Using the asymptotic forms of $\phi_n, \bar{\phi}_n, \psi_n, \bar{\psi}_n$, one gets
\begin{equation}\begin{split}
                 \lim_{n\rightarrow -\infty} W_n(\bar{\phi}, \phi)& = - \bar{\phi}_{2,n}\phi_{1,n} = -(\text{-}z^{-n})(z^{n}) = 1\\ \lim_{n\rightarrow +\infty} W_n(\bar{\psi}, \psi)& = - \bar{\psi}_{1,n}\psi_{2,n} = (z^{n})(z^{-n}) = 1. 
\
                \end{split}
 \end{equation}
 By induction, one can show that the Wronskian of the eigenfunctions obey
\begin{equation}
 W_n(\bar{\phi}, \phi) = \prod_{i=-\infty}^n (1-R_iQ_i), \qquad 
W_n(\bar{\psi}, \psi) = \prod_{i=n+1}^\infty (1-R_iQ_i)^{-1}
\end{equation}
for $z$ on the unit circle. For $R_n = -Q_n^\ast$, $W_n$ is positive-definite so that $\bar{\phi}_n, \phi_n$ are linearly independent. Otherwise assume that initially $R_n$ and $Q_n$ are less than one. Likewise one can show that $\bar{\psi}_n, \psi_n$ are linearly independent. Thus one can define the scattering data by
\begin{equation}\label{eqn:phi_vs_psi}
 \begin{split}
  \phi_n =&\, a(z,t)\bar{\psi}_n + b(z,t)\psi_n\\
  \bar{\phi}_n =&\, -\bar{a}(z,t)\psi_n + \bar{b}(z,t)\bar{\psi}_n
 \end{split}
\end{equation}
where $t$ is a parameter.
%
\paragraph{Time Dependence}
The eigenfunctions $\phi,\bar{\phi},\psi,\bar{\psi}$ satisfy the generalized eigenvalue problem, but not the time evolution equations (\ref{eqn:time_evol}).  In the asymptotic limit (\ref{eqn:time_evol}) becomes
\begin{equation}
 \begin{matrix} 
  \dot{V}_{1,n} & = & A_{\pm}V_{1,n}\\
  \dot{V}_{2,n} & = & D_{\pm}V_{2,n}
 \end{matrix}
\end{equation}
where 
\begin{equation}
 \begin{split} 
  A_{\pm} & = \lim_{n \rightarrow \pm \infty}(A_n)  = z^2A\_^{(2)} + A\_^{(0)}\\
  D_{\pm} & = \lim_{n \rightarrow \pm \infty}(D_n) = (1/z^2)D\_^{(2)} + D\_^{(0)}.
 \end{split}
\end{equation}
Define a new set of eigenfunctions which will satisfy both the generalized eigenvalue problem and the time-dependence
\begin{equation}\label{eqn:def_phi_psi_t}
 \begin{matrix}
  &\phi_n^{(t)}  = \phi_n \exp(A\_ t) \qquad &\psi_n^{(t)} = \psi_n \exp(D_+t) \\
  &\bar{\phi}_n^{(t)}  = \bar{\phi}_n \exp(D\_ t) \qquad &\psi_n^{(t)} = \psi_n \exp(A_+t) 
 \end{matrix}
\end{equation}
Both $\{\bar{\phi}_n^{(t)}, \phi_n^{(t)}\}$ and $\{\bar{\psi}_n^{(t)}, \psi_n^{(t)}\}$ are linearly independent, so we may write
\begin{equation}\label{eqn:phi_t_vs_psi_t}
 \begin{split}
  \phi_n^{(t)} =&\, a_0\bar{\psi}^{(t)}_n + b_0\psi^{(t)}_n\\
  \bar{\phi}_n^{(t)} =&\, -\bar{a}_0\psi^{(t)}_n + \bar{b}_0\bar{\psi}^{(t)}_n
 \end{split}
\end{equation}
where $a_0,\bar{a}_0,b_0,\bar{b}_0$ equal the scattering coefficients, $a,\bar{a},b,\bar{b}$ at t=0. Substituting (\ref{eqn:def_phi_psi_t}) into (\ref{eqn:phi_t_vs_psi_t}) gives
\begin{equation}
 \begin{split}
  \phi_n =&\, a_0e^{[(A_+ - A_{-})t]}\bar{\psi} + b_0e^{[(D_+ - A_-)t]}\psi_n\\
  \bar{\phi}_n =&\, -\bar{a}_0e^{[(D_+ - D_{-})t]}\psi_n + \bar{b}_0e^{[(A_+ - D_{-})t]}\bar{\psi}_n
 \end{split}
\end{equation}
Comparing with (\ref{eqn:phi_vs_psi}) and noting that $A_+ = A_-$ and $D_+ = D_-$, the scattering coefficients are
\begin{equation}
 \begin{split}
  a = a_0, \qquad & b = b_0 e^{[(D_+ - A_{-})t]}\\
  \bar{a} = \bar{a}_0, \qquad & \bar{b} = \bar{b}_0 e^{[(D_+ - A_{-})t]}.
 \end{split}
\end{equation}
The most significant result here, for our purposes, is that the coefficients $a, \bar{a}$ are constant.  In the next section, we use this to derive the conserved quantities of the AL equation.
%
%
\subsection{Conservation Laws}
The conservation laws can be derived by considering the asymptotic form of $\bar{a}(z)$.  
\begin{equation}
 \bar{\phi}_n \sim - \bar{a}(z)\begin{pmatrix} 0 \\ 1 \end{pmatrix}z^{-n} +  \bar{b}(z)\begin{pmatrix} 1 \\ 0 \end{pmatrix} z^n \qquad \Rightarrow  \qquad  \bar{a}(z)  \sim -  z^n\bar{\phi}_{2,n}
\end{equation}
Next substitute $\bar{\phi}_n$ into the eigenvalue problem (\ref{eqn:gev}),
\begin{subequations}
\begin{align}
 z^{n-1}\bar{\phi}_{1,n+1} &= z^{n} \bar{\phi}_{1,n}+ z^{n-1}Q_n \bar{\phi}_{2,n} \label{eqn:asymtot_ev_1}\\
 z^n\bar{\phi}_{2,n+1} &= z^{n-1} \bar{\phi}_{2,n}+ z^n R_n \bar{\phi}_{1,n} \label{eqn:asymtot_ev_2}.
\end{align}
\end{subequations}
Solve (\ref{eqn:asymtot_ev_2}) for $z^n\bar{\phi}_{1,n}$ and substitute into (\ref{eqn:asymtot_ev_1}), to eliminate $\bar{\phi}_{1,n}$
\begin{equation}
\begin{split}
z^{n}\bar{\phi}_{1,n} = \frac{1}{R_n}\left(z^{n}\bar{\phi}_{2,n+1} - z^{n-1} \bar{\phi}_{2,n}\right) = \frac{1}{zR_n}\Delta_n \left(z^{n}\bar{\phi}_{2,n}\right)\\
z^{n-2} \frac{1}{R_{n+1}}\Delta_n \left(z^{n+1}\bar{\phi}_{2,n+1}\right) =  \frac{1}{zR_n}\Delta_n \left(z^{n}\bar{\phi}_{2,n}\right) + z^{n-1}Q_n \bar{\phi}_{2,n}\\
\frac{1}{z^{2n+3}R_{n+1}}\Delta_n \left(z^{n+1}\bar{\phi}_{2,n+1}\right) - \frac{1}{z^{2n+1}R_n}\Delta_n \left(z^{n}\bar{\phi}_{2,n}\right) = Q_n z^{-2n-1}\left(z^{n}\bar{\phi}_{2,n}\right),
\end{split}
\end{equation}
which can be written as,
\begin{equation}\label{eqn:phi2bar}
 \Delta_n\left( \frac{\Delta_n (z^n\bar{\phi}_{2,n})}{z^{2n+1}R_n}\right) = Q_n z^{-2n-1} (z^n\bar{\phi}_{2,n}).
\end{equation}
Define 
\begin{equation}
-z^n\bar{\phi}_{2,n} = \prod_{k=-\infty}^n g_k
\end{equation}
such that $\Delta_n ( -z^n\bar{\phi}_{2,n}) = \Delta_n(\prod_{k=-\infty}^n g_k ) = (g_{n+1} - 1)\prod_{k=-\infty}^n g_k $ and substitute into (\ref{eqn:phi2bar}) to get the recursion relation
\begin{equation}
 g_{n+1} (g_{n+2} - 1 ) - z^2 \frac{R_{n+1}}{R_{n}}(g_{n+1} - 1 ) = z^2R_{n+1}Q_n,
\end{equation}
which can be re-written as
\begin{equation}
 R_{n-1}g_n(g_{n+1} - 1 ) - z^2 R_{n}(g_{n} - 1 ) = z^2 R_{n}R_{n-1}Q_{n-1}.
\end{equation}
Expand $g_n$ and $g_{n+1}$ in powers of $z^2$,
\begin{equation}
\begin{split}
  g_n &= g_n^{(0)} + z^2 g_n^{(1)}+ z^4 g_n^{(2)} + ... = \sum_{m=0} g_n^{(m)} z^{2m}\\
  g_{n+1} &= g_{n+1}^{(0)} + z^2 g_{n+1}^{(1)}+ z^4 g_{n+1}^{(2)} + ... = \sum_{m=0} g_{n+1}^{(m)} z^{2m}.
\end{split}
\end{equation}
Substitute into the recursion relation for $g_n$, $g_{n+1}$,
\begin{equation}
\begin{split}
 R_{n-1}\sum_{m=0} g_n^{(m)} z^{2m}(\sum_{l=0} g_{n+1}^{(l)} z^{2l} - 1 ) - z^2 R_{n}(\sum_{m=0} & g_n^{(m)} z^{2m} - 1 )\\
&= z^2 R_{n}R_{n-1}Q_{n-1}\\
 R_{n-1}\left[\sum_{m,l=0} z^{2(m+l)} g_n^{(m)} g_{n+1}^{(l)}  -  \sum_{m=0} z^{2m}g_n^{(m)} \right] - & R_n\sum_{m=0}  z^{2(m+1)} g_n^{(m)}\\
 &= z^2 R_{n}\left( R_{n-1}Q_{n-1}  - 1 \right).
\end{split}
\end{equation}

Solving recursively for $g_n^{(m)}$ in orders of $z^2$, gives
 
\begin{equation}
\begin{split}
\mathcal{O}(z^0 ):  \qquad g_n^{(0)} = &\, 1\\
\mathcal{O}(z^2 ):  \qquad g_n^{(1)} = &\, R_{n-1}Q_{n-2}\\
\mathcal{O}(z^4 ):  \qquad g_n^{(2)} = &\, R_{n-1}Q_{n-3}\left(1 - R_{n-2}Q_{n-2} \right)
\end{split}
\end{equation}

For orders of $z^{2p}$, $p \ge 2$, this recursion relation reduces to 
\begin{equation}
\begin{split}
R_{n-1}\left[\sum_{m=0}^{p} g_n^{(m)} g_{n+1}^{(p-m)}  -  g_n^{(p)} \right] =
 R_{n-1}\left[ \sum_{m=1}^{p-1} g_n^{(m)} g_{n+1}^{(p-m)}  +  g_{n+1}^{(p)} \right] =  R_n  g_n^{(p-1)}
\end{split}
\end{equation}

or 
\begin{equation}
\boxed{ g_{n+1}^{(p)}= \frac{R_n}{R_{n-1} }  g_n^{(p-1)} - \sum_{m=1}^{p-1} g_n^{(m)} g_{n+1}^{(p-m)} . }
\end{equation}
which gives expressions for higher orders of the coefficients,
\begin{equation}
\begin{split}
\mathcal{O}(z^6 ):  \qquad g_n^{(3)} = &\, \frac{R_{n-1}}{R_{n-2}}g_{n-1}^{(2)} - g_{n-1}^{(1)}g_{n}^{(2)} - g_{n-1}^{(2)}g_{n}^{(1)}\\
= &\, R_{n-1} Q_{n-4}\big(1 - R_{n-3}Q_{n-3} - R_{n-2}Q_{n-2}\\
& \phantom{ R_{n-1} Q_{n-4}\big(} + R_{n-3}Q_{n-3}R_{n-2}Q_{n-2}\big) \\
&\, - R_{n-1} R_{n-2} Q_{n-3}^2\left(1 - R_{n-2}Q_{n-2}\right)\\
\mathcal{O}(z^8 ):  \qquad g_n^{(4)} = &\, \frac{R_{n-1}}{R_{n-2}}g_{n-1}^{(3)} - g_{n-1}^{(1)}g_{n}^{(3)} -
g_{n-1}^{(2)}g_{n}^{(2)}- g_{n-1}^{(3)}g_{n}^{(1)}\\
\mathcal{O}(z^{10}):  \qquad g_n^{(5)} = &\, \frac{R_{n-1}}{R_{n-2}}g_{n-1}^{(4)} - g_{n-1}^{(1)}g_{n}^{(4)} -
g_{n-1}^{(2)}g_{n}^{(3)}- g_{n-1}^{(3)}g_{n}^{(2)} - g_{n-1}^{(4)}g_{n}^{(1)} 
\end{split}
\end{equation}

Next use the expansion
\begin{equation}
 \log(1+x) = \sum_{n=1}^\infty \frac{x^n}{n}(-1)^{n+1}
\end{equation}
to write out the series expansions,
\begin{equation}
\begin{split}
 \log[\bar{a}(z)] = &\lim_{n \rightarrow \infty} \log \left( \prod_{k=-\infty}^n g_k\right) = \sum_{n=-\infty}^\infty \log(g_n) = \sum_{n=-\infty}^\infty \log\left[\sum_{m=0}^\infty g_n^{(m)}z^{2m}\right]\\
=&\, \sum_{n=-\infty}^\infty \log\left[1 + \sum_{m=1}^\infty g_n^{(m)}z^{2m}\right]= \sum_{n=-\infty}^\infty \sum_{p=1}^{\infty}\frac{(-1)^{p+1}}{p}\left[\sum_{m=1}^\infty g_n^{(m)}z^{2m}\right]^p. 
\end{split}
\end{equation}
Since $\bar{a}(z)$ is a constant of motion, each coefficients of $z^{2\alpha}$ in the above expansion of the must also be time-independent.  These coefficients are the conserved quantities.
Expanding the first few terms gives,
\begin{equation}
\begin{split}
 \log[\bar{a}(z)] = & \sum_{n=-\infty}^\infty \sum_{p=1}^{\infty}\frac{(-1)^{p+1}}{p}\big[ g_n^{(1)}z^{2} + g_n^{(2)}z^{4} + g_n^{(3)}z^{6} + g_n^{(4)}z^{8}\\
& \phantom{\sum_{n=-\infty}^\infty \sum_{p=1}^{\infty}\frac{1}{p}\big[ g_n^{(1)}z^{2} + g_n^{(2)}z^{4} + g_n^{(3)}z^{6} } + g_n^{(5)}z^{10} + \mathcal{O}(z^{12})\big]^p\\
= & \sum_{n=-\infty}^\infty g_n^{(1)}z^{2} + \left(g_n^{(2)} - \frac{1}{2}\left[g_n^{(1)}\right]^2 \right)z^{4} \\
& \phantom{\sum_{n=-\infty}^\infty} + \left( g_n^{(3)}  - g_n^{(1)}g_n^{(2)} + \frac{1}{3} \left[g_n^{(1)}\right]^3\right)z^{6}\\
& \phantom{\sum_{n=-\infty}^\infty} + \left(g_n^{(4)} - \frac{1}{2} \left[g_n^{(2)}\right]^2 - g_n^{(1)} g_n^{(3)} +  \left[g_n^{(1)}\right]^2g_n^{(2)}-  \frac{1}{4} \left[g_n^{(1)}\right]^4 \right)z^{8} \\
& \phantom{\sum_{n=-\infty}^\infty} + \bigg(g_n^{(5)}- g_n^{(4)}g_n^{(1)}- g_n^{(3)}g_n^{(2)} +
\left[g_n^{(2)}\right]^2g_n^{(1)}\\ 
& \phantom{\sum_{n=-\infty}^\infty + \bigg(g_n^{(5)}- g_n^{(4)}g_n^{(1)}}
- g_n^{(3)}\left[g_n^{(1)}\right]^2+g_n^{(2)}\left[g_n^{(1)}\right]^3 - \frac{1}{5}g_n^{(5)}\bigg)z^{10}\\
 & \phantom{\sum_{n=-\infty}^\infty}+ \mathcal{O}(z^{12}).
\end{split}
\end{equation}

\subsection{Conserved Quantities}
The first few conserved quantities are:
\begin{equation}
 \begin{split}
  C_1 = \sum_n 	&\, g_n^{(1)} 
      = \sum_n 	\, R_n Q_{n-1}\\
  C_2 = \sum_n 	&\, g_n^{(2)} + \frac{1}{2}\left[g_n^{(1)}\right]^2
      = \sum_n 	\, R_n Q_{n-2}\left(1 - R_{n-1}Q_{n-1} \right) + \frac{1}{2}R_n^2Q_{n-1}^2\\
  C_3 = \sum_n 	&\, g_n^{(3)}+g_n^{(1)}g_n^{(2)} + \frac{1}{3} \left[g_n^{(1)}\right]^3\\
      = \sum_n 	&\, R_n Q_{n-3}\left(1 - R_{n-2}Q_{n-2} - R_{n-1}Q_{n-1} + R_{n-2}Q_{n-2}R_{n-1}Q_{n-1}\right)  \\
        	&\, - R_{n}R_{n-1}Q_{n-2}^2 \left(1 - R_{n-1}Q_{n-1}\right) + R_{n}^2Q_{n-1}Q_{n-2}\left(1 - R_{n-1}Q_{n-1}\right)\\
		& \, + \frac{1}{3} \left( R_nQ_{n-1} \right)^3\\
  C_4 = \sum_n &\, g_n^{(4)}+ \frac{1}{2} \left[g_n^{(2)}\right]^2 + g_n^{(1)} g_n^{(3)} +  \left[g_n^{(1)}\right]^2g_n^{(2)}+ \frac{1}{4} \left[g_n^{(1)}\right]^4 \\
  C_5 = \sum_n &\, g_n^{(5)}+ g_n^{(4)}g_n^{(1)}+g_n^{(3)}g_n^{(2)} +
\left[g_n^{(2)}\right]^2g_n^{(1)}+ g_n^{(3)}\left[g_n^{(1)}\right]^2+g_n^{(2)}\left[g_n^{(1)}\right]^3 + \frac{1}{5}g_n^{(5)}
 \end{split}
\end{equation}
Further conserved quantities can be determined by using the recursion relations and expansions given. From looking at these equations, we notice a few patterns.  First of all, the leading order in $\alpha$ of $C_m $ is $\sum_n \psi_{n}^\ast\psi_{n-m}$, and the largest order of $\alpha$ is $\alpha^m$.

\section{AL and BHM}
To obtain the AL, \ref{eqn:AL} we let $R_n=\alpha Q_n^\ast$. Furthermore, to put it in a more familiar form, we write it in terms of the real-space fields, $\tilde{\psi}_n=Q_n$, and $\tilde{\psi}_n^\ast=Q_n^\ast$. 
\begin{equation}
\pd{}{t}\tilde{\psi}_n = i \left[ \left(\tilde{\psi}_{n+1} + \tilde{\psi}_{n-1}- 2\tilde{\psi}_n\right)
- \alpha|\tilde{\psi}_n|^2(\tilde{\psi}_{n+1} + \tilde{\psi}_{n-1})\right]\\
\end{equation}
Recall that in real-space the BHM has the form 
\begin{equation}
\pd{}{t}\tilde{\psi}_n = -i \left[ -J\left(\tilde{\psi}_{n+1} + \tilde{\psi}_{n-1}- 2\tilde{\psi}_n\right)
+ \mu_0N_{\mbox{\scriptsize s}}|\tilde{\psi}_n|^2 \tilde{\psi} \right]\\
\end{equation}
By defining a new time scale, $\tau = Jt$, 
\begin{equation}
\pd{}{\tau}\tilde{\psi}_n = i \left[ \left(\tilde{\psi}_{n+1} + \tilde{\psi}_{n-1}- 2\tilde{\psi}_n\right)
- \frac{\mu_0N_{\mbox{\scriptsize s}}}{J}|\tilde{\psi}_n|^2 \tilde{\psi}_n \right]\\
\end{equation}
Note that here $n$ is an index of real-space fields, following the notation of \citet{ablowitz_nonlinear_1976}, while in earlier chapters it was used as a momentum index. From these two forms, it can be seen that these two equations are mathematically close for $\alpha = 2\mu_0N_s/J \equiv 2\kappa N_s$ and identical in the limit $\psi_{n+1} + \psi_{n-1} \rightarrow 2\psi_n$. 

The first few coefficients of $g_n$ are
\begin{equation}
 \begin{split}
g_n^{(1)} = &\, \alpha\tilde{\psi}^\ast_{n-1}\tilde{\psi}_{n-2}\\
g_n^{(2)} = &\, \alpha\tilde{\psi}^\ast_{n-1}\tilde{\psi}_{n-3}\left(1 - \alpha|\tilde{\psi}_{n-2}|^2 \right)\\
g_n^{(3)} = &\, \frac{\alpha\tilde{\psi}^\ast_{n-1}}{\alpha\tilde{\psi}^\ast_{n-2}}g_{n-1}^{(2)} - g_{n-1}^{(1)}g_{n}^{(2)} - g_{n-1}^{(2)}g_{n}^{(1)}\\
= &\, \alpha\tilde{\psi}^\ast_{n-1} \tilde{\psi}_{n-4}\left(1 - \alpha|\tilde{\psi}_{n-3}|^2 - \alpha|\tilde{\psi}_{n-2}|^2 + \alpha^2|\tilde{\psi}_{n-3}|^2|\tilde{\psi}_{n-2}|^2\right) \\
&\, - \alpha^2\tilde{\psi}^\ast_{n-1}\tilde{\psi}^\ast_{n-2} \tilde{\psi}_{n-3}^2\left(1 - \alpha|\tilde{\psi}_{n-2}|^2\right).
 \end{split}
\end{equation}

In terms of the real-space fields $\{\widetilde{\psi}_n, \widetilde{\psi}_n^\ast\}$, the first few conserved quantities are 
\begin{equation}
 \begin{split}
  C_1 &= \sum_n \alpha \tilde{\psi}^\ast_n \tilde{\psi}_{n-1} \\
  C_2 &= \sum_n  g_n^{(2)} + \frac{1}{2}\left[g_n^{(1)}\right]^2
      = \sum_n \alpha\tilde{\psi}^\ast_n \tilde{\psi}_{n-2}\left(1 - \alpha|\tilde{\psi}_{n-1}|^2 \right) + \frac{1}{2}\alpha^2(\tilde{\psi}^\ast_n)^2\tilde{\psi}_{n-1}^2\\
  C_3 &= \sum_n g_n^{(3)}+g_n^{(1)}g_n^{(2)} + \frac{1}{3} \left[g_n^{(1)}\right]^3\\
      &= \sum_n  \alpha\tilde{\psi}^\ast_n \tilde{\psi}_{n-3}\left(1 - \alpha|\tilde{\psi}_{n-2}|^2 - \alpha|\tilde{\psi}_{n-1}|^2 + \alpha^2|\tilde{\psi}_{n-2}|^2|\tilde{\psi}_{n-1}|^2\right)  \\
       &\phantom{=\sum_n} - \alpha^2\tilde{\psi}^\ast_{n}\tilde{\psi}^\ast_{n-1}\tilde{\psi}_{n-2}^2 \left(1 - \alpha|\tilde{\psi}_{n-1}|^2\right)\\
	&\phantom{=\sum_n}+ \alpha[\tilde{\psi}^\ast_{n}]^2\tilde{\psi}_{n-1}\tilde{\psi}_{n-2}\left(1 - \alpha|\tilde{\psi}_{n-1}|^2\right) \\
        &\phantom{=\sum_n}+\frac{1}{3}\left( \alpha\tilde{\psi}^\ast_n\tilde{\psi}_{n-1} \right)^3.
 \end{split}
\end{equation}
 In terms of the momentum-space fields $\{\psi_p,\psi_p^\ast\}$, $C_1$ and $C_2$ are  
\begin{equation}
 \begin{split}
  C_1 	= \alpha \sum_p &\, |\psi_p|^2 e^{-2\pi ip/N_s}\\
	= \alpha \sum_p &\, |\psi_p|^2 \cos\left(\frac{2\pi p}{N_s}\right) - i \alpha \sum_p | \psi_p|^2 \sin\left(\frac{2\pi p}{N_s}\right)\\
  C_2 = \alpha \sum_p &\, |\psi_p|^2 e^{-4 \pi ip/N_s} 
        - \frac{\alpha^2}{N_s}\sum_{p,q,r} \psi^\ast_p \psi_q^\ast \psi_r \psi_{p+q-r}e^{-2 \pi i(p+r)/N_s}\\
	& + \frac{\alpha^2}{2 N_s}\sum_{p,q,r} \psi^\ast_p \psi_q^\ast \psi_r \psi_{p+q-r}e^{-2 \pi i(p+q)/N_s}.
 \end{split}
\end{equation}

As seen in earlier chapters, there is a threshold for chaos in the BHM.  Below this threshold there are initial states that do not thermalize, but nevertheless relax to a steady-state.  For fully integrable systems it is known that this steady-state can be described by a constrained thermodynamic ensemble that accounts for all of the integrals of motion. Anecdotal evidence suggests that the initial total quasi-momentum plays a role in the dynamics of the BHM for small nonlinearities, $\kappa$. Is it possible that the other integrals of motion of the AL affect the dynamics?  Is the nonthermal steady-state governed by a constained ensemble that takes into account the conserved quantities of nearby integrable models? There are other nearby integrable models, which include the noninteracting case and the continuum limit. The imaginary part of the first conserved quantity of AL, Im $C_1$, is analogous to the total quasi-momentum and Re $C_1$ to the total kinetic energy. In the noninteracting limit the momentum distribution is conserved, so any higher moments of the momentum distribution is also conserved. If the mapping between the BHM and AL fields corresponded to simply equating the fields ($\psi_{n,\text{BHM}} = \psi_{n,\text{AL}}$), then conservation of $C_1$ in BHM would not distinguish between being close to the noninteracting case and close to AL.  In contrast, the second and third terms of $C_2$ are not conserved in the noninteracing model.  However, the mapping between the BHM fields and the AL is nontrivial and the nonlinear corrections to the mapping are expected to allow one to distinguish between effects of quasi-conserved quantities of AL and of the noninteracting case.  A derivation of the mapping between the BHM and AL fields and a study of the role of the conserved quantities of AL in the dynamics of BHM is the subject of proposed future work.

\chapter{Outlook and Conlusion}
\label{chap:conclusion}


\section{Summary of Results}

One of the fundamental assertions of statistical mechanics is that the time average of a physical observable is equivalent to the average over phase-space, with microcanonical measure. A system for which this is true is said to be ergodic and one can calculate dynamical properties of the system from static phase-space averages. Dynamics of a system which is fully integrable, that is has as many conserved quantities as degrees of freedom, is constrained to a reduced phase space and thus not ergodic, although it may relax to a modified equilibrium.  What happens as one moves away from the fully integrable case? 

In this work we have studied the relationship between chaos and thermalization in the one-dimensional Bose-Hubbard model in the classical-field approximation.  We have compared two quantitative measures of chaos and thermalization: (1) the finite-time maximal Lyapunov exponent, averaged over a microcanonical ensemble and (2) the normalized spectral entropy, which is a measure of equipartition of a modified energy in the independent mode approximation. There is a strong correspondence between the Lyapunov exponents and normalized spectral entropy.

We find a threshold for chaos and a corresponding broad transition from incomplete to complete thermalization. The stochasticity threshold is governed by two parameters: the strength of the nonlinearity $\kappa$ and the average energy-per-particle, $\epsilon_T$.  Both of these parameters are finite in the thermodynamic limit and suggest that the threshold will survive in that limit. Far above the threshold, in the strongly chaotic regime, relaxation to the thermal state is complete. In this region, the fluctuations in kinetic energy scale as $\sqrt{N_{\mbox{\scriptsize s}}}$ confirming their thermal nature. We study the size scaling of the Lyapunov exponent and find that it is universal with respect to the size of the lattice. For small nonlinearities, the stochasticity and thermalization thresholds are finite for the range of energies studied and don't show tendencies to vanish at high energies. 

In the vicinity of the threshold the relationship between chaos and thermalization is complex. There is a transient regime supporing both chaotic and regular trajectories, so that the Lyapunov exponent is non-zero, but full thermalization does not occur. Remarkably, in this region individual initial states with larger Lyapunov exponent tend to relax closer to the thermal state. There is also a region where although the Lyapunov exponent is zero, there is significant relaxation towards the thermal state. We conjecture that this redistribution in the phase-space is due to the quasi-regular dynamics governed by the nearby Ablowitz-Ladik lattice, which is integrable.
Above $\epsilon_T \simeq 0.6 J$, both the stochasticity threshold and thermalization threshold overlap very closely and appear to depend only on the nonlinearity strength. This suggests that there is a critical nonlinearity, $\kappa_c \simeq 0.2$, such that the system is regular for $\kappa < \kappa_c$ independent of the energy of the system. In the opposite limit of small energy and large nonlinearity, there is a separation of thresholds, where almost complete relaxation to the thermal distribution is observed in the absence of chaos.

An analysis of resonances of the BHM gives a Chirikov's criterion for the chaos threshold that depends on parameters which vanish in the thermodynamic limit.  This conclusion is confirmed on dimensional grounds.  The criterion predicted by the Chrikov criterion is different from the one inferred from numerical calculations, signifying the failure of the standard Chirikov's approach.

Anecdotal evidence suggests that the total quasi-momentum may play a role in the relaxation dynamics.  The quasi-momentum is strictly conserved in the nearby fully integrable Ablowitz-Ladik model, as well as in the non-interacting and continuum limits.  

There are at least three known near-by integrable models: the Ablowitz-Ladik lattice, the continuous nonlinear Schr\"odinger equation and the noninteracting model.  We outline the method of Inverse Scattering Transform and generate all of the integrals of motion of the closely related, fully integrable model of Ablowitz-Ladik.  We conjecture that the presense of quasi-conserved quantities may alter the scaling of the chaos criterion.

\section{Open Questions}
These observations lead to many questions that deserve further investigation.

\begin{itemize}
 \item What is the reason for the failure of the Chirikov criterion to accurately predict the chaos threshold?  Is it related to interference between resonances due to near-by integrable systems?  Could the proper scaling be recovered in a multiple-resonance model?
\item What is the underlying theory that governs the threshold?
\item For $\kappa \gtrsim 1$, how does the chaos threshold scale in the thermodynamic limit?  Is the number of modes involved relevant?
 \item What governs the slow relaxation times where they appear?  Is it related to the conserved quantities of the near-by integrable systems?  If this is the case, which near-by integrable system?
\item For those states that show no signs of thermalization, what governs the steady-state?  Can these states be described by a constrained ensemble?  Which \ldq quasi-conserved quantities'' are the relevant for the constrained ensemble?
 
\end{itemize}


\bibliographystyle{apsrmp}
\bibliography{amys_bibs/ablowitz-ladik.sub,amys_bibs/chaos_mfbh.sub,amys_bibs/interferometry.sub,amys_bibs/ultracold_atoms.sub}

\appendix

%

\chapter{Thermodynamic Distribution within Hartree-Fock}\label{app:hartree_fock}
\section{Hartree Fock}
In this section, the following integrals are used:
\begin{equation}\begin{split}
 \int_0^{\infty} dx e^{-\alpha x} &= \frac{1}{\alpha}\\
 \int_0^{\infty} dx x e^{-\alpha x} &= \frac{1}{\alpha^2}\\
\int_0^{\infty} dx x^2 e^{-\alpha x} &= \frac{2}{\alpha^3}
 \end{split}\end{equation}
Note that $N$ is the number of degrees of freedom and $a=L/N$ is the lattice spacing.

Within Hartree-Fock the form of the density distribution function is taken to be Gaussian and the thermal expectation value of the Grand Potential,
\begin{equation}
 \langle F\rangle=\langle H\rangle-T\langle S \rangle-\mu \langle N_a \rangle,
\end{equation}
is minimized, where $N_a$ is the norm. The density distribution function with two-body interactions has the form, 
\begin{equation}
 \sigma_{HF}=\frac{1}{Z}\exp\left( \sum_{n,n^\prime} - \alpha_{n,n^\prime} \psi_n\psi^\ast_{n^\prime} \right)=\frac{1}{Z}\exp\left(- \sum_n \alpha_n I_n \right).
\end{equation}
In the second step, we assume that off-diagonal terms are zero. The ${\alpha_n}$ coefficients are unknown and are determined by the condition of minimizing the grand potential. The partition function Z, normalizes $\sigma_{HF}$ so that the integration of $\sigma_{HF}$ over all of phase space is 1. Throughout the sums run from 1 to N, where N is the number of lattice sites.
\begin{equation}\begin{split}
 Z=&\frac{1}{(2\pi\hbar)^N}\int d^N \theta \int d^N I e^{-\sum_n \alpha_n I_n}\\
  =&\frac{1}{(2\pi\hbar)^N}\int_0^{2\pi}d \theta_1\int_0^{2\pi}d\theta_2\cdots\int_0^{2\pi}d \theta_N\int_0^\infty dI_1 \int_0^\infty dI_2\cdots \int_0^\infty dI_N e^{-\sum_n \alpha_n I_n}\\
  =& \frac{1}{\hbar^N} \int_0^\infty dI_1 e^{-\alpha_1I_1}\int_0^\infty dI_2 e^{-\alpha_2I_2}\cdots \int_0^\infty dI_N e^{-\alpha_N I_N}\\
  =& \prod_{i=1}^N \frac{1}{\hbar\alpha_i}
\end{split}\end{equation}
The expectation values of each term in the Grand Potential is calculated, using the Hartree-Fock density distribution function, $\sigma_{HF}$.  The expectation value of a generic observable is given by
\begin{equation}\begin{split}
 \langle O \rangle =&\frac{1}{(2\pi\hbar)^N}\int d^N \theta \int d^N I O(\lbrace I_n, \theta_n \rbrace) \sigma_{HF}\\ 
=& \frac{1}{(2\pi)^N}\prod_{i=1}^N \alpha_i\int d^N \theta \int d^N I O e^{ -\sum_n \alpha_n I_n}
\end{split}\end{equation}
The expectation values of the relevant observable are calculated below.

\begin{equation}\begin{split}
&\text{Norm}\\
\langle Norm\rangle=&\frac{1}{(2\pi\hbar)^N}\prod_{i=1}^N \hbar\alpha_i\int d^N \theta \int d^N I
\left[\frac{1}{\hbar}\sum_n  I_n \right] e^{ -\sum_m \alpha_m I_m }\\
=&\prod_{i=1}^N \alpha_i\sum_n \frac{1}{\hbar} \int_0^\infty dI_1 e^{-\alpha_1I_1}\cdots \int_0^\infty dI_n I_n e^{-\alpha_n I_n}\cdots\int_0^\infty dI_N e^{-\alpha_N I_N}\\
=&\sum_{n=1}^N \frac{1}{\hbar\alpha_n}\\
&\text{Hamiltonian - Kinetic Term}\\
\langle H_0\rangle=&\frac{1}{(2\pi\hbar)^N}\prod_{i=1}^N \hbar\alpha_i\int d^N \theta \int d^N I\left[\sum_n I_n\omega_n\right]  e^{-\sum_m \alpha_m I_m }\\
=&\sum_{n=1}^N\frac{\omega_n}{\alpha_n}\\
&\text{Hamiltonian - Interaction Term}\\
 \langle H_I \rangle =&\frac{1}{(2\pi\hbar)^N}\prod_{i=1}^N \hbar\alpha_i\int d^N \theta \int d^N I\\
&\left[\frac{\mu_0}{2\hbar^2}\sum_{m,p,q,r} \left(I_m I_p I_q I_{r}\right)^{1/2}\delta_{m+p,q+r} e^{-i(\theta_m + \theta_p - \theta_q -\theta_{r})}\right] e^{ -\sum_n \alpha_n I_n}\\
=&\frac{1}{(2\pi)^N}\prod_{i=1}^N \alpha_i (2\pi)^N \int d^N I \frac{\mu_0}{2\hbar^2}\left(\sum_{m} I_m^{2} + 2 \sum_{m \neq p} I_m I_p \right) e^{\left( -\sum_n \alpha_n I_n \right)}\\
=&\frac{\mu_0}{2\hbar^2}\prod_{i=1}^N \alpha_i \left( \prod_{j=1}^N \frac{1}{\alpha_j}\sum_{m} \frac{2}{\alpha_m^2} + 2 \sum_{m \neq p} \frac{1}{\alpha_m}\frac{1}{\alpha_p} \right)\\
=&\frac{\mu_0}{\hbar^2}\left(\sum_m \frac{1}{\alpha_m^2} + \sum_{m \neq p} \frac{1}{\alpha_m}\frac{1}{\alpha_p}\right)\\
=&\frac{\mu_0}{\hbar^2}\sum_{m,p} \frac{1}{\alpha_m}\frac{1}{\alpha_p} 
\end{split}
\end{equation}
\paragraph{Entropy}
\begin{equation}\begin{split}
 S=&-\frac{1}{(2\pi\hbar)^N}\int d^N \theta \int d^N \sigma_{HF}\log\sigma_{HF}\\
=&\frac{1}{(2\pi\hbar)^N}\int d^N \theta \int d^N \frac{1}{Z}e^{-\sum_n \alpha_n I_n} \left(-\sum_n\alpha_nI_n+\log Z\right)\\
=&\frac{1}{\hbar^N} \prod_{i=1}^N (\hbar \alpha_i)\int d^N e^{-\sum_n \alpha_n I_n} \left(-\sum_n\alpha_n I_n + \log \prod_{j=1}^N \frac{1}{\hbar \alpha_j}\right)\\
=& \sum_n\alpha_n \frac{1}{\alpha_n} - \sum_j\log(\hbar \alpha_j)\\
=& N - \sum_j\log(\hbar \alpha_j)
\end{split}\end{equation}

\section{Minimization of the Grand Potential}
The thermal expectation value of the Grand Potential within Hartree-Fock is given by 
\begin{equation}
\langle F\rangle=\sum_m \frac{\omega_m}{\alpha_m} +\frac{\mu_0}{\hbar^2} \sum_{m,p} \frac{1}{\alpha_m}\frac{1}{\alpha_p} - T\left( N - \sum_m \log(\hbar\alpha_m) \right) -\mu \sum_m \frac{1}{\hbar\alpha_m}
\end{equation}
Taking the variation with respect to $\alpha_n$, and setting it equal to zero gives
\begin{equation}
 \frac{\delta \langle F\rangle}{\delta \alpha_n}= -\frac{\omega_n}{\alpha_n^2} -2\frac{\mu_0}{\hbar^2} \frac{1}{\alpha_n^2}\sum_m\frac{1}{\alpha_m} + \frac{T}{\alpha_n} + \frac{\mu}{\hbar\alpha_n^2} = 0
\end{equation}
Using $\sum_m \alpha_m^{-1} = \hbar N_a $ and solving for $\alpha_n$,
\begin{equation}
\alpha_n = \frac{1}{\hbar T} \left[\hbar\omega_n+2\mu_0 N_a -\mu\right]
\end{equation}
The thermal expectation values of the occupation of momentum mode $n$ become
\begin{equation}
\langle I_n \rangle = \frac{1}{\alpha_n} = \frac{\hbar T}{\hbar\omega_n+2\mu_0 N_a -\mu}.
\end{equation}
In general, the coefficients $\mu$ and $T$ are unknown and are determined by imposing constraints on the norm and energy, which come from the dynamical code.  The constraints are 
\begin{equation}\begin{split}
N_a =& \langle N_a \rangle = \frac{1}{\hbar}\sum_n \langle I_n \rangle\\
E_T = & \langle H \rangle = \sum_n \omega_n\langle I_n \rangle + \frac{\mu_0}{\hbar^2}\sum_{m,n} \langle I_m \rangle \langle I_n \rangle = \sum_n \omega_n\langle I_n \rangle + \mu_0 N_a^2
\end{split}\end{equation}
Beginning with the expression for $\langle I_n \rangle$, we can solve for $T$ in terms of $\mu$, $N_a$ and energy.
\begin{equation}
T = \omega_n \langle I_n \rangle + 2\frac{\mu_0}{\hbar} N_a \langle I_n \rangle -\frac{\mu}{\hbar} \langle I_n \rangle
\end{equation}
Summing over $n$ 
\begin{equation}
T = \frac{1}{N}\left[E_{k} + 2\mu_0 N_a^2 -\mu N_a \right]
\end{equation}
where $E_k \equiv \sum_n \omega_n \langle I_n \rangle$. This expression for $T$ can be substituted back into the constraints to reduce the system to two equations with two unknowns.  Using the expression for temperature and normalization condition, a single constraint remains to be solved,
\begin{equation}
\frac{1}{N}\sum_n\frac{\left[E_k + 2\mu_0 N_a^2 -\mu N_a \right]}{\left[\hbar\omega_n + 2\mu_0 N_a -\mu \right]}- N_a  = 0
\end{equation}


\end{document}